\documentclass[12pt, a4paper, twoside, openright]{book}

\usepackage{vuwthesis} 

\usepackage{palatino} 

\usepackage{url} 

\usepackage[utf8]{inputenc}
\usepackage{amsmath}
\usepackage{amsfonts}
\usepackage{amssymb}
\usepackage{graphicx}
\usepackage{multicol}
\usepackage[hidelinks]{hyperref}
\usepackage{float}
\usepackage{subcaption}
\usepackage{cite}
\usepackage{amssymb}
\usepackage{xcolor}
\usepackage{array}
\usepackage{wrapfig}
\usepackage{multirow}
\usepackage{tabularx}
\usepackage{enumitem}

\newcommand{\R}{\mathbb{R}}
\newcommand{\M}{\mathcal{M}}
\newcommand{\pd}{\partial}
\newcommand{\n}{\nabla}
\newcommand{\be}{\begin{equation}}
\newcommand{\ee}{\end{equation}}
\newcommand{\la}{\mathcal{L}}
\newcommand{\OO}{\mathcal{O}}
\newcommand{\Ronumb}[1]{\MakeUppercase{\romannumeral #1}}
\newcommand{\specialcell}[2][c]{%
\begin{tabular}[#1]{@{}c@{}}#2\end{tabular}}
\newcommand{\C}{\mathcal{C}}

\numberwithin{equation}{section}

\begin{document}

\frontmatter


\title{Unit-Lapse Forms of Various Spacetimes}
\author{Joshua Baines}

\subject{Mathematics}
\abstract{Every spacetime is defined by its metric, the mathematical object which further defines the spacetime curvature. From the relativity principle, we have the freedom to choose which coordinate system to write our metric in. Some coordinate systems, however, are better than others. In this text, we begin with a brief introduction into general relativity, Einstein's masterpiece theory of gravity. We then discuss some physically interesting spacetimes and the coordinate systems that the metrics of these spacetimes can be expressed in. More specifically, we discuss the existence of the rather useful unit-lapse forms of these spacetimes. Using the metric written in this form then allows us to conduct further analysis of these spacetimes, which we discuss.\\

Overall, the work given in this text has many interesting mathematical and physical applications. Firstly, unit-lapse spacetimes are quite common and occur rather naturally for many specific analogue spacetimes. In an astrophysical context, unit-lapse forms of stationary spacetimes are rather useful since they allow for very simple and immediate calculation of a large class of timelike geodesics, the rain geodesics. Physically these geodesics represent zero angular momentum observers (ZAMOs), with zero initial velocity that are dropped from spatial infinity and are a rather tractable probe of the physics occurring in the spacetime. Mathematically, improved coordinate systems of the Kerr spacetime are rather important since they give a better understanding of the rather complicated and challenging Kerr spacetime. These improved coordinate systems, for example, can be applied to the attempts at finding a ``Gordon form" of the Kerr spacetime and can also be applied to attempts at upgrading the ``Newman-Janis trick" from an ansatz to a full algorithm. Also, these new forms of the Kerr metric allows for a greater observational ability to differentiate exact Kerr black holes from ``black hole mimickers". }

\mscthesisonly


\maketitle

\chapter*{Acknowledgments}\label{C:ack} 

I would like to firstly thank my supervisor, Professor Matt Visser. Your guidance, advice and oversight has been greatly appreciated. From talks I have had with my friends who have also completed work with a supervisor, I realise how lucky I am. How lucky I am to have had a supervisor who invested so much time and effort into helping me, whether it be by constantly giving me pages upon pages of hand typed notes to guide me, by having constant and regular meetings, by answering all my questions you can, by proofreading my work in a timely manner and returning very useful feedback. All of these things have been crucial to my success and to the development of not only this thesis but also to the development of myself as an academic. It is easy to take all of these actions for granted, but in the grand scheme of things, I realise how lucky I was. Thank you Matt, thank you for all of your time and effort that you pour into helping all of your students. I for one, greatly appreciate it and hope that my work here and future work reflects that help that you have given me. 

I would like to thank my mum, Sharron. Thank you for all the support you have given me, not only over the last year, but over my entire life. You have always supported me and encouraged me to succeed. You always celebrate my successes and are always there for support when I stumble or fail. Especially over the last year where my life was upturned and my future was uncertain for many different reasons, you did everything you could to help and support me. Through everything, you have supported me and encouraged me to do my best, even if you haven't the slightest idea what I am actually doing, you still help motivate me to do my best. Thank you for everything mum!

To my dad John and step-mother Steph, thank you for your support. You have always been there when I call for aid, and I know you will always be there to help. This has been a great source of relief and support over this last year. You are deeply invested in how I am doing, you are proud of my successes, you are always supportive of my decisions and are always more than happy to lend a hand whenever possible. I greatly appreciate and am thankful for everything you do for me. Thank you John and Steph.

Thank you to my sister Michelle, like with mum, you are always there to celebrate my successes and there when I need help or a different point of view on things. You help me see things realistically and as they really are, you help me get out of my own head when I am stuck on a particularly difficult issue. All of your help hasn't gone unnoticed, thank you Michelle. 

To my physics pal, slightly turned engineer, Martin Markwitz, thank you for your friendship. I have greatly enjoyed our physics talks, even if you think black holes have no interior and hence talk of anything inside a black hole has no physical relevance. (Martin, the geodesics of Schwarzschild do not terminate at the event horizon when the metric is expressed in Painlev\'{e}-Gullstrand coordinates! But I digress). Our talks have helped me understand the fundamentals of general relativity better and helped contribute to some of my explanations in this thesis. Thanks Martin.


To my colleges at space place at Carter observatory, you all have been a great source of joy, not only over the last year, but the entire time we have all worked together. Even if some of you recoil in disgust at the sight of my work, you are all supportive and interested in what I am doing. Whether it be by asking me for some spicy space facts, asking me interesting space questions or just your supportive comments, general friendship and great space place memes, I appreciate every single one of you. 

To the other members of the unofficial Victoria university of Wellington gravity squad, Thomas and Alex, thanks for your input and help over the last year. It has been a pleasure working and writing papers with you. You have served me as guides to follow, examples of great aspiring academics that I am happy to have worked alongside. Thank you both. 

Lastly, I would like to thank everyone who have helped me but who I haven't mentioned. Unfortunately, I have a word limit for my thesis, hence I cannot thank every single person and also this section is already getting long enough! But to everyone else who have helped me along the way, you all know who you are, thank you.

I would also like to acknowledge that I was supported by a MSc scholarship funded by the Marsden Fund, via a grant administered by the Royal Society of New Zealand.

Thank you to every person who has helped me over the last year, I appreciate it more than I let on and I cannot thank you all enough.

\addtocontents{toc}{\protect\enlargethispage{\baselineskip}}

\tableofcontents


\mainmatter



\chapter{Introduction}\label{C:intro}

\section{Why use the concept of a curved spacetime?} \label{intro}
Pre-relativity we believed that we lived a 3-dimensional Euclidean space that evolved with time, that is to say: a 3-dimensional flat space where time and space were viewed as separate quantities entirely. Not only that, but we also believed that many quantities we measured were observer independent such as distance and acceleration. For these quantities, what one person measured would be exactly the same as anyone else's measurement. Velocities were relative however, but the Galilean velocity transformation law was very trivial (and also not an accurate description of physical reality). However, Einstein was able to use electromagnetism to show that some of these quantities (such as distance and acceleration) are observer dependent, each observer will measure a quantity relative to their frame of reference, but these different values are equivalent; if transformed correctly between frames.  Thus, we find that there exists no preferred frame, no preferred observer. Around the same time, Hermann Minkowski proposed the concept of spacetime. Instead of viewing space and time to be separate objects, we view them to be both part of a larger object, spacetime. A 4-dimensional Euclidean (for now) space, with 3 spatial dimensions and one temporal, where time and space are now on, more or less, equal footing. These ideas are the basis of special relativity.

However, it turns out that that there do exist some observer independent quantities in special relativity, the most important being the spacetime interval between two events in spacetime, defined as follows:
\begin{equation} \label{flat_line_element}
I=-(\Delta t)^2+(\Delta x)^2+(\Delta y)^2+(\Delta z)^2 \quad .
\end{equation}
We notice that this looks like the 3-dimensional form of Pythagoras' theorem if we neglect the first term on the right hand side of the equation. Indeed this is the line element which defines the metric of flat spacetime (the idea of a metric will be expanded on later). Thus, equation \eqref{flat_line_element} gives the squared `distance' between two separated events in flat spacetime. Since equation \eqref{flat_line_element} is observer independent, this quantity must be a fundamental property of the spacetime itself. Hence, we can use the structure of spacetime itself to define quantities that can be measured by observers in that spacetime. This is one of the ideas that led Einstein to formulate general relativity, using the structure of a curved spacetime to define gravitational fields in that spacetime. 

So, we can use fundamental properties of a spacetime to define measurable quantities in our spacetime, but how does considering a curved spacetime correlate with gravity? This connection comes from \emph{the Equivalence Principle}. The Equivalence principle can be stated as follows: \emph{"one cannot distinguish between gravitational and inertial forces"}. The building blocks for this principle can even be seen in the Newtonian theory of gravity, which states that the gravitational force an object experiences is proportional to its inertial mass, this is called the universality of free fall. Einstein then took this idea and the relativity principle to form the equivalence principle. We can visualise this principle through the following thought experiment: if we were in a small, closed box with no windows, we wouldn't be able to tell if we were being accelerated towards the `floor' of the box due to the gravity of a massive object or if we were in a vacuum and experiencing acceleration due to rocket boosters on the box which accelerates us at the same rate that the gravitational field of the object would. Hence, these two physical systems are \emph{equivalent}. Now the interesting part comes when we consider the propagation of light in these two systems. 

Consider we are in a box being accelerated due to rocket boosters, furthermore, imagine we have a laser in this box. We turn on the laser and we find that the path that the light takes, as viewed by us in our accelerating box, is a curved line. Hence, in a gravitational field, light must also follow curved lines. However, from Fermat's principle, we know that the path that light follows between two points is the path of least time connecting those two points. In Euclidean space, the path of least time between any two points is a straight line. But, in a gravitational field, we see that light does not follow a ``straight line" as observed by us in the box. Hence, we must conclude that space (and hence spacetime) is curved in regions where observers experience a gravitational field. 

The ideas that spacetime is curved in regions where observers experience a gravitational field and that we can define measurable quantities via intrinsic properties of the spacetime itself are the core tenets of general relativity. We use the properties of a curved spacetime, namely the metric, to define `gravitational fields' on our spacetime. More specifically, the curvature of spacetime causes the paths that particles travel along (geodesics) to be curved, which is how we view particles to move within a gravitational field. 

We now focus on the mathematical framework required to properly express these ideas. This framework is the topic of differential geometry. 
\chapter{Fundamentals}\label{C:Fundamentals}

\section{Spacetime as a 4-dimensional manifold}

Before we can construct the interesting quantities that allow us to define gravitational fields, we must begin by giving a precise notion of what we mean by \emph{'spacetime'}. We start by defining a topological space.\\

\noindent \textbf{Definition 2.1.1} (Topological Space)

\noindent A topological space $(\mathcal{E}, \mathcal{T})$ is a set $\mathcal{E}$ with a collection $\mathcal{T}$ of open subsets of $\mathcal{E}$ (called the topology on $\mathcal{E}$) such that:

\begin{enumerate}
\item The union of an arbitrary number of subsets in $\mathcal{T}$, is in $\mathcal{T}$. i.e. If $O_{\alpha} \in \mathcal{T}$ for all $\alpha$, then $\bigcup\limits_{\alpha} O_{\alpha} \in \mathcal{T}$.

\item The intersection of a finite number of subsets in $\mathcal{T}$ is in $\mathcal{T}$. i.e. If $O_1, O_2,...,O_n \in \mathcal{T}$, then $\bigcap\limits_{i=1}^{n} O_{i} \in \mathcal{T}$.

\item The set $\mathcal{E}$ and the empty set $\emptyset$ is in $\mathcal{T}$.
\end{enumerate}

\noindent Next we define a Hausdorff topological space.\\

\newpage

\noindent \textbf{Definition 2.1.2} (Hausdorff)

\noindent A topological space $(\mathcal{E}, \mathcal{T})$ is Hausdorff if for every pair of distinct points $p, q \in \mathcal{E}$, $p \neq q$, we can find open sets $O_p, O_q \in \mathcal{T}$, such that: $p \in O_p$, $q \in O_q$ and $O_p \cap O_q = \emptyset$. That is to say, for every distinct pair of points $p, q \in \mathcal{E}$, there exists two open sets each containing one (and only one) of the points which do not overlap.\\

\noindent We now define the notion of a locally Euclidean space.\\

\noindent \textbf{Definition 2.1.3} (Locally Euclidean Space)

\noindent A topological space $(\mathcal{E}, \mathcal{T})$ is locally Euclidean if the following condition is satisfied: $\forall x \in \mathcal{E}$, $\exists O \in \mathcal{T}$ and $\exists n \in \mathbb{Z}^{+}$ such that: $x \in O$, $\exists X \subset \mathbb{R}^{n}$ and $\exists$ homeomorphism $f:O \leftrightarrow X$. That is to say, for every point in $\mathcal{E}$ there exists an open neighbourhood around it which can be mapped 1 to 1 and bi-continuously to a subset of $\mathbb{R}^n$, i.e. there exists a region surrounding each point in $\mathcal{E}$ that `looks like' a segment of an n-dimensional Euclidean space.\\

\noindent We are now very close to being able to properly define a manifold, but we must first define charts, atlases and the notion of a connected topological space.\\

\noindent \textbf{Definition 2.1.4} (Chart)

\noindent A chart $(O, f, U)$ on an open subset $O \in \mathcal{T}$ is a set $U \in \mathbb{R}^n$, with a homeomorphism $f:O \leftrightarrow U=f(O)$.\\

\noindent Chart is a mathematical term but physicists typically call these objects a \emph{coordinate system}. Charts are simply maps between open sets $O \in \mathcal{T}$ and subsets $U$ of $\mathbb{R}^n$, in this way we can identify $O$ as a segment of $\mathbb{R}^n$.\\

\newpage

\noindent \textbf{Definition 2.1.5} (Atlas)

\noindent An atlas is a collection of charts that covers the entire locally Euclidean space $\mathcal{E}$.\\

\begin{figure}
\centering
\includegraphics[width=0.71\textwidth]{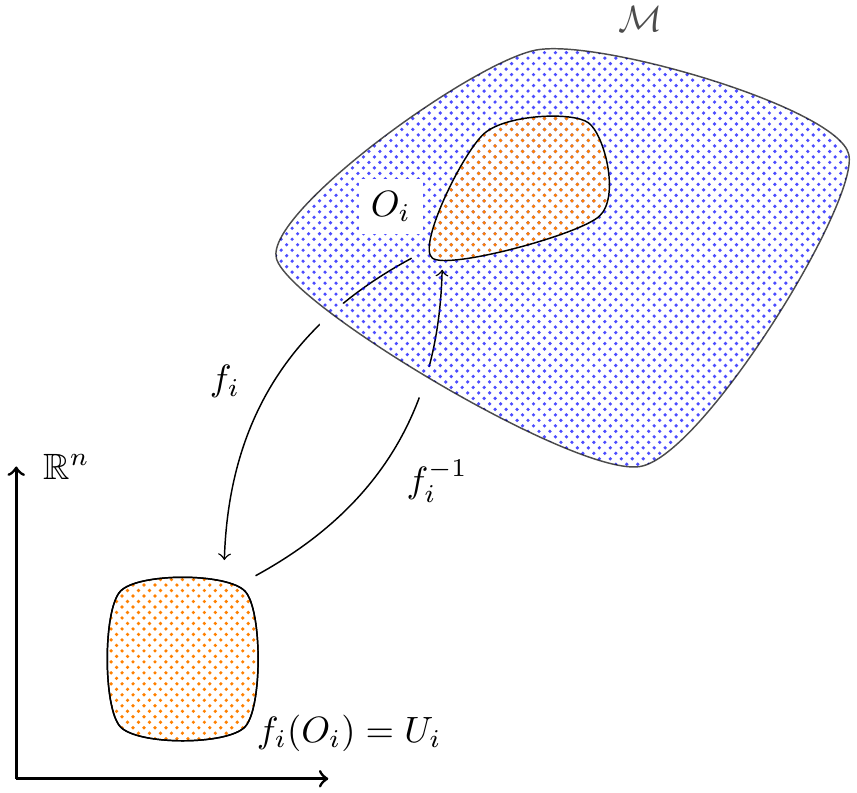}
\caption{The map $f_i$ maps an open subset $O_i \in \mathcal{T}$ to an open subset $U_i \subset \mathbb{R}^n$.}
\end{figure}

\noindent \textbf{Definition 2.1.6} (Connected Topological Space)

\noindent A topological space $(\mathcal{E}, \mathcal{T})$ is connected iff $\mathcal{E}$ and $\emptyset$ are the only sets in $\mathcal{T}$ that are both open and closed. \\

\enlargethispage{20pt}

\noindent Alternatively, a connected space is a space that is not the union of two or more disjoint open spaces. We shall assume that spacetime is connected, this is because if spacetime were disconnected, then the only segment of spacetime that we would be interested in is the segment we live in. Hence, we can disregard the other segments, then we are left with a connected spacetime. \\

\noindent We are now ready to define a manifold.\\

\noindent \textbf{Definition 2.1.6} (Manifold)

\noindent A manifold $\mathcal{M}$ is a locally Euclidean space such that:

\begin{enumerate}
\item $\mathcal{M}$ is connected. 

\item $\mathcal{M}$ has the same dimension everywhere.

\item $\mathcal{M}$ is Hausdorff.

\item $\mathcal{M}$ has at least one countable atlas.
\end{enumerate}

\noindent By \emph{countable atlas}, we mean that every chart in the atlas can be put into a 1 to 1 correspondence with the set of natural numbers, $\mathbb{N}$.

We can now see that spacetime can be defined as a 4-dimensional manifold. This structure of spacetime allows us to define vectors, tensors and later the notion of curvature on the spacetime itself, which is paramount to the formulation of general relativity. 

\section{Vectors, dual vectors and tensors}

Let $\mathcal{M}$ be a manifold of dimension $n$. We define a curve $h(\lambda)$ on the manifold $\mathcal{M}$ to be a map $h:\mathbb{R} \rightarrow \mathcal{M}$. Notice, $h(\lambda)$ is a parametric curve, parametrised by $\lambda$, i.e. different values of $\lambda$ represent different points along the curve $h$. Given a chart $(O, f, U)$ we can construct the map $f \circ h(\lambda): \mathbb{R} \rightarrow \mathbb{R}^n$, such that $x^a(\lambda)=f \circ h(\lambda)=f(h(\lambda)) \in f(U) \subset \mathbb{R}^n$. The upper index $a$ in $x^a(\lambda)$ denotes the various components of $x^a(\lambda)$ and ranges $a \in \mathbb{N}: 0 \leq a \leq n$ (since $x^a(\lambda)$ is simply just a vector in $\mathbb{R}^n$). The object $x^a(\lambda)$ can be seen as the `collection of coordinates corresponding to points along the curve $h(\lambda)$'.

\begin{figure}[H]
\centering
\includegraphics[width=0.78\textwidth]{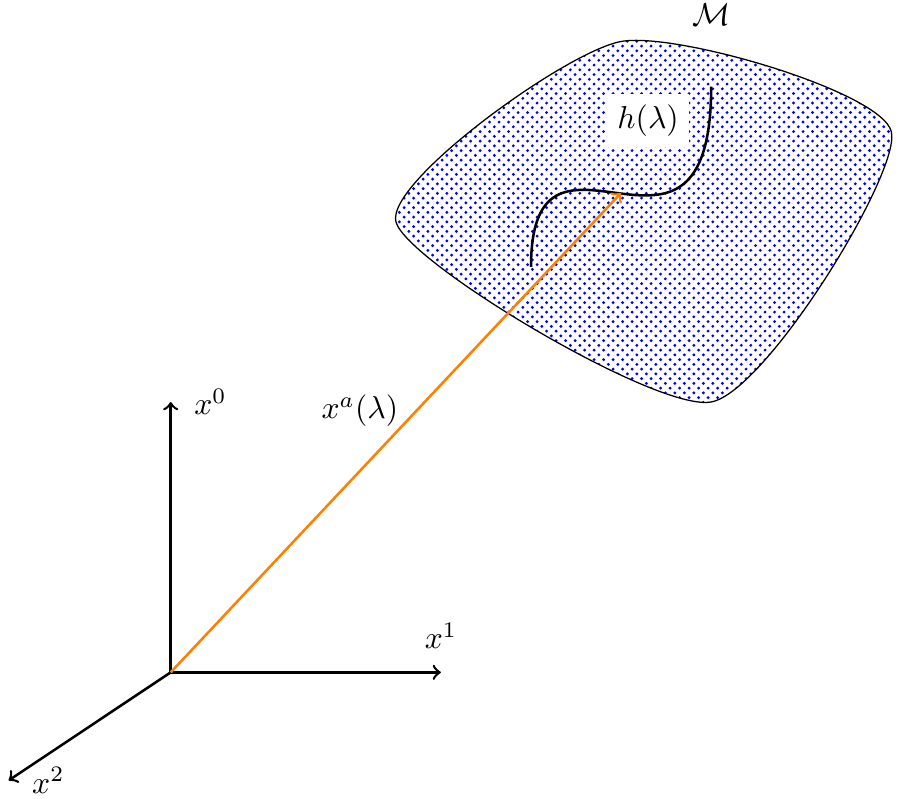}
\caption{Given a coordinate system, the object $x^a(\lambda)$ can be seen as the `collection of coordinates corresponding to points along the curve $h(\lambda)$'.}
\end{figure}

\noindent Consequently, the tangent vector of $x^a(\lambda)$ is denoted as:
\begin{equation}
T^a=\frac{dx^a}{d\lambda} \quad .
\end{equation}

Our choice of parametrisation is arbitrary, we may choose to reparametrise our curve $h(\lambda) \rightarrow h(\bar{\lambda})$. If we do so, then by the chain rule we find:
\begin{equation}
\bar{T}^a=\frac{dx^a}{d\bar{\lambda}}=\frac{d\lambda}{d\bar{\lambda}} \frac{dx^a}{d\lambda}=\frac{d\lambda}{d\bar{\lambda}}\, T^a \quad .
\end{equation}
If we now wish to change our coordinate patch $U$ to $\bar{U}$ (via $\bar{f} \circ f^{-1} : f^{-1}(U) \rightarrow \bar{f}(U)$), then via the multi-variable chain rule, we find:
\begin{equation} \label{vec_trans}
\bar{T}^a=\frac{d\bar{x}^a}{d\lambda}=\sum_{b=1}^n \frac{\partial \bar{x}^a}{\partial x^b} \frac{dx^b}{d\lambda} = \sum_{b=1}^n \frac{\partial \bar{x}^a}{\partial x^b}\, T^b \quad . 
\end{equation}
Hence, once given a vector in one particular coordinate system, we can transform it to any other coordinate system (this is similar to transforming vectors under a change in basis). We can also define vectors via equation \eqref{vec_trans}, any mathematical object that transforms via equation \eqref{vec_trans} under a change in coordinates, is a vector.

Tangent vectors are defined at one point $p \in \mathcal{M}$, the set of all tangent vectors at the point $p$ called the tangent space at $p$ is denoted as $V_p$.

We can also make the correspondence between vectors and directional derivatives. Let $\mathcal{F}$ denote the set of all $C^{\infty}$ functions \label{F_def} (i.e. continuously differentiable functions) from $\mathcal{M}$ to $\mathbb{R}$. We can define the vector $v$ at point $p \in \mathcal{M}$ to be a map $v : \mathcal{F} \rightarrow \mathbb{R}$ which satisfy the following properties:

\begin{enumerate}
\item Linearity: $v(af+bg)=av(f)+bv(g)$ for all $f, g \in \mathcal{F}$ and $a, b \in \mathbb{R}$.

\item Leibnitz rule: $v(fg)=f(p)v(g)+g(p)v(f)$ for all $f, g \in \mathcal{F}$.
\end{enumerate}
Hence, we can write $v=\sum_{a=1}^n v^a\partial_a$ \label{direc_deriv_vec} (where $\partial_a \equiv \partial / \partial x^a$). Furthermore, the vectors $\partial_a$ form a basis of $V_p$, therefore we see that $V_p$ has the structure of a vector space via the conditions stated above. Notice that since $\partial_a$ are basis vectors, we denote their components with subscripts.\\

\noindent For every vector space $V$, there exists a dual vector space $V^*$ whose elements are maps from $V$ into $\mathbb{R}$, i.e. for $v^a \in V$ and $w_a \in V^*$, $w_a v^a=w(v) \in \mathbb{R}$. We now look at a couple simple examples of dual vectors. For $\mathbb{R}^n$ the dual to a column vector is a row vector, via matrix multiplication, the product of a column vector and a row vector produces a scalar. In quantum mechanics vectors in a Hilbert space define the possible states of our physical system, we typically denote these vectors by kets $|\alpha \rangle$ where $\alpha$ is some parameter which defines that possible state. The corresponding dual vector to this ket is a bra $\langle \alpha|$, these objects are the foundations of bra-ket notation (or bracket notation, to this day we have no idea where the c went...). We notice that for each vector there is a corresponding dual vector, in fact there exists an isomorphism between any vector space $V$ and its corresponding dual vector space $V^*$, for example in $\mathbb{R}^n$ we can transform a column vector to a row vector (and vice-versa) via transposition and for bra-ket notation, we can transform a ket into a bra (and vice-versa) by taking the Hermitian conjugate. 

Taking the tangent space at point $p \in \mathcal{M}$, $V_p$, we can construct the dual vector space, ${V_p}^*$, whose elements are denoted as $w_a$ where the lower index $a$ again denotes the various components of $w_a$ and ranges $a \in \mathbb{N}: 0 \leq a \leq n$. Using upper and lower indices allow us to differentiate between vectors and dual vectors (note: the isomorphism between $V_p$ and $V_p^*$ will be shown in due time). We now give more precise definitions of dual vectors in the dual vector space ${V_p}^*$. Let $\phi(p)$ be a map from $\mathcal{M}$ into $\mathbb{R}$. Then given a chart $(O, f, U)$ we can write $\phi(x^a)$ which means $\phi \circ f^{-1} : f(O)=U \subseteq \mathbb{R}^n \rightarrow \mathbb{R}$. Then we can write the components of the dual vector $w_a$ as:
\begin{equation}
w_a = \frac{\partial \phi}{\partial x^a} \quad .
\end{equation}
If we now wish to change our coordinate patch $U$ to $\bar{U}$ (via $\bar{f} \circ f^{-1} : f^{-1}(U) \rightarrow \bar{f}(U)$), then via the multi-variable chain rule, we find:
\begin{equation} \label{dvec_trans}
\bar{w}_a=\frac{\pd \phi}{\pd \bar{x}^a}=\sum_{b=1}^n \frac{\pd x^b}{\partial \bar{x}^a} \frac{\pd \phi}{\pd x_b} = \sum_{b=1}^n \frac{\pd x^b}{\partial \bar{x}^a} w_b \quad . 
\end{equation}
Hence, once given a dual vector in one particular coordinate system, we can transform it to any other coordinate system. Similar to before we can also define dual vectors via equation \eqref{dvec_trans}, any mathematical object that transforms via equation \eqref{dvec_trans} under a change in coordinates, is a dual vector.\\

\enlargethispage{20pt}

\noindent Now that we have introduced vectors and dual vectors we can now consider maps on vectors and dual vectors, more specifically we can now define tensors. Let $V$ be a finite dimensional vector space and let $V^*$ be its corresponding dual vector space. Then a tensor of type $(k, l)$ is a multilinear map
\begin{equation*}
T : \underbrace{V^* \times ... \times V^*}_{k} \times \underbrace{V \times ... \times V}_{l} \rightarrow \R \quad .
\end{equation*}
That is, given $l$ vectors and $k$ dual vectors, $T$ maps these into real numbers and if we fix all but one vector or dual vector, $T$ is a linear map on this variable. We will denote a tensor of type $(k, l)$ as ${T^{a_1 a_2...a_k}}_{b_1 b_2...b_l}$ here (as before) the indices denote the various components of the tensor and range from $0$ to $n$ (the dimension of our vector space). Tensors can be viewed as a generalisation of scalars, vectors and matrices. A tensor of type $(0, 0)$ is a scalar, a tensor of type $(1, 0)$ is a vector, a tensor of type $(0,1)$ is a dual vector and a tensor of type $(1, 1)$, $(2,0)$ and $(0,2)$ are matrices (note: while these all represent matrices, they all transform differently under a change in coordinate basis as we shall see). Similar to vectors and matrices, we can define some operations on tensors. Firstly, outer products, given two tensors say $T$ of type $(k, l)$ and $T'$ of type $(k', l')$ we can construct a new tensor $S$ via the outer product $S=T \otimes T'$ denoted as:
\begin{equation}
{S^{a_1...a_{k+k'}}}_{b_1...b_{l+l'}}={T^{a_1 a_2...a_k}}_{b_1 b_2...b_l}{T'^{a_{k+1} a_{k+2}...a_{k+k'}}}_{b_{l+1} b_{l+2}...b_{l+l'}} \quad . 
\end{equation}
The second operation we can perform is contraction. Let $\mathcal{T} (k, l)$ denote the set of all tensors of type $(k, l)$, then contraction is the map $C : \mathcal{T} (k, l) \rightarrow \mathcal{T} (k-1, l-1)$ defined as
\begin{equation}
CT={T^{a_1...a_{k-1}}}_{b_1...b_{l-1}}=\sum_{i=0}^n {T^{a_1... i ...a_k}}_{b_1... i ...b_l}
\end{equation}
here we choose one upper index and one lower index (not necessarily in the same index location, i.e. we can contract the $i$th upper index with the $j$th lower index for all $0 \leq i \leq k$ and $0 \leq j \leq l$) then we sum over all tensors with their corresponding indices evaluated as component $i$. To get a better understanding of this, we look at the simple example where $T$ is just a matrix, a tensor of type $(1,1)$ ${T^a}_b$, if we contract over $a$ and $b$ we get 
\begin{equation} \label{mat_tr}
CT={T^a}_a= \sum_{i=0}^n {T^i}_i = {T^0}_0+{T^1}_1+...+{T^n}_n \equiv \text{Tr}[T] \quad . 
\end{equation}
Hence, we see that contraction on this tensor is the same as taking the trace of the matrix. So, we can generalise this to say that contraction on a general tensor ${T^{a_1 a_2...a_k}}_{b_1 b_2...b_l}$ is similar to taking the trace over some upper index $a_i$ and some lower index $b_j$. Notice in equation \eqref{mat_tr} we have used the notation ${T^a}_a= \sum_{i=0}^n {T^i}_i$, this is known as the Einstein summation convention. When we see the same symbol in the top index and lower index of a tensor (e.g.  ${T^a}_a$) or tensor product (e.g. $w_av^a$), it is assumed that we are summing over that variable.

If we now wish to change our coordinate patch $U$ to $\bar{U}$ (via $\bar{f} \circ f^{-1} : f^{-1}(U) \rightarrow \bar{f}(U)$), then we know that vectors transform as 
\begin{equation*}
\bar{v}^a=\frac{\partial \bar{x}^a}{\partial x^b} v^b 
\end{equation*}
while dual vectors transform as
\begin{equation*}
\bar{w}_a=\frac{\pd x^b}{\partial \bar{x}^a} w_b \quad . 
\end{equation*}
hence, we can ``bootstrap" this to tensors of type $(k, l)$ as
\begin{equation} \label{ten_trans}
\bar{T}^{a_1...a_k}{}_{b_1...b_l}={T^{a'_1...a'_k}}_{b'_1...b'_l}\frac{\pd \bar{x}^{a_1}}{\pd x^{a'_1}}...\frac{\pd \bar{x}^{a_k}}{\pd x^{a'_k}} \frac{\pd x^{b'_1}}{\pd \bar{x}^{b_1}}...\frac{\pd x^{b'_l}}{\pd \bar{x}^{b_l}} \quad . 
\end{equation}
Like with vectors and dual vectors, equation \eqref{ten_trans} can be used as the definition of a tensor, i.e. any mathematical object that transforms via equation \eqref{ten_trans} under a change in coordinates, is a tensor.  

Much like with matrices, we can think of tensors as being symmetric or anti-symmetric. As a reminder, a symmetric matrix satisfies the property $M_{ij}=M_{ji}$ while an anti-symmetric matrix satisfies the property $M_{ij}=-M_{ji}$. For simplicity let us first consider a tensor of type $(0, 2)$, we can define symmetric parts and anti-symmetric parts of the tensor as follows, for the symmetric part
\begin{equation} \label{sym02}
T_{(ab)}=\frac{1}{2}(T_{ab}+T_{ba})
\end{equation}
while for the anti-symmetric part
\begin{equation} \label{asym02}
T_{[ab]}=\frac{1}{2}(T_{ab}-T_{ba}) \quad . 
\end{equation}
A totally symmetric tensor of type $(0, 2)$ satisfies the property $T_{(ab)}=T_{ab}$ or equivalently $T_{[ab]}=0$ and via equation \eqref{asym02} $T_{ab}=T_{ba}$, while a totally anti-symmetric tensor of type $(0, 2)$ satisfies the property $T_{[ab]}=T_{ab}$ or equivalently $T_{(ab)}=0$ and via equation \eqref{sym02} $T_{ab}=-T_{ba}$ as stated above. More generally, for a tensor of type $(0, l)$ we have for the symmetric part
\begin{equation}
T_{(a_1...a_l)}=\frac{1}{l!}\sum_{\pi} T_{a_{\pi (1)}...a_{\pi (l)}}
\end{equation}
while for the anti-symmetric part
\begin{equation}
T_{[a_1...a_l]}=\frac{1}{l!}\sum_{\pi} \delta_{\pi} T_{a_{\pi (1)}...a_{\pi (l)}} 
\end{equation}
where we are summing over all permutations ($\pi$) of $1,...,l$ and $\delta_{\pi}$ is 1 for every even permutation and $-1$ for every odd permutation.\\

\noindent We are now ready to define arguably the most important tensor in general relativity, the metric tensor. The mathematical definition of a metric and the physicists' definition of a metric (which we will use here) differ slightly. \\

\noindent In mathematics a metric is a function $d : X \times X \rightarrow \R^{\geq 0}$ which satisfies the following conditions:
\begin{enumerate}
\item $d(x, y)=0 \Longleftrightarrow x=y$

\item $d(x, y)=d(y, x)$

\item $d(x, y) \leq d(x, z) + d(z, y)$
\end{enumerate}
we see that this implies that $d(x, y) \geq 0$ for all $x, y \in X$. However, in general relativity we consider metrics where the metric can be less than zero and hence the first and third condition stated above doesn't necessarily hold for all $x, y \in X$, however the second condition, the symmetry condition, still holds. Physically, the metric defines the distance between any two events in our curved spacetime. If we consider a Euclidean 3-space (i.e. no time) via Pythagoras' theorem we have
\begin{equation}
ds^2=dx^2+dy^2+dz^2=\delta_{ab}\, dX^adX^b
\end{equation}
where $dX^a=(dx, dy, dz)^T$ and written as an array
\begin{equation}
\delta_{ab}=
\begin{bmatrix}
1 & 0 & 0\\
0 & 1 & 0\\
0 & 0 & 1
\end{bmatrix}_{ab} \quad .
\end{equation}
If we consider a Euclidean spacetime, via equation \eqref{flat_line_element} we have 
\begin{equation}
ds^2=-dt^2+dx^2+dy^2+dz^2=g_{ab}\, dX^adX^b
\end{equation}
where $dX^a=(dt, dx, dy, dz)^T$ and written as an array
\begin{equation}
g_{ab}=
\begin{bmatrix}
-1 & 0 & 0 & 0\\
0 & 1 & 0 & 0\\
0 & 0 & 1 & 0\\
0 & 0 & 0 & 1
\end{bmatrix}_{ab} \quad .
\end{equation}
The type $(0, 2)$ and symmetric tensor $g_{ab}$ is called the metric tensor and is defined by the equation
\begin{equation}
ds^2=g_{ab}\, dX^adX^b
\end{equation}
and in the examples above we see once given the infinitesimal line element of any spacetime, we can ``read off" the components of the metric tensor. Once we have a metric, we can define the inverse metric denoted as $g^{ab}$, where $g_{ab}\, g^{ab}=n$ (where $n$ is the dimension of the manifold) and $g_{ab}\, g^{bc}=\delta_a{}^c$ (where $\delta_a{}^c$ is the Kronecker delta) . We can calculate the inverse metric by writing the metric as an array and then finding the matrix inverse. 

The metric not only gives us the infinitesimal squared distance between any two events in our spacetime, but also is the isomorphism that takes vectors in our tangent space $V_p$ to dual vectors in our dual tangent space $V_p^*$, and vice-versa, via tensor product. For example, we can transform a vector into a dual vector via $v_a=g_{ab}\, v^b$ and we can transform a dual vector into a vector via $w^a=g^{ab}\, w_b$. Furthermore, we can use the metric tensor to raise and lower indices of tensors of arbitrary type. For example ${T^{ab}}_{c}=g^{ad}\, g_{ce}\, {T_d}^{be}$.

\section{Curvature}

We now begin to discuss the notion of curvature in our spacetime. We normally find the curvature of a line or surface by embedding it in a higher dimensional space. For example, we can view the curvature of a 2-dimensional surface by embedding it in a 3-dimensional space, this is the way that curvature is found in usual multi-variable calculus. However in general relativity, our spacetimes are usually not embedded within a higher dimensional space. We could take that route if we wanted to, but this would prove to be more complicated than necessary since we would have to construct more mathematical objects in higher dimensions (that are not physical objects nor have any known physical relevance). Hence, we define curvature in our spacetime by constructing mathematical objects within the spacetime itself. This is an interesting problem, how can observers in a curved space measure the curvature of that space without relying on higher dimensions? We define curvature by looking at how vectors transform when parallel propagated along curves within our spacetime.

To see how parallel propagated vectors can define curvature, imagine a flat 2-dimensional plane with a circle on its surface. If we have a tangent vector to the circle at any point and parallel propagate it around the circle (this means moving the vector without changing its direction, we can think of this as `picking up' the `base' of the vector and moving it along the curve), then we find that when the vector returns to its original position, it is pointing in the same direction, i.e. the vector is unchanged when parallel propagated around the curve. 

\begin{figure}[H]
    \centering
    \begin{subfigure}[b]{0.37\textwidth}
        \includegraphics[width=\textwidth]{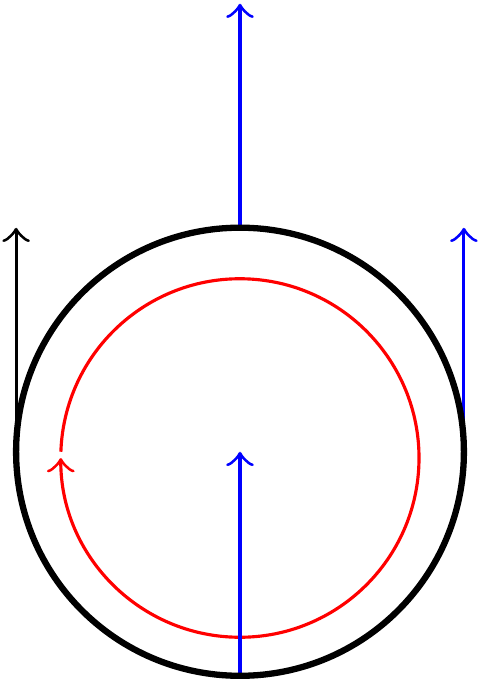}
        \caption{Parallel propagation on a flat surface}
        \label{fig:Flat_Para_Prop}
    \end{subfigure}
    ~ 
    \begin{subfigure}[b]{0.39\textwidth}
        \includegraphics[width=\textwidth]{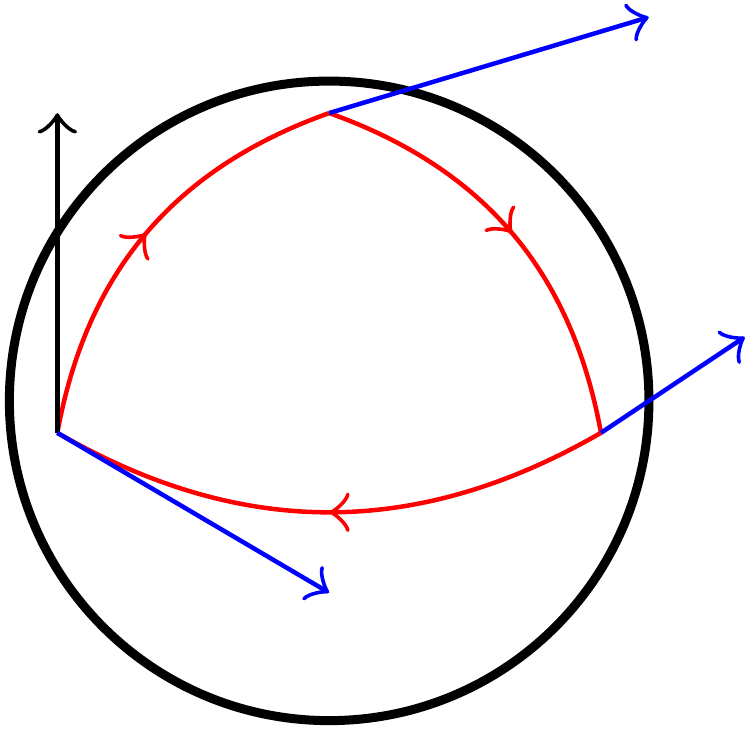}
        \caption{Parallel propagation on the surface of a sphere}
        \label{fig:Curved_Para_Prop}
    \end{subfigure}
    \caption{Examples of parallel propagation of vectors on various surfaces.}\label{fig:Para_Prop_Ex}
\end{figure}

But now imagine a 3-dimensional sphere like the Earth for example (the Earth is really an oblate spheroid, but in this case we will ignore this fact), furthermore, imagine a tangent vector at the equator of the sphere pointing towards one of the poles as shown in figure \ref{fig:Para_Prop_Ex}. Now if we parallel propagate this vector to the pole, then we parallel propagate the vector back to the equator along a curve that is perpendicular to the first curve of propagation and finally propagate the vector along the equator back to its starting location on the sphere, we will find that the vector is now perpendicular to the original vector. Hence, curvature on a manifold can be viewed as a measure of the failure of a vector to return to its original value when parallel propagated along a closed loop in the manifold. However, an equivalent description of curvature is that curvature can be viewed as the failure of successive derivative operators to commute. Hence, we will define curvature in this method, which we will then see is intrinsically linked to parallel propagation.\\

\noindent We start by defining a derivative operator on our manifold. A covariant derivative operator $\n_a$ (note: although we use a lower index, $\n_a$ is not really a dual vector, however it is convention to write it in this manner) is an operator which satisfies the following conditions:
\begin{enumerate}
\item Linearity: $$\n_c(\alpha {A^{a_1...a_k}}_{b_1...b_l} + \beta {B^{a_1...{a_{k'}}}}_{b_1...b_{l'}})=\alpha \n_c{A^{a_1...a_k}}_{b_1...b_l} + \beta \n_c{B^{a_1...{a_{k'}}}}_{b_1...b_{l'}}$$ for all $A \in \mathcal{T}(k, l)$, $B \in \mathcal{T}(k', l')$ and $\alpha, \beta \in \R$.

\item Leibnitz rule: 
\begin{equation*} \nonumber
 \begin{split} 
\n_e({A^{a_1...a_k}}_{b_1...b_l} {B^{c_1...{c_{k'}}}}_{d_1...d_{l'}}) = & \n_e({A^{a_1...a_k}}_{b_1...b_l}) {B^{c_1...{c_{k'}}}}_{d_1...d_{l'}}\\ 
& + {A^{a_1...a_k}}_{b_1...b_l} (\n_e{B^{c_1...{c_{k'}}}}_{d_1...d_{l'}}) 
\end{split}
\end{equation*}
for all $A \in \mathcal{T}(k, l)$ and $B \in \mathcal{T}(k', l')$. 

\item Commutativity with contraction: $$\n_d CT= \n_d ({A^{a_1...c...a_k}}_{b_1...c...b_l}) = \n_d {A^{a_1...c...a_k}}_{b_1...c...b_l}= C(\n_dT)$$ for all $T \in \mathcal{T}(k, l)$

\item \label{Deriv_Prop_4}  Consistency with directional derivatives: $$v(f)=v^a \n_a f$$ for all $f \in \mathcal{F}$ (see page \pageref{F_def}) and all $v^a \in V_p$.

\item \label{Deriv_Prop_5} Torsion free: $$\n_a \n_b f = \n_b \n_a f$$ for all $f \in \mathcal{F}$ (note: there is no direct physical need for our theory to include torsion. However, in string theory and some other alternate/modified theories of gravity, this condition is not imposed which implies the existence of a tensor ${T^a}_{bc}$ which is anti-symmetric in $b$ and $c$ such that $\n_b \n_c f - \n_c \n_b f = {T^a}_{bc}\n_a f$ called the torsion tensor. But here we will assume this tensor to be zero). 
\end{enumerate}

\noindent We now derive the action of the covariant derivative operator $\n_a$ on a vector in our spacetime. We use the notation as shown on page \pageref{direc_deriv_vec} where we represent a vector by $v=v^a\mathbf{e}_a \in V_p$, where $\mathbf{e}_a$ is the collection of basis vectors of $V_p$ and $v^a$ is a collection of scalar functions. Acting on $v$, we have
\be 
\n_b v = \n_b(v^a\mathbf{e}_a)=(\n_bv^a)\mathbf{e}_a+v^a\n_b\mathbf{e}_a \quad .
\ee
However, when acting on a scalar function $f$, we define 
\be
\n_a f \equiv \pd_a f \quad . 
\ee
Hence we have
\be
\n_b v = (\pd_bv^a)\mathbf{e}_a+v^a\n_b\mathbf{e}_a \quad .
\ee
Now, in our curved spacetime, the tangent space at $p\in\M$, $V_p$, is a distinct vector space from the tangent space at $q\in\M$, $V_q$ where $q\neq p$. Hence, the basis vectors in $V_p$ will be different from those in $V_q$, they change throughout the manifold. However, they change in a very precise manner, they change under the action of parallel transport. Hence, it is sufficient to know the basis vectors of $V_p$ at some point $p\in\M$ then via parallel transport we can find the basis vectors of $V_q$ for any other point $q\in\M$. This is all well and good, but how do we mathematically formulate this notion? Given a curve $h(\lambda)$ with tangent $T^a$, a vector $v^a$ is parallel transported along the curve $h(\lambda)$ if the following condition is satisfied
\be
T^a\n_av^b=0 \quad .
\ee
This shows us if a vector is parallel transported but does not give the components of the transported vector. For our basis vectors, we can transport these vectors along $v$, or equivalently we can transport these vectors along each coordinate and sum these transformations. That is, we calculate
\be \label{Basis_Vec_CD}
\n_b\mathbf{e}_a = \Gamma^c{}_{ab}\mathbf{e}_c \quad .
\ee
Where $\Gamma^c{}_{ab}$ is the Christoffel symbol, which can be used to calculate basis vectors after they have been parallel propagated throughout the manifold. 

\begin{figure}[H]
    \centering
    \includegraphics[width=\textwidth]{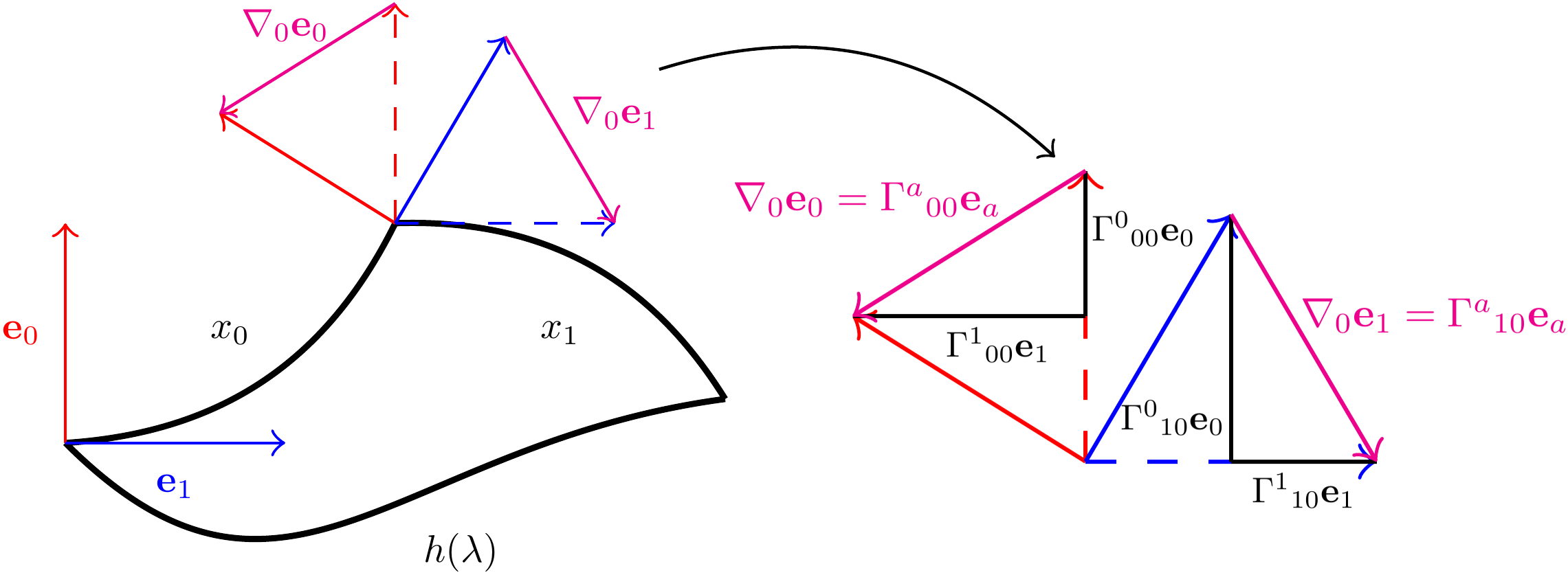}
    \caption{We define the Christoffel symbol by how basis vectors transform when parallel propagated along a coordinate curve in our manifold.}
    \label{fig:Basis_Vec_Transport}
\end{figure}

\noindent As shown in figure \ref{fig:Basis_Vec_Transport}, the first index of the Christoffel symbol denotes the various components of vector $\n_b\mathbf{e}_a$, the second index tells us which basis vector is being transported and the third index tells us along which coordinate the basis vectors are being transported along. 

Now that we know how basis vectors transform when parallel propagated along some curve in our manifold, and moreover how to calculate the covariant derivative of our basis vectors, we can define how the derivative operator acts on a vector $v$. We have 
\be \label{v_deriv}
\begin{split}
\n_b v & = (\pd_bv^a)\mathbf{e}_a+v^a\n_b\mathbf{e}_a \\
& = (\pd_bv^a)\mathbf{e}_a+v^a\Gamma^c{}_{ab}\mathbf{e}_c \\
& = (\pd_bv^a)\mathbf{e}_a+\Gamma^a{}_{cb}v^c\mathbf{e}_a
\end{split}
\ee
where to get from the first line to the second we used the definition of how $\n_b$ acts on our basis vectors, equation \eqref{Basis_Vec_CD}, and from the second to the third line we made the index substitution $a \leftrightarrow c$. Now, to recast this in our usual index notation, we notice that $\n_b v$ is itself a vector. Hence we write $\n_b v$ as $(\n_b v)^a=\n_b v^a \otimes \mathbf{e}_a = (\n_b v^a)\mathbf{e}_a$. Hence, we have
\be
(\n_b v^a)\mathbf{e}_a = (\pd_bv^a)\mathbf{e}_a+\Gamma^a{}_{cb}v^c\mathbf{e}_a
\ee
therefore
\be \label{dv}
\n_b v^a = \pd_bv^a+\Gamma^a{}_{cb}v^c \quad .
\ee
Which now gives the action of the derivative operator on a vector $v^a \in V_p$.

Now to calculate the action of the derivative operator on some dual vector $w_a$ we calculate the action of $\n_b$ on the scalar function $v^aw_a$ (for some arbitrary vector $v^a$). That is
\be
\n_b(v^aw_a)=\n_b(v^a)w_a+v^a\n_bw_a \quad .
\ee
But since $v^aw_a$ is a scalar function, we also have
\be
\n_b(v^aw_a)\equiv\pd_b(v^aw_a)=\pd_b(v^a)w_a+v^a\pd_bw_a
\ee
hence
\be
\n_b(v^a)w_a+v^a\n_bw_a=\pd_b(v^a)w_a+v^a\pd_bw_a \quad .
\ee
Recall that $\n_bv^a=\pd_b v^a + \Gamma^a{}_{bc} v^c$, so we find
\be
(\pd_b v^a + \Gamma^a{}_{bc} v^c)w_a+v^a\n_bw_a=\pd_b(v^a)w_a+v^a\pd_bw_a
\ee
hence
\be
v^a\n_bw_a=v^a\pd_bw_a-\Gamma^a{}_{bc}v^cw_a \quad .
\ee
In the second term on the right hand side of the equation above the $a$ index is being summed over and hence can be replaced with any other index label. So we will make the index substitution $a \leftrightarrow c$ and also use the fact that $\Gamma^a{}_{bc}$ is symmetric in its lower two indices. Hence
\begin{gather}
v^a\n_bw_a=v^a\pd_bw_a-\Gamma^c{}_{ab}v^aw_c \\
\Rightarrow \n_bw_a=\pd_bw_a-\Gamma^c{}_{ab}w_c \quad .
\end{gather}

We can then ``bootstrap" this construction to define the action of the derivative operator on a tensor of arbitrary rank 
\be \label{dT}
\begin{split}
\n_c {T^{a_1...a_k}}_{b_1...b_l} =  \pd_c {T^{a_1...a_k}}_{b_1...b_l} & + \sum_{i=1}^k {\Gamma^{a_i}}_{mc} {T^{a_1...a_{i-1}m\; a_{i+1}...a_k}}_{b_1...b_l}\\
& - \sum_{j=1}^l {\Gamma^m}_{b_j c} {T^{a_1...a_k}}_{b_1...b_{j-1}m\; b_{j+1}...b_l} \quad . 
\end{split}
\ee

Now we have defined the action for the derivative operator in terms of the partial derivative operator and the Christoffel symbol. However, we don't yet know the components of the Christoffel symbol, making our definitions useless at this stage. However, given a conjecture (which shall be proven later in the text), we can relate the components of the Christoffel symbol to various components of the metric (more so, the derivatives of the components).\\ 

\noindent \textbf{Conjecture 2.3.1} 
\be \label{dg=0}
\n_a \, g_{bc} = 0 \quad .
\ee

\noindent Using this conjecture and using equation \eqref{dT}, we find
\be
\n_a\,  g_{bc} = \pd_a\, g_{bc} - {\Gamma^d}_{ab}\, g_{dc} - {\Gamma^d}_{ac}\, g_{bd} = 0
\ee
hence
\be \label{cab_1}
\Gamma_{cab} + \Gamma_{bac} = \pd_a\, g_{bc} \quad . 
\ee 
We are free to relabel indices, hence by using the following index substitution $a \rightarrow b$ and $b \rightarrow a$, we get
\be \label{cab_2}
\Gamma_{cba} + \Gamma_{abc} = \pd_b\, g_{ac}
\ee
and using the index substitution $a \rightarrow c$, $b \rightarrow a$ and $c \rightarrow b$, we get
\be \label{cab_3}
\Gamma_{bca} + \Gamma_{acb} = \pd_c\, g_{ab} \quad . 
\ee
If we add equations \eqref{cab_1} and \eqref{cab_2} then subtract equation \eqref{cab_3}, and using the fact that ${\Gamma^a}_{bc}$ is symmetric in its lower two indices, we get
\be
2\Gamma_{cab} = \pd_a\, g_{bc} + \pd_b\, g_{ac} - \pd_c\, g_{ab}
\ee
hence
\be \label{Christoffel_def}
{\Gamma^c}_{ab} =\frac{1}{2} g^{cd} (\pd_a\, g_{bd} + \pd_b\, g_{ad} - \pd_d\, g_{ab}) \quad . 
\ee
Therefore, given a metric, we can calculate the components of the Christoffel symbol and hence the action of the derivative operator on any tensor. \\

\noindent As stated above, curvature can be viewed as the failure of successive covariant derivative operators to commute. Now that we have properly defined derivative operators on our spacetime, we are now ready to properly define curvature in our spacetime. 

Let $\n_a$ be a derivative operator and let $w_a$ be a dual vector, then we have
\be
[\n_a, \n_b]w_c = (\n_a\n_b - \n_b\n_a)w_c = \n_a\n_b w_c - \n_b\n_a w_c
\ee
via equation \eqref{dT} (and after simplifying) this is equivalent to
\be \label{R_def_1}
[\n_a, \n_b]w_c = (\pd_a \Gamma^d{}_{cb}-\pd_b \Gamma^d{}_{ca} +\Gamma^d{}_{ma}\Gamma^m{}_{cb}-\Gamma^d{}_{mb}\Gamma^m{}_{ca})w_d \quad .
\ee
We notice that the object inside the brackets is an algebraic operator, not a differential operator. Also, it can be verified that this object transforms via equation \eqref{ten_trans}, so this object is indeed a tensor of type $(1,3)$ called the Riemann curvature tensor, denoted ${R^a}_{bcd}$, defined as
\be
[\n_a, \n_b]w_c = {R^d}_{cab}\, w_d \quad .
\ee 
Using equation \eqref{R_def_1} and realising that this equation holds for all $w_d$ we find
\be
{R^a}_{bcd} = \pd_c {\Gamma^a}_{bd} - \pd_d {\Gamma^a}_{bc} + {\Gamma^a}_{mc}{\Gamma^m}_{bd} - {\Gamma^a}_{md}{\Gamma^m}_{bc}
\ee
hence, given a spacetime with a metric we can calculate the curvature of the spacetime itself. We are beginning to realise our goal to define gravitational fields via intrinsic properties of the spacetime. However, we have more formalism to introduce before we can fully realise our goal. \\

\noindent The Riemann tensor has a few useful symmetry properties:
\begin{enumerate}
\item $R_{abcd}=-R_{bacd}=-R_{abdc}$ \quad .

\item \label{Rie_Sy_2} $R_{abcd}=R_{cdab}$ \quad .

\item $R_{a[bcd]}=0$ \quad .

\item \label{Rie_Sy_4} $\n_{[e}{R^a}_{|b|cd]}={R^a}_{b[cd;e]}=0$ \quad .
\end{enumerate}
We have introduced some new notation in the fourth property above. The vertical bars around the $b$ index in $\n_{[e}{R^a}_{|b|cd]}$ indicate that we anti-symmetrise over the $e$, $c$ and $d$ index, but not the $b$ index. Another new piece of notation is $\n_c {T^{a_1...a_k}}_{b_1...b_l} = {T^{a_1...a_k}}_{b_1...b_l;c}$ , this is just notation, there is no new mathematics going on here. We also note that we will also sometimes use the notation $\pd_c {T^{a_1...a_k}}_{b_1...b_l} = {T^{a_1...a_k}}_{b_1...b_l,c}$ to denote a partial derivative acting on a tensor. 

\begin{figure}[H]
    \centering
    \includegraphics[width=0.94\textwidth]{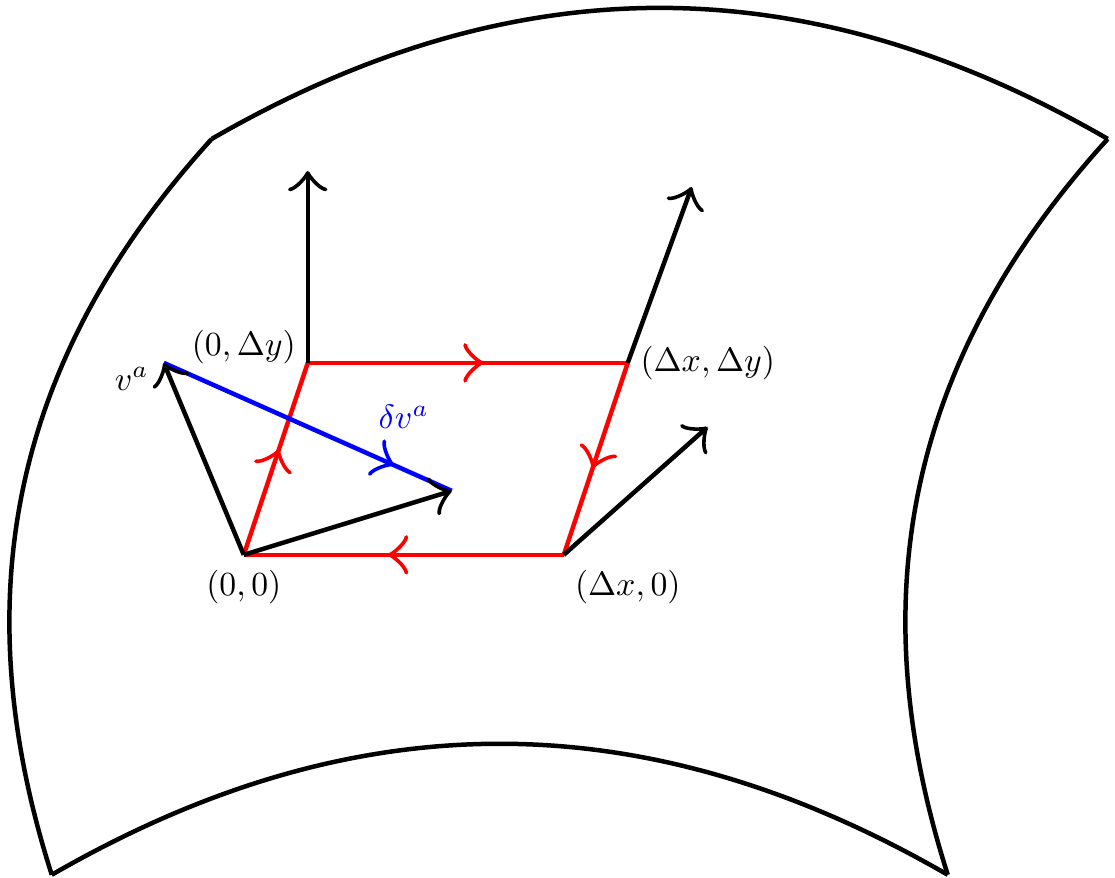}
    \caption{We see that a vector will fail to return to its initial value when parallel propagated along a closed loop in a curved manifold.}
    \label{fig:Vec_fail_to_return}
\end{figure}

We now show that the Riemann tensor is directly related to the failure for a vector to return to its initial value when parallel transported along a closed loop in our manifold. Let $p \in \M$ and let $s$ be a 2-dimensional surface through $p$ with coordinates $x$ and $y$. Let $p$ be at the coordinate values $(0, 0)$ then let $v^a$ be a vector in $V_p$ and we now parallel transport that vector along the curve given by the coordinate values $(0, 0) \rightarrow (0, \Delta y) \rightarrow (\Delta x, \Delta y) \rightarrow (\Delta x, 0) \rightarrow (0, 0)$ for $\Delta x > 0$ and $\Delta y > 0$ as shown in figure \ref{fig:Vec_fail_to_return}. Now let $w_a$ be some arbitrary dual vector field and we now calculate the change in $v^a w_a$ along the curve. In the first part of the curve, given by the change in coordinate values $(0, 0) \rightarrow (0, \Delta y)$ we have 
\be
\delta (v^a w_a)_1 = \Delta y\, \pd_y (v^a w_a) |_{(0, \Delta y/2)}
\ee
here we have evaluated the derivative at the midway point so that our expression is valid to second order in $\Delta y$. 
\noindent Via the fourth property of the derivative operator (see page \pageref{Deriv_Prop_4}), we can write
\be
\begin{split}
\delta (v^a w_a)_1 & = \Delta y\, Y^b\n_b (v^a w_a) |_{(0, \Delta y/2)}\\
& = \Delta y\, v^a Y^b\n_b w_a |_{(0, \Delta y/2)}
\end{split}
\ee where $Y^a$ is the tangent vector to curves of constant $y$ and since $v^a$ is being parallel propagated along the curve $(0, 0) \rightarrow (0, \Delta y)$ with tangent $Y^a$ by definition $Y^b \n_b v^a = 0$. We can do very similar constructions to find $\delta (v^a w_a)_2$, $\delta (v^a w_a)_3$ and $\delta (v^a w_a)_4$ along the entire curve. However, if we now sum the first and third contribution, we find
\be
\delta (v^a w_a)_1 + \delta (v^a w_a)_3 = \Delta y\, (v^a Y^b\n_b w_a |_{(0, \Delta y/2)}-v^a Y^b\n_b w_a |_{(\Delta x, \Delta y/2)})
\ee
which goes to zero as $\Delta x \rightarrow 0$. We can do a similar construction when we sum the second and fourth contributions. This shows that the total change in $v^aw_a$, $\delta(v^aw_a)$, along the curve is zero to first order in $\Delta x$ and $\Delta y$. So to find the change in $\delta(v^aw_a)$ along the curve, we must calculate the second order contributions. We shall look at the $\delta (v^a w_a)_1 + \delta (v^a w_a)_3$ contribution. To find the second order term, we firstly consider the curve $x=\Delta x/2$, now we parallel transport both $v^a$ and $Y^b\n_b w_a$ along this curve from $(0, \Delta y/2)$ to $(\Delta x, \Delta y/2)$. To first order in $\Delta x$, $v^a$ is invariant under this transformation. However, to first order in $\Delta x$, $Y^b\n_b w_a$ will differ by the amount $\Delta x\, X^c(Y^b\n_b w_a)$. Hence, to second order in $\Delta x$ and $\Delta y$, we have 
\be
\delta (v^a w_a)_1 + \delta (v^a w_a)_3 = - \Delta x\; \Delta y\; v^aX^c(Y^b\n_b w_a) \quad .
\ee
Doing a similar construction for $\delta (v^a w_a)_2 + \delta (v^a w_a)_4$ and then summing all contributions, we find the total change to be 
\be
\begin{split}
\delta (v^a w_a) & = \Delta x\, \Delta y\, v^a[Y^c\n_c(X^b\n_bw_a)-X^c\n_c(Y^b\n_bw_a)]\\
& = \Delta x\, \Delta y\, v^aY^cX^b(\n_c\n_b-\n_b\n_c)w_a\\
& =  \Delta x\, \Delta y\, v^aY^cX^b{R^d}_{abc}w_d \quad .
\end{split}
\ee 
Now, given that we initially chose a specific vector $v^a$, to have this hold for all $w_a$ we must assert that
\be
\delta v^a=\Delta x\, \Delta y\, v^dY^cX^b{R^a}_{dbc} \quad . 
\ee
So, we see that a vector will fail to return to its initial value when parallel propagated along a closed curve in our manifold if ${R^a}_{bcd} \neq 0$, as we originally postulated at the beginning of our discussion on curvature. \\

\noindent We can also construct some other useful quantities from the Riemann tensor. If we contract over the first and third indices we construct the Ricci tensor, $R_{ab}$ , given by 
\be
R_{ab}={R^c}_{acb}
\ee
via the second symmetry property of the Riemann tensor (see page \pageref{Rie_Sy_2}), we see that $R_{ab}=R_{ba}$, i.e. the Ricci tensor is symmetric. Now if we were to contract the Ricci tensor, we produce the Ricci scalar
\be
R={R^a}_a \quad . 
\ee
We can also produce the Weyl tensor, $C_{abcd}$, which is defined as (for manifolds with dimension $n \geq 3$)
\be
C_{abcd} = R_{abcd} + \frac{2}{(n-1)(n-2)} R\, g_{a[c}g_{d]b} - \frac{2}{n-2}(g_{a[c}R_{d]b} - g_{b[c}R_{d]a}) .
\ee
We note that $C_{abcd}$ is trace free, that is to say, if we contract over any of the indices in $C_{abcd}$ we get zero.

\section{Geodesics}
In section \ref{intro} we briefly discussed Fermat's principle, which states that when light travels from one point to another, it will always take the path of least time. We can think of this path as the ``straightest possible path" between those two points, this type of path is called a \emph{geodesic}. In flat space, the path of least time between any two points is clearly just a straight line connecting those two points. However, in curved space, this is no longer the case. We now discuss how to calculate geodesics in curved space.

In the last section, we discussed that when a vector $v^a$ is parallel transported along a curve with tangent $T^a$, the following condition holds by definition 
\be
T^a \n_a v^b = 0
\ee
using equation \eqref{dv} we find
\be
T^a \pd_a v^b + T^a {\Gamma^b}_{ac} v^c = 0
\ee
and if the curve is parameterised by the affine parameter $\lambda$, then
\be \label{geo_1}
\frac{d v^b}{d\lambda} + {\Gamma^b}_{ac} T^a v^c = 0  \quad . 
\ee
Aside: an affine parameter is a parameter such that $\frac{d v^b}{d\lambda} + {\Gamma^b}_{ac} T^a v^c = 0$.\break If we did not choose an affine parameter then we would instead have \break $\frac{d v^b}{d\lambda} + {\Gamma^b}_{ac} T^a v^c \propto v^b$. For simplicity we choose to use an affine parameter. If the curve is traced out by the motion of a massive particle, then the affine parameter we use is the proper time measured by the particle along its trajectory, $\tau$. If the curve is traced out by the motion of a massless particle, we shall see, that we cannot use proper time to parameterise the curve, we instead use an arbitrary parameter $\lambda$, which then we shall pick $\lambda$ such that it is an affine parameter. 

Now, a geodesic is defined as a curve whose tangent, $T^a$, is parallel transported along itself, that is to say
\be \label{Geo_Eqn_Kill}
T^a \n_a T^b = 0 
\ee
then by equation \eqref{geo_1}
\be
\frac{d T^b}{d\lambda} + {\Gamma^b}_{ac} T^a T^c = 0 \quad . 
\ee
But, as previously stated, the components of a tangent vector of a curve parameterised by the parameter $\lambda$ is given by
\be
T^a = \frac{d x^a}{d\lambda}
\ee
hence
\be \label{geo_eqn}
\frac{d^2 x^b}{d\lambda^2} + {\Gamma^b}_{ac} \frac{d x^a}{d\lambda} \frac{d x^c}{d\lambda} = 0 \quad . 
\ee
Equation \eqref{geo_eqn} is called the geodesic equation. A curve with position vector $x^a(\lambda)$ is a geodesic iff it satisfies equation \eqref{geo_eqn}.

Given a curve with tangent vector $T^a$, we can calculate the length of the curve by the following equation
\be \label{geo_len}
l = \int \sqrt{g_{ab} T^a T^b}\,  d\lambda \quad .
\ee
However, since the metric is not positive definite, we cannot guarantee that $g_{ab} T^a T^b \geq 0$, but this does not pose a problem as we will show, in fact we can categorise geodesics by the sign of $g_{ab} T^a T^b$. A geodesic is timelike if $g_{ab} T^a T^b < 0$ everywhere along the curve, a geodesic is null if $g_{ab} T^a T^b = 0$ everywhere along the curve and a geodesic is spacelike if $g_{ab} T^a T^b > 0$ everywhere along the curve. Now, if the geodesic is a curve traced out by particle in our spacetime, then the tangent vector $T^a=dx^a/d\lambda$ is just the 4-velocity of the particle $v^a$. From special relativity, we know that the norm of the 4-velocity is given by 
\be \label{T_N_Con}
g_{ab}\, v^a v^b = v^a v_a = \gamma(||\vec{v}||) (-c^2 + \vec{v} \cdot \vec{v})
\ee
where $\vec{v}$ is the 3-velocity of the particle and we have temporally not set $c=1$. Now, for timelike curves $g_{ab} T^a T^b < 0 \Rightarrow v < c$ , so the types of particles that travel along timelike geodesics are massive particles. For null curves $g_{ab} T^a T^b = 0 \Rightarrow v = c$ , so the types of particles that travel along null geodesics are photons, hence null geodesics are light rays. For spacelike curves $g_{ab} T^a T^b > 0 \Rightarrow v > c$ , so the types of particles that travel along spacelike geodesics are tachyons, which as of this writing have no solid evidence to support their existence. Hence, we will only consider timelike and null geodesics. 

For null geodesics, the length of the geodesic is zero. This is equivalent to saying that photons experience no change in their proper time along their trajectory. For timelike geodesics, equation \eqref{geo_len} is undefined, hence we cannot define the length of a timelike curve. However, we can define the proper time elapsed along the curve
\be \label{geo_pt}
\tau = \int \sqrt{-g_{ab} T^a T^b}\,  d\lambda
\ee
hence the above equation gives the the amount of the particle's proper time that has elapsed along the particle's trajectory. 

Now, our choice of parametrisation is arbitrary, we may choose to reparametrise our curve by changing parameter $\lambda \rightarrow \bar{\lambda}$. Then by equation \eqref{geo_pt} we find
\be
\begin{split}
\tau & = \int \sqrt{-g_{ab} \bar{T}^a \bar{T}^b}\,  d\bar{\lambda} \\
& = \int \sqrt{-g_{ab} \frac{d\lambda}{d\bar{\lambda}} T^a \frac{d\lambda}{d\bar{\lambda}} T^b}\,  d\bar{\lambda} \\
& = \int \sqrt{-g_{ab} T^a T^b}\, \frac{d\lambda}{d\bar{\lambda}}\, d\bar{\lambda} \\
& = \int \sqrt{-g_{ab} T^a T^b}\,  d\lambda \quad . 
\end{split}
\ee
Hence, the proper time elapsed along the curve is invariant under a change in parametrisation, which is exactly what we expect. It does not matter how our trajectory is mathematically formulated, the proper time we would experience along our trajectory will never change as long as the trajectory itself does not change. 

Above we alluded to the fact that the paths that inertial particles travel along are indeed geodesics, we now prove this statement. This is really just a generalisation of Fermat's principle and the principle of least action to curved space. We wish the extremise the proper time taken along a curve, to do so we make use of the variational method using the Euler-Lagrange equation. Consider the Lagrangian
\be
\la = \sqrt{-g_{ab}\frac{dx^a}{d\lambda}\frac{dx^b}{d\lambda}}
\ee
and the following Euler-Lagrange equation 
\be \label{E_L_eqn}
\frac{d}{d\lambda}\frac{\pd\la}{\pd (dx^c/d\lambda)} - \frac{\pd\la}{\pd x^c} = 0 \quad . 
\ee
Now
\be \label{dl_dxc}
\frac{\pd\la}{\pd x^c} = \frac{1}{\sqrt{-g_{ab}v^av^b}} \left( \frac{\pd g_{ab}}{\pd x^c} \right)\frac{dx^a}{d\lambda}\frac{dx^b}{d\lambda}
\ee
also
\be \label{dl_dxadlambda}
\begin{split}
\frac{\pd\la}{\pd (dx^c/d\lambda)} & = \frac{1}{\sqrt{-g_{ab}v^av^b}} \left(g_{ab} \frac{dx^a}{d\lambda} + g_{ab} \frac{dx^b}{d\lambda}\right) \\
& = \frac{2}{\sqrt{-g_{ab}v^av^b}}\, g_{ab} \frac{dx^a}{d\lambda}
\end{split}
\ee
here we have used the fact that $g_{ab}$ is symmetric.
Now we differentiate equation \eqref{dl_dxadlambda} with respect to $\lambda$
\be \label{Geo_la_1}
\def\arraystretch{2.2}
\begin{array}{cl}
\dfrac{d}{d\lambda}\left[  \dfrac{2}{\sqrt{-g_{ab}v^av^b}}\, g_{ab} \dfrac{dx^a}{d\lambda} \right] & = \dfrac{d}{d\lambda}\left[\dfrac{2}{\sqrt{-g_{ab}v^av^b}}\right]g_{ab} \dfrac{dx^a}{d\lambda}  \\
 & \hspace{4.5mm} + \dfrac{2}{\sqrt{-g_{ab}v^av^b}}\dfrac{d}{d\lambda}\left[g_{ab} \dfrac{dx^a}{d\lambda}\right]\\
 & = \dfrac{2}{(-g_{ab}v^av^b)^{3/2}}\dfrac{d}{d\lambda}\left[ g_{ab}\dfrac{dx^a}{d\lambda}\dfrac{dx^b}{d\lambda}\right] g_{ab}\dfrac{dx^a}{d\lambda}\\
 & \hspace{4.5mm} + \dfrac{2}{\sqrt{-g_{ab}v^av^b}}\dfrac{d}{d\lambda}\left[g_{ab} \dfrac{dx^a}{d\lambda}\right] \quad .
\end{array}
\ee
However, we are deriving the equation of motion for inertial particles, hence these particles are non-accelerating. Because of this, the norm of the tangent vector to a geodesic is invariant along the curve. That is to say
\be
\frac{d}{d\lambda}(g_{ab}v^av^b)=0
\ee
where $v^a$ is the tangent vector to the trajectory of our particle. We also have
\be \label{dl_big_boi}
\begin{split}
\dfrac{d}{d\lambda}\left[g_{ab} \dfrac{dx^a}{d\lambda}\right] & = 2\, g_{ab} \frac{d^2x^a}{d\lambda^2} + 2\,\frac{d g_{ab}}{d\lambda} \frac{dx^a}{d\lambda}\\
& = 2\, g_{ab} \frac{d^2x^a}{d\lambda^2} + 2\, \frac{\pd g_{ab}}{\pd x^c} \frac{dx^a}{d\lambda}\frac{dx^c}{d\lambda}\\
& = 2\, g_{ab} \frac{d^2x^a}{d\lambda^2} + 2\, \frac{\pd g_{ac}}{\pd x^b} \frac{dx^a}{d\lambda}\frac{dx^b}{d\lambda}\\
& = 2\, g_{ab} \frac{d^2x^a}{d\lambda^2} + \frac{\pd g_{ac}}{\pd x^b} \frac{dx^a}{d\lambda}\frac{dx^b}{d\lambda} + \frac{\pd g_{ac}}{\pd x^b} \frac{dx^a}{d\lambda}\frac{dx^b}{d\lambda}\\
& = 2\, g_{ab} \frac{d^2x^a}{d\lambda^2} + \frac{\pd g_{ac}}{\pd x^b} \frac{dx^a}{d\lambda}\frac{dx^b}{d\lambda} + \frac{\pd g_{bc}}{\pd x^a} \frac{dx^a}{d\lambda}\frac{dx^b}{d\lambda}\\
& = 2\, g_{ab} \frac{d^2x^a}{d\lambda^2} + \left(\frac{\pd g_{ac}}{\pd x^b}+ \frac{\pd g_{bc}}{\pd x^a}\right) \frac{dx^a}{d\lambda}\frac{dx^b}{d\lambda}\\
\end{split}
\ee
here we have been extremely explicit and clear about what we are doing, we are only relabelling indices here. So, equation \eqref{Geo_la_1} simplifies to
\be
\begin{split}
\dfrac{d}{d\lambda}\left[  \dfrac{2}{\sqrt{-g_{ab}v^av^b}}\, g_{ab} \dfrac{dx^a}{d\lambda} \right]=\dfrac{1}{\sqrt{-g_{ab}v^av^b}} & \left( 2\, g_{ab} \frac{d^2x^a}{d\lambda^2} \right. \\
& \hspace{0.23cm} \left. + \left(\frac{\pd g_{ac}}{\pd x^b}+ \frac{\pd g_{bc}}{\pd x^a}\right) \frac{dx^a}{d\lambda}\frac{dx^b}{d\lambda}\right)
\end{split}
\ee
\enlargethispage{50pt}
then combining equations \eqref{E_L_eqn}, \eqref{dl_dxc} and \eqref{dl_big_boi}, we find
\be \label{geo_deri_EL}
\begin{split}
0 & = \frac{d}{d\lambda}\frac{\pd\la}{\pd (dx^c/d\lambda)} - \frac{\pd\la}{\pd x^c}\\
& = \dfrac{1}{\sqrt{-g_{ab}v^av^b}}\left(2\, g_{ab} \frac{d^2x^a}{d\lambda^2} + \left(\frac{\pd g_{ac}}{\pd x^b}+ \frac{\pd g_{bc}}{\pd x^a}\right) \frac{dx^a}{d\lambda}\frac{dx^b}{d\lambda} - \left( \frac{\pd g_{ab}}{\pd x^c} \right)\frac{dx^a}{d\lambda}\frac{dx^b}{d\lambda}\right)\\
& = g_{ab} \frac{d^2x^a}{d\lambda^2} + \frac{1}{2}\left(\frac{\pd g_{ac}}{\pd x^b}+ \frac{\pd g_{bc}}{\pd x^a} - \frac{\pd g_{ab}}{\pd x^c}\right) \frac{dx^a}{d\lambda}\frac{dx^b}{d\lambda}\\
& = g_{ab} \frac{d^2x^a}{d\lambda^2} + \Gamma_{cab}\frac{dx^a}{d\lambda}\frac{dx^b}{d\lambda}\\
& = g_{cd} \frac{d^2x^c}{d\lambda^2} + \Gamma_{dab}\frac{dx^a}{d\lambda}\frac{dx^b}{d\lambda}\\
& = \frac{d^2x^c}{d\lambda^2} + g^{cd} \Gamma_{dab}\frac{dx^a}{d\lambda}\frac{dx^b}{d\lambda}\\
& = \frac{d^2x^c}{d\lambda^2} + {\Gamma^c}_{ab}\frac{dx^a}{d\lambda}\frac{dx^b}{d\lambda}\\
\end{split}
\ee
which is just the geodesic equation (equation \eqref{geo_eqn}). To get from line 3 to 4 of equation \eqref{geo_deri_EL} we just used the definition of ${\Gamma^c}_{ab}$ , equation \eqref{Christoffel_def}. Hence, we see that geodesics extremise the proper time of a curve in spacetime. By Fermat's principle and the principle of least action, we see that all curves traced out by the motion of inertial particles in our spacetime must be geodesics.

At this stage, we can now prove equation \eqref{dg=0}. All inertial particles travel along geodesics as shown above, so by definition we have
\be
a^a=v^b\n_b v^a=0
\ee
since the tangent to the world line of a particle is the 4-velocity of the particle. Hence, particles traveling along geodesics have zero 4-acceleration, i.e. they are non-accelerating. Because of this, as above, we have
\be
\frac{d}{d\lambda}(g_{ab}v^av^b)=0
\ee
where $v^a$ is the tangent vector to some arbitrary geodesic. Expanding, we get
\be
\left(\frac{d g_{ab}}{d\lambda}\right)v^av^b+g_{ab}\left(\frac{dv^a}{d\lambda}\right)v^b+g_{ab}v^a\left(\frac{dv^b}{d\lambda}\right)=0 \quad .
\ee
Via the chain rule, we have
\be
\frac{dg_{ab}}{d\lambda}=\frac{\pd g_{ab}}{\pd x^c}\frac{dx^c}{d\lambda} \equiv (\pd_c \, g_{ab}) v^c
\ee
and from the geodesic equation we also have 
\be
\frac{dv^a}{d\lambda}=-\Gamma^a{}_{bc}v^bv^c \quad .
\ee
Therefore, we now have the consistency condition
\be
\pd_c (g_{ab})v^av^bv^c-g_{db}\Gamma^d{}_{ac}v^av^bv^c-g_{ad}\Gamma^d{}_{bc}v^av^bv^c=0 \quad .
\ee
hence
\be
\pd_c\, g_{ab}-\Gamma^d{}_{ac}\, g_{db} - \Gamma^d{}_{bc}\, g_{ad} =0
\ee
where by comparing this to equation \eqref{dT}, we see this condition is equivalent to
\be
\n_c\, g_{ab} = 0
\ee
which is equation \eqref{dg=0}.\\

\noindent We now give another example showing that if the Riemann tensor is non-zero, our manifold is indeed non-Euclidean and hence curved. Consider a family of geodesics $\gamma_s(\lambda)$, where the parameter ``s" allows us to differentiate between the different geodesics in the family. We can construct a 2-dimensional sub-manifold which is spanned by the geodesics in $\gamma_s(\lambda)$, we denote this sub-manifold as $\Sigma$. Furthermore, we can construct a basis of $\Sigma$ by the vector field $T^a=\frac{\pd}{\pd \lambda}$ which is tangent to the geodesics and the vector field $X^a=\frac{\pd}{\pd s}$ which physically represents the infinitesimal displacement between nearby geodesics, $X^a$ is also called the deviation vector. Recall that since $T^a$ is tangent to the family of geodesics, is satisfies the equation
\be
T^a \n_a T^b = 0 \quad . 
\ee
Now, since $T^a$ and $X^a$ are coordinate vector fields, they commute. That is to say that $T^a$ and $X^a$ satisfy the following condition
\be
[X, T]^a\equiv X^a\n_aT^b-T^a\n_aX^b=0
\ee
hence
\be
T^a \n_a X^b = X^a \n_a T^b \quad .
\ee
Physically, the quantity $v^a = T^b \n_b X^a$ represents the change in the deviation vector along the geodesics, i.e. the relative velocity of a nearby geodesic. Hence the quantity
\be 
a^a = T^c \n_c v^a = T^c \n_c (T^b \n_b X^a)
\ee
physically represents the relative acceleration of a nearby geodesic. We can show that the relative acceleration of nearby geodesics is proportional to the Riemann tensor: 
\be \label{Geo_Div_Eqn}
\begin{split}
a^a & = T^c \n_c (T^b \n_b X^a)\\
& = T^c \n_c (X^b \n_b T^a)\\
& = (T^c \n_c X^b)(\n_b T^a) + X^b T^c \n_c \n_b T^a\\
& = (X^c \n_c T^b)(\n_b T^a) + X^b T^c \n_c \n_b T^a\\
& = (X^c \n_c T^b)(\n_b T^a) + X^b T^c \n_b \n_c T^a - {R^a}_{dcb} X^b T^c T^d\\
& = X^c \n_c (T^b \n_b T^a) - {R^a}_{dcb} X^b T^c T^d\\
& = - {R^a}_{dcb} X^b T^c T^d \quad .
\end{split}
\ee
Where we have used the Leibnitz rule in lines 3 and 6 and we have used the definition of the Riemann tensor is line 5 of the above equation. Equation \eqref{Geo_Div_Eqn} is known as the geodesic deviation equation or the Jacobi equation. Hence we see that if the Riemann tensor is non-zero, the geodesic deviation equation implies that initially parallel geodesics fail to remain parallel, i.e. Euclid's fifth postulate fails. Since Euclid's fifth postulate fails, our spacetime is non-Euclidean, i.e. curved. This is yet another example showing that the Riemann tensor does indeed correlate to curvature in our spacetime.

\section{Einstein's equation}
In the previous sections we found that space is curved where we observe a gravitational field and we introduced the mathematical framework required to describe curvature in our spacetime. We are now ready to use this framework to describe Einstein's theory of gravity. Einstein's theory basically states that spacetime is a 4-dimensional manifold with a metric $g_{ab}$ (or multiple metrics if the manifold is split into distinct patches) with Lorentzian signature, which satisfies Einstein's equation. We now motivate Einstein's equation, which relates the curvature of spacetime with the mass-energy present within the spacetime.\\

\noindent Firstly, we require a tensorial definition of the mass-energy present within the spacetime, this is because curvature is defined via tensors, hence equality between the curvature and mass-energy would require that the mass-energy be defined as a tensor quantity. An obvious candidate for this quantity is the stress-energy-momentum tensor $T_{ab}$ , or the stress-energy tensor for short. The stress-energy tensor is defined as follows. Consider an observer with 4-velocity $v^a$, then $T_{ab}v^av^b$ physically represents the energy density, $\rho$, as measured by this observer. If the vector $x^a$ is orthogonal to $v^a$ then $-T_{ab}x^av^b$ physically represents the flux of mass along $x^a$ (i.e. the momentum density along $x^a$), the quantity $T_{ab}x^ax^b$ represents the normal stress (or pressure) and finally if $y^a$ is also orthogonal to $v^a$ and $y^a \neq x^a$, then $T_{ab}x^ay^b$ represents the shear stress. 

Now to relate the curvature to the mass-energy, we appeal to Poisson's equation. Given a scalar potential $\phi$, Poisson's equation states 
\be 
\n^2\phi=4\pi\rho \quad .
\ee 
Writing this with our tensor notation, we get 
\be
\pd^a\pd_a\phi = 4\pi T_{ab}v^av^b \quad .
\ee
However, we can use Newtonian theory to relate our curvature to our potential $\phi$. In the Newtonian theory of gravity, the tidal acceleration of two nearby objects is given by $-(\Vec{x}\cdot\Vec{\n})\Vec{\n}\phi$, or in tensor notation
\be 
a^a=-x^b\pd_b\pd^a\phi 
\ee
but from equation \eqref{Geo_Div_Eqn} we see that 
\be
a^a=-{R^a}_{dcb}x^bv^cv^d \quad .
\ee
Hence equality of these two equations yields
\be
\pd^a\pd_a\phi={R^c}_{acb}v^av^b=R_{ab}v^av^b
\ee
therefore yielding the tentative field equation
\be
\begin{split}
R_{ab}v^av^b & =4\pi T_{ab}v^av^b\\
\Rightarrow R_{ab} & =4\pi T_{ab} \quad .
\end{split}
\ee
Historically, this was the first field equation that Einstein proposed. However, this field equation has a problem when one considers energy conservation in the spacetime. Firstly, consider the stress-energy tensor of a perfect fluid in a spacetime with metric $g_{ab}$ given by 
\be
T_{ab}=\rho v_av_b + P(g_{ab}+v_av_b)
\ee
the equation of motion of the perfect fluid is 
\be \label{En_Con}
\n^aT_{ab}=0
\ee
which imply the following equations:
\begin{gather}
v^a\n_a\rho + (\rho+P)\n^av_a=0\\
(\rho+P)v^a\n_av_b+(g_{ab}+v_av_b)\n^aP=0 \quad .
\end{gather}
If we go to flat space and take the non-relativistic limit where $P\ll \rho$, $v^a=(1, \Vec{v})$ and $v\, dP/dt\ll |\Vec{\n}P|$, the above equations reduce to
\begin{gather}
\frac{\pd\rho}{\pd t} + \Vec{\n}\cdot(\rho\Vec{v})=0\\
\rho \left[ \frac{\pd \Vec{v}}{\pd t} + (\Vec{v}\cdot\Vec{\n})\Vec{v}\right]=-\Vec{\n} P \quad ,
\end{gather}
which is just conservation of mass and Euler's equation respectively. Hence, we can interpret equation \eqref{En_Con} as the energy conservation condition. 

Since equation \eqref{En_Con} always holds, we should expect that $\n^aR_{ab}=0$. However, examination of the Bianchi identity (which is the fourth symmetry property of the Riemann tensor; see page \pageref{Rie_Sy_4}), shows this is not the case. The Bianchi identity states that
\be
\n_{[e}{R^a}_{|b|cd]}=0
\ee
if we contract the first and third index of the Riemann tensor we find
\be
\n_{[c}{R^a}_{|b|ad]}=\n_a{R^a}_{bcd}+\n_dR_{bc}-\n_cR_{bd}=0
\ee
then if we raise the index $d$ via the metric and then contract over $b$ and $d$, we get the twice contracted Bianchi identity
\be
\n_a{R^a}_{c}+\n_b{R^b}_c-\n_c{R^b}_b=\n_a(2{R^a}_b-R)=0
\ee
or equivalently (and most commonly written as)
\be
\n^aG_{ab}=0
\ee
where
\be
G_{ab}=R_{ab}-\frac{1}{2}R g_{ab}
\ee
is the Einstein tensor.

Finally, we see that while $\n^aG_{ab}=0$, $\n^aR_{ab}=-\frac{1}{2}\n_bR$. Hence the field equation $R_{ab}=4\pi T_{ab}$ implies that $\n_bR=0$, or equivalently that $T={T^a}_a$ is constant throughout the universe, which implies that density is constant throughout the universe. However, there is a simple fix to this problem. If we consider the field equation
\be \label{Ein_Eq}
G_{ab}=R_{ab}-\frac{1}{2}R g_{ab} = 8\pi T_{ab}
\ee
then when we take the covariant derivative, both sides both give zero, hence upholding energy conservation. Equation \eqref{Ein_Eq} is called Einstein's equation and is the equation we were looking for which relates the mass-energy within the spacetime to the curvature (note that here we have set the speed of light, $c$, and the Newtonian gravitational constant, $G_N$, equal to $1$). If we rearrange this for $R_{ab}$ specifically in terms of the stress-energy tensor, we find
\be \label{Ein_Eq_2}
R_{ab}=8\pi \left( T_{ab}-\frac{1}{2}T g_{ab} \right)
\ee
where $T={T^a}_a$ is the trace of the stress-energy tensor. For this text, we will focus on solutions to the vacuum Einstein equation. That is, Einstein's equation with no mass-energy present within the spacetime. In this case, equation \eqref{Ein_Eq_2} reduces to
\be \label{Vac_Ein_Eq}
R_{ab}=0 \quad .
\ee

\noindent We have now introduced the fundamentals of Einstein's theory of general relativity. We now begin our analysis of various spacetimes and their properties by looking at the Schwarzschild solution.
\chapter{The Schwarzschild solution}\label{C:Schwarzschild}

\section{Introduction to the Schwarzschild solution}

The Schwarzschild solution was derived by Karl Schwarzschild in 1916, just one year after Einstein published his formulation of general relativity in 1915. One can show via Birkhoff's theorem that the Schwarzschild solution is the unique spherically symmetric solution to the vacuum Einstein equation. In this text we will not derive the Schwarzschild solution, if one wishes to view the full derivation, we refer the reader to General Relativity, Wald, 1984. Here, we will just state the metric. In Hilbert's coordinates, the metric is given by the line element 
\be \label{Sch_Sch_Coords}
ds^2=-\left( 1-\frac{2m}{r} \right) dt^2 +\frac{1}{1-2m/r} dr^2 +r^2d\theta^2 +r^2\sin^2{\left(\theta\right)}\, d\phi^2
\ee
here $m$ is a parameter proportional to the mass of the central object, given by 
\be
m=\frac{2G_N M}{c^2}
\ee
where $M$ is the physical mass of the central object and $G_N$ is the Newtonian gravitational constant. Note that $m$ has units of length. This spacetime physically represents a spherically symmetric (i.e. non-rotating) object with mass parameter $m$ and no other matter present within the spacetime (if we wanted to include other matter sources, we would have to instead solve the full Einstein equation, equation \eqref{Ein_Eq}, which dramatically increases the complexity of the problem). Hence, the above metric, equation \eqref{Sch_Sch_Coords}, can be used to model non-rotating planets, stars and black holes. This solution is most often correlated with black holes however. To see this correlation, we conduct further examination of this metric.

By examining equation \eqref{Sch_Sch_Coords} we see the metric is singular at two distinct values of $r$, one at $r=0$ and the other at $r=2m$. However, we are interested in points that also reside within the radial position $r=2m$. However, the singularity at $r=2m$ means that all geodesics will terminate at $r=2m$. So if we wish to probe what happens when $r<2m$ by constructing geodesics that begin with radial component values $r>2m$ then allowing these geodesics to pass $r=2m$ into the region $r<2m$, we have to rid ourselves of the singularity at $r=2m$ first, which we shall now do.   

\section{Coordinate systems of Schwarzschild} \label{S:Sch_Coord}

In the previous section we saw that the Schwarzschild metric is singular at radial coordinate values of $r=0$ and $r=2m$. If we compute a curvature scalar such as:
\be
R^{abcd}R_{abcd}=\frac{48m^2}{r^6}
\ee
we see that this scalar is infinite as $r \rightarrow 0$ but is finite for $r=2m$. This shows that there exists a true physical singularity at $r=0$ but the singularity at $r=2m$ is but a coordinate singularity. We can rid ourselves of this coordinate singularity by a coordinate transformation.

Let $\Bar{t}=t+f(r)$, then $d\Bar{t}=dt+\frac{df(r)}{dr}dr$. Substituting this into the original metric, equation \eqref{Sch_Sch_Coords}, we get
\be
ds^2=-\left( 1-\frac{2m}{r} \right) \left( d\Bar{t}-\frac{df(r)}{dr}dr\right)^2 +\frac{1}{1-2m/r} dr^2 +r^2d\theta^2 +r^2\sin^2{\left(\theta\right)}\, d\phi^2.
\ee
Let 
\be
\frac{df(r)}{dr}=\frac{1}{1-2m/r} \sqrt{\frac{2m}{r}}
\ee
then the new metric becomes (dropping the bar from $\Bar{t}$):
\be \label{Sch_PG_Coords}
ds^2=-\left( 1-\frac{2m}{r} \right) dt^2 +2\sqrt{\frac{2m}{r}}dt dr + dr^2 + r^2 d\theta^2 + r^2\sin^2(\theta) d\phi^2 \quad .
\ee
Here we see that the metric components are no longer singular at $r=2m$, the only singularity occurs at $r=0$ where the physical singularity is. This new set of coordinates we have adopted here is called Painlev\'{e}-Gullstrand coordinates, which then expresses the metric in Painlev\'{e}-Gullstrand form. 

Now, we consider radial geodesics where we limit movement of our test particle to purely radial movement. That is $d\theta=d\phi=0$, so our metric simplifies to
\be
ds^2{}_{\text{rad}}=-\left( 1-\frac{2m}{r} \right) dt^2 +2\sqrt{\frac{2m}{r}}dt dr + dr^2 \quad .
\ee
If we further consider null geodesics where $ds^2=0$, then we find
\be
-\left( 1-\frac{2m}{r} \right) dt^2 +2\sqrt{\frac{2m}{r}}dt dr + dr^2=0
\ee
that is
\be
\left(\frac{dr}{dt}\right)^2+2\sqrt{\frac{2m}{r}}\frac{dr}{dt}-\left(1-\frac{2m}{r}\right)=0
\ee
solving for $dr/dt$ we have
\be
\frac{dr}{dt}=-\sqrt{\frac{2m}{r}}\pm 1 \quad .
\ee
We have two solutions here, if we look at points where $r>2m$ we see that if we take the positive root, we have $dr/dt>0$ so this represents outgoing light rays, whereas if we take the negative root, we have $dr/dt<0$ so this represents incoming light rays. Now we look at points where $r<2m$, for outgoing light rays we see that $dr/dt<0$ for $r<2m$ and for incoming light rays we also see that $dr/dt<0$ for $r<2m$. This physically means that when light passes the radial coordinate $r=2m$, it will continue to fall towards the physical singularity at $r=0$ no matter if it is outgoing or incoming. That is, light cannot go above the radial component value $r=2m$ once it has passed it, that light is then trapped within the region where $r<2m$. We can physically interpret this as a sphere surrounding the singularity which once an observer passes through, cannot return above this sphere. We call this the event horizon. Hence, we see that this metric can be used to represent a sphere with radius $2m$ which no object, not even light, can escape once it enters. This is, by definition, a black hole.\\

\noindent The new form of the metric given above, equation \eqref{Sch_PG_Coords}, has some other rather useful qualities. Firstly, the spatial part of the metric (i.e. the 3-dimensional metric where we ``throw away" the time components, also called the 3-metric) is just flat space written in spherical coordinates. This makes analysis of the 3-dimensional spatial hyper-surfaces of this spacetime trivial. Furthermore, the metric is given by 
\be
g_{ab}=
\begin{bmatrix}
-\left(1-\frac{2m}{r}\right) & \sqrt{\frac{2m}{r}} & 0 & 0 \\
\sqrt{\frac{2m}{r}} & 1 & 0 & 0 \\
0 & 0 & r^2 & 0 \\
0 & 0 & 0 & r^2\sin^2(\theta)
\end{bmatrix}
_{ab}
\ee
while the inverse metric is given by
\be
g^{ab}=
\begin{bmatrix}
-1 & \sqrt{\frac{2m}{r}} & 0 &0\\
\sqrt{\frac{2m}{r}} & -\left(1-\frac{2m}{r}\right) & 0 & 0\\
0 & 0 & \frac{1}{r^2} & 0\\
0 & 0 & 0 & \frac{1}{r^2\sin^2(\theta)}
\end{bmatrix}
^{ab} \quad .
\ee
Notice that both the metric and inverse metric are non-singular matrices with finite elements for $r\neq 0$ and $\theta\neq n \pi$ where $n\in \mathbb{N}$. Here we also see that the $tt$ component of the inverse metric is $g^{tt}=-1$, we call this condition \emph{unit lapse}. This condition is rather useful when calculating geodesics of our spacetime. More notably, the ``rain" geodesics of the Schwarzschild geometry become rather easy to calculate. Consider the vector field
\be
V^a=-g^{ab}\n_b t=-g^{ta}=\left(1; -\sqrt{\frac{2m}{r}}, 0, 0\right)
\ee
where the corresponding dual vector field is 
\be
V_a=-\n_a t=(-1;0,0,0) \quad .
\ee
Hence, $g_{ab}V^aV^b=V^aV_a=-1$, so $V^a$ is a timelike vector field with unit norm. This vector field has zero 4-acceleration
\be
A^a=V^b\n_bV^a=-V^b\n_b\n_at=-V^b\n_a\n_bt=V^b\n_aV^b=\frac{1}{2}\n_a(V^bV_b)=0
\ee
therefore the integral curves of $V^a$ are timelike geodesics. More specifically, the integral curves given by 
\be
\frac{dx^a}{d\tau}=\left(\frac{dt}{d\tau};\frac{dr}{d\tau},\frac{d\theta}{d\tau},\frac{d\phi}{d\tau}\right)=\left(1; -\sqrt{\frac{2m}{r}}, 0, 0\right)
\ee
are timelike geodesics. We can trivially integrate 3 of these equations
\be
t(\tau)=\tau; \quad \theta(\tau)=\theta_{\infty}; \quad \phi(\tau)=\phi_{\infty}
\ee
thus $t$ is just the proper time of these geodesics, while $\theta_{\infty}$ and $\phi_{\infty}$ are the original (and permanent) values of the $\theta$ and $\phi$ coordinate respectively for these geodesics. As for the last equation, we have
\be
\frac{1}{2}\left(\frac{dr}{dt}\right)^2=\frac{m}{r} \quad .
\ee
Therefore, these geodesics mimic a particle with zero initial velocity falling towards a mass ($m$) from spatial infinity.

\section{ISCOs and photon orbits of Schwarzschild} \label{S:Sch_ISCOs}

We now focus our attention on the innermost stable circular orbits, or ISCOs, of the Schwarzschild solution. Here we will use the Schwarzschild metric in Hilbert's coordinates, equation \eqref{Sch_Sch_Coords}, this is because while the Painlev\'{e}-Gullstrand coordinates did rid us of the singularity at $r=2m$, it introduced another metric component, the $dtdr$ term as seen in equation \eqref{Sch_PG_Coords}. Since, we are only considering geodesics above $r=2m$, we will use the original coordinates since in those coordinates the metric is diagonal, which will make the analysis easier in this case.\\

\noindent There are two symmetries present in the Schwarzschild geometry: time translation symmetry and spherical symmetry. These symmetries give rise to two Killing vectors (see appendix \ref{C:Killing}): the time translation Killing vector $\xi^a=(-1;0,0,0)$ and the spherical symmetry Killing vector $\psi^a=(0;0,0,1)$. As shown in appendix \ref{C:Killing}, we can construct conserved quantities from these Killing vectors. Here our two conserved quantities are, energy
\be \label{Sch_E}
E=g_{ab}\xi^a\frac{dx^b}{d\lambda}=-\left(1-\frac{2m}{r}\right)\frac{dt}{d\lambda}
\ee
and angular momentum
\be \label{Sch_L}
L=g_{ab}\psi^a\frac{dx^b}{dt}=r^2\sin^2(\theta)\frac{d\phi}{d\lambda} \quad .
\ee
With completely no loss of generality (since our spacetime is spherically symmetric), we may choose to conduct our analysis at the equator. That is to say, at the fixed angular coordinate $\theta=\pi/2\Rightarrow d\theta/d\lambda=0$. The timelike/null condition (equation \eqref{T_N_Con}), gives
\be \label{Sch_TN}
g_{ab}\frac{dx^a}{d\lambda}\frac{dx^b}{d\lambda}=-\left( 1-\frac{2m}{r} \right) \left(\frac{dt}{d\lambda}\right)^2 +\frac{1}{1-2m/r} \left(\frac{dr}{d\lambda}\right)^2 +r^2 \left(\frac{d\phi}{d\lambda}\right)^2=\epsilon
\ee
where
\begin{equation}
    \epsilon = \left\{
    \begin{array}{rl}
    -1 & \qquad\mbox{massive particle, \emph{i.e.} timelike geodesic} \\
     0 & \qquad\mbox{massless particle, \emph{i.e.} null geodesic} .
    \end{array}\right. 
\end{equation}
Treating equations \eqref{Sch_E}, \eqref{Sch_L} and \eqref{Sch_TN} as a linear system of equations, we can solve for $dr/d\lambda$:
\be
\frac{dr}{d\lambda}=\pm \sqrt{\left(1-\frac{2m}{r}\right)\left(\epsilon-\frac{L^2}{r^2}\right)+E} \quad .
\ee
This equation is what we would expect for a particle with unit mass in usual 1-dimensional, non-relativistic mechanics, under the influence of the following potential:
\be \label{Sch_ISCO_Pot}
V(r)=E^2-\left(\frac{dr}{d\lambda}\right)^2=\left(1-\frac{2m}{r}\right)\left(\frac{L^2}{r^2}-\epsilon\right)
\ee
we can then use the features of this potential to find the location of the timelike ISCO and innermost circular photon orbit (or photon ring).\\

\noindent In the null case where $\epsilon=0$, the potential reduces to
\be
V_0(r)=\frac{L^2}{r^2}\left(1-\frac{2m}{r}\right) \quad .
\ee
The photon ring will occur at the extrema of the potential, when $dV_0(r)/dr=0$. That is, when
\be
\frac{dV_0(r)}{dr}=\frac{(6m-2r)L^2}{r^4}=0
\ee
which has solution
\be
r=3m \quad .
\ee
To check the stability of this orbit, we calculate $d^2V_0(r)/dr^2$ and evaluate it at the photon ring location at $r=3m$. Which gives
\be
\left. \frac{d^2V_0(r)}{dr^2} \right|_{r=3m}=\left. \frac{(6r-24m)L^2}{r^5}\right|_{r=3m}=-\frac{2L^2}{81m^4}
\ee
which is always negative. Hence null geodesics at the photon ring location are unstable. Therefore, any massless particle traveling along a geodesic which falls below $r=3m$, will spiral in towards the physical singularity at $r=0$.\\

\noindent In the timelike case where $\epsilon=-1$, the potential reduces to
\be
V_{-1}(r)=\left(1-\frac{2m}{r}\right)\left(\frac{L^2}{r^2}+1\right) \quad .
\ee
Taking the derivative of the potential we find
\be
\frac{dV_{-1}(r)}{dr}= \frac{(6m-2r)L^2+2mr^2}{r^4} \quad .
\ee
Here we cannot simply set $dV_{-1}(r)/dr=0$ and solve directly for $r$ since this would yield a function in terms of $L$, i.e. $r(L)$. In order to find the ISCO location in terms of the properties of the spacetime itself, we instead solve $dV_{-1}(r)/dr=0$ for $L$ instead. This yields
\be \label{Sch_L(r)}
L(r)=\pm \frac{r\sqrt{(r-3m)m}}{r-3m} \quad .
\ee
The ISCO will occur when $dL/dr=0$, more explicitly, when
\be
\frac{dL(r)}{dr}=-\frac{\sqrt{m}}{2}\frac{(r-6m)}{(r-3m)^{3/2}}=0
\ee
which is satisfied when
\be
r=6m \quad .
\ee
To check the stability of this orbit, we calculate $d^2V_{-1}(r)/dr^2$
\be
\frac{d^2 V_{-1}(r)}{dr^2}=\frac{6(r-4m)L^2-4mr^2}{r^5}
\ee
using our expression for $L(r)$, equation \eqref{Sch_L(r)}, we have
\be
\frac{d^2 V_{-1}(r)}{dr^2}=\frac{2m(r-6m)}{r^3(r-3m)} \quad .
\ee
Now, at $r=6m$ we see that the second derivative is zero, however for points just above $6m$, the second derivative is positive and for points just below $6m$, the second derivative is negative. This shows that the ISCO is like a saddle point, if the particle is perturbed to point above $6m$, it will tend to move back to the ISCO location, whereas if it is perturbed to a point below $6m$, it will tend to spiral towards the event horizon and further towards the physical singularity at $r=0$.

\chapter{The Kerr solution}\label{C:Kerr}

\section{Introduction to the Kerr solution}
The Kerr solution is a generalisation to the Schwarzschild solution, where we now allow our central mass to rotate. This generalisation however, is highly non-trivial and this solution is arguably one of the most complex, physically relevant, exact solutions to the vacuum Einstein equation. It was discovered in 1963, that is 47 years after the discovery of the Schwarzschild solution in 1916. In the original coordinates that Kerr wrote this metric, it reads:
\be 
\begin{split}
ds^2 = & (r^2+a^2\cos^2(\theta))(d\theta^2+\sin^2(\theta)d\phi^2)\\
& +2(du-a\sin^2(\theta)d\phi)(dr-a\sin^2(\theta)d\phi)\\
& -\left(1-\frac{2mr}{r^2+a^2\cos^2(\theta)}\right)(du-a\sin^2(\theta)d\phi)^2
\end{split}
\ee 
where $a=J/m$ where $J$ is the angular momentum of the central mass ($m$). However, this coordinate system is rather difficult to physically analyse. The most common coordinate system that this metric is written in is the Boyer-Lindquist coordinate system
\be \label{Kerr_BL}
ds^2=-\left(1-\frac{2mr}{\rho^2}\right) dt^2 - \frac{4mar\sin^2(\theta)}{\rho^2} dtd\phi +\frac{\rho^2}{\Delta} dr^2 + \rho^2 d\theta^2 + \Sigma \sin^2(\theta) d\phi^2
\ee
where $\rho=\sqrt{r^2+a^2\cos^2(\theta)}$, $\Delta=r^2+a^2-2mr$ and $\Sigma=r^2+a^2+2mra^2\sin^2(\theta)/\rho^2$. Note that our definition of $\Sigma$ differs from the most common definition which is $\Sigma=r^2+a^2\cos(\theta)$, however we find the definition above more useful. As we can see, this metric is considerably more complex than the Schwarzschild metric. Hence, some physical quantities derived from this metric will be stated as opposed to derived in this text, with reference to the relevant derivation. 

\section{Unit-lapse versions of the Kerr spacetime}

In section \ref{S:Sch_Coord} we showed that the Schwarzschild solution admits a unit-lapse form of the spacetime metric. Recall, a metric is unit lapse if the following condition is satisfied: $g^{tt}=-1$. A typical metric can be written in the following form (using the ADM foliation):
\be
g_{ab}=\left[
\begin{array}{c|c}
-N^2+(h^{ij}v_iv_j) & -v_j \\
\hline
-v_i & h_{ij}
\end{array}\right]_{ab} ; \qquad
g^{ab}=\left[
\begin{array}{c|c}
-N^{-2} & -v^jN^{-2} \\
\hline
-v^iN^{-2} & h^{ij}-v^iv^jN^{-2}
\end{array}\right]^{ab}
\ee
Now, the general form of a metric which is unit-lapse can be written as follows:
\be
g_{ab}=\left[
\begin{array}{c|c}
-1+(h^{ij}v_iv_j) & -v_j \\
\hline
-v_i & h_{ij}
\end{array}\right]_{ab} ; \qquad
g^{ab}=\left[
\begin{array}{c|c}
-1 & -v^j \\
\hline
-v^i & h^{ij}-v^iv^j
\end{array}\right]^{ab}
\ee
where $h^{ij}=[h_{ij}]^{-1}$ and $v^i=h^{ij}v_j$. Also note that spacetime indices $a, b, c, d$ run $0...3$ while spatial indices $i, j$ run $1...3$. We call $h_{ij}$ the spatial metric or the 3-metric and physically represents the metric of the spatial hypersurfaces of constant $t$. $v_{i}$ is called the flow vector and is the negative of what is typically called the shift vector. So here we can see that the unit-lapse condition is equivalent to the condition that $N\rightarrow 1$. Written as a line element, a unit-lapse form of a spacetime can be written as
\be
ds^2=-dt^2+h_{ij}(dx^i-v^i dt)(dx^j-v^j dt) \quad .
\ee
As we saw for the Schwarzschild solution in section \ref{S:Sch_Coord}, once the metric has been put into unit-lapse form the rain geodesics of the spacetime can be easily calculated.\\

\noindent Before we perform any coordinate transformations on the Kerr metric, we must know which transformations we can perform and how these transformations affect the metric. The Kerr spacetime contains two symmetries: time translation symmetry and axial symmetry. Hence from appendix \ref{C:Killing}, these symmetries give rise to two Killing vectors:
\be
\xi^a=(-1;0,0,0); \qquad \psi^a=(0;0,0,1) \quad .
\ee
Since our spacetime contains symmetries, to be useful, our coordinate transformations should preserve these symmetries. This restricts our choice of transformations to be of the form
\be
t\rightarrow \bar{t}=t+T(r, \theta); \qquad \phi\rightarrow\bar{\phi}=\phi+\Phi(r, \theta);
\ee
\be
(r, \theta)\rightarrow(\bar{r},\bar{\theta})=(\bar{r}(r, \theta),\bar{\theta}(r, \theta)) \quad .
\ee
However, for now, we will only consider transformations on the $t$ and $\phi$ coordinates. Then we find
\be
dt\rightarrow d\bar{t}=dt+T_r dr + T_{\phi} d\phi; \qquad d\phi\rightarrow d\bar{\phi}=d\phi+\Phi_r dr + \Phi_{\phi} d\phi \quad .
\ee
We then find our Jacobi matrix to be 
\be
J^a{}_b=\frac{\pd \bar{x}^a}{\pd x^b}=
{\begin{bmatrix}
1 & T_r & T_{\phi} & 0\\
0 & 1 & 0& 0\\
0 & 0 & 1 & 0\\
0 & \Phi_r & \Phi_{\phi} & 1
\end{bmatrix}
^a}_b \, ; \qquad \text{det}(J^a{}_b)=1 \quad .
\ee

\subsection{Temporal-only transformations}

For now, we'll only consider temporal-\emph{only} coordinate transformations. Our Jacobi matrix reduces to
\be
J^a{}_b=\frac{\pd \bar{x}^a}{\pd x^b}=
{\begin{bmatrix}
1 & T_r & T_{\phi} & 0\\
0 & 1 & 0& 0\\
0 & 0 & 1 & 0\\
0 & 0 & 0 & 1
\end{bmatrix}
^a}_b \quad .
\ee
Transforming the inverse metric into the new coordinate system, we get
\be
\bar{g}^{ab}=J^a{}_c\, J^b{}_d\, g^{cd} \quad .
\ee
More specifically, the $g^{tt}$ component transforms as
\be
\bar{g}^{tt}=J^t{}_c\, J^t{}_d\, g^{cd}=g^{tt}+2T_i\,g^{ti}+T_i\,T_j\,g^{ij}=-N^{-2}(1+v^iT_i)^2+h^{ij}T_i\,T_j \quad .
\ee
Hence, to force the unit-lapse condition ($\bar{g}^{tt}=-1$), we have to solve the partial differential equation (PDE):
\be
-1=g^{tt}+2T_i\,g^{ti}+T_i\,T_j\,g^{ij}
\ee
or equivalently
\be
-1=-N^{-2}(1+v^iT_i)^2+h^{ij}T_i\,T_j \quad .
\ee
While the unit-lapse condition does simplify the process of calculating the rain geodesics of a given spacetime, there is a price to pay. Enforcing unit-lapse typically complicates the flow vector and the 3-metric. For the flow vector we have
\be
\bar{v}^i=-\bar{g}^{ti}=-J^t{}_c\,J^i{}_d\,g^{cd}=-J^t{}_t\,J^i{}_t\,g^{tt}-J^t{}_t\,J^i{}_j\,g^{tj}-J^t{}_k\,J^i{}_t\,g^{kt}-J^t{}_k\,J^i{}_l\,g^{kl} \quad .
\ee
However for temporal only transformations we have $J^t{}_t=1$, $J^i{}_t=0$, $J^t{}_i=T_i$ and $J^i{}_j=\delta^i{}_j$, then our flow vector reduces to:
\be
\bar{v}^i=-g^{ti}-J^t{}_j\,g_{ij}=\frac{v^i}{N^2}-T_j\left( h^{ij}-\frac{v^iv^j}{N^2}\right) =v^i\left(\frac{1+T_jv^j}{N^2}\right) -h^{ij}T_j \quad .
\ee
Furthermore, the 3-metric transforms as:
\be
\bar{h}_{ij}=\bar{g}_{ij}=J^a{}_i\,J^b{}_j\,g_{ab}=g_{ij}+g_{it}\,T_j+g_{jt}\,T_i+g_{tt}\,T_i\,T_j
\ee
which implies
\be
\bar{h}_{ij}=h_{ij}-v_i\,T_j-T_iv_j-(N^2-(h_{kl}v^kv^l))T_i\,T_j \quad .
\ee
Hence, we see that while enforcing the unit-lapse condition is useful in some situations, it carries the cost of complicating the flow vector and 3-metric significantly. 

\subsection{Azimuthal-only transformations} \label{Sec:Azi_Trans}

Given that by enforcing the unit-lapse condition, the 3-metric and flow vector has now been significantly complicated, we now wish to conduct further transformations to simplify these objects while retaining unit-lapse. Let us assume our metric is in unit-lapse form, we now use our freedom to transform the $\phi$ coordinate in order to simplify the 3-metric and flow vector. Leaving $t$ fixed and transforming the $\phi$ coordinate, we have
\be
J^a{}_b=\frac{\pd \bar{x}^a}{\pd x^b}=
{\begin{bmatrix}
1 & 0 & 0 & 0\\
0 & 1 & 0 & 0\\
0 & 0 & 1 & 0\\
0 & \Phi_r & \Phi_{\theta} & 1
\end{bmatrix}^a}_b \quad .
\ee
Now, the $tt$ component of the inverse metric has not changed during this process since
\be
\bar{g}^{tt}=J^t{}_c\, J^t{}_d\, g^{cd}=g^{tt} \quad .
\ee
The flow vector is not invariant under this transformation
\be
\bar{v}^i=-\bar{g}^{ti}=-J^t{}_c\,J^i{}_d\,g^{cd}=-J^t{}_t\,J^i{}_t\,g^{tt}-J^t{}_t\,J^i{}_j\,g^{tj}-J^t{}_k\,J^i{}_t\,g^{kt}-J^t{}_k\,J^i{}_l\,g^{kl} \quad .
\ee
However, here we have $J^i{}_t=0=J^t{}_i$, hence
\be
\bar{v}^i=-J^i{}_j\,g^{tj}=J^i{}_j\,v^j=v^i+(0, 0, \Phi_rv^r + \Phi_{\theta}v^{\theta})^i \quad .
\ee
That is, the $r$ and $\theta$ components on the flow vector are invariant, but the $\phi$ component is not $\bar{v}^{\phi}=v^{\phi} + \Phi_rv^r + \Phi_{\theta}v^{\theta}$. Hence, we can use the freedom in choosing our $\phi$ coordinate such to simplify the $\phi$ component of the flow vector. However, as we saw for enforcing unit-lapse, simplifying the metric in one regard typically complicates the metric in another regard. In this case, the 3-metric is transformed too in this process and, as a consequence, is typically complicated. For the inverse 3-metric we have:
\be \label{Kerr_UL_Phi_3_Met_1}
\bar{g}^{rr}=g^{rr}; \qquad \bar{g}^{r\theta}=g^{r\theta}; \qquad \bar{g}^{\theta\theta}=g^{\theta\theta};
\ee
\be \label{Kerr_UL_Phi_3_Met_2}
\bar{g}^{r\phi}=g^{r\phi}+g^{rr}\Phi_r+g^{r\theta}\Phi_{\theta}; \qquad \bar{g}^{\theta\phi}=g^{\theta\phi}+g^{\theta r}\Phi_r+g^{\theta\theta}\Phi_{\theta};
\ee
\be \label{Kerr_UL_Phi_3_Met_3}
\bar{g}^{\phi\phi}=g^{\phi\phi}+(g^{\phi r}\Phi_r+g^{\phi\theta}\Phi_{\theta})+(g^{rr}\Phi^2_r+2g^{r\theta}\Phi_r\Phi_{\theta}+g^{\theta\theta}\Phi^2_{\theta}) .
\ee
Hence, we see that simplifying the flow vector, does indeed complicate 3-metric. However, our arguments here have been quite general, we now look at applying these techniques to the Kerr spacetime specifically to find unit-lapse forms of the metric.

\subsection{Boyer-Lindquist-rain metric}
In Boyer-Lindquist coordinates the Kerr metric reads
\be 
(ds^2)_{\text{BL}}=-\left(1-\frac{2mr}{\rho^2}\right) dt^2 - \frac{4mar\sin^2(\theta)}{\rho^2} dtd\phi +\frac{\rho^2}{\Delta} dr^2 + \rho^2 d\theta^2 + \Sigma \sin^2(\theta) d\phi^2
\ee
where, as before, $\rho=\sqrt{r^2+a^2\cos^2(\theta)}$, $\Delta=r^2+a^2-2mr$ and $\Sigma=r^2+a^2+2mra^2\sin^2(\theta)/\rho^2$. Written as an array
\be
(g_{ab})_{\text{BL}}=
\left[ \begin{array}{c|cc|c}
-1+\frac{2mr}{\rho^2} & 0 & 0 & -\frac{2mar\sin^2(\theta)}{\rho^2}\\
\hline
0 & \frac{\rho^2}{\Delta} & 0 & 0\\
0 & 0 & \rho^2 & 0\\
\hline
-\frac{2mar\sin^2(\theta)}{\rho^2} & 0 & 0 & \Sigma\sin^2(\theta)
\end{array}\right]_{ab} \quad .
\ee
where 
\be
\text{det}[(g_{ab})_{\text{BL}}]=-\rho^4\sin^2(\theta)
\ee
as we would expect. The inverse metric as written as an array is given as
\be \label{Kerr_BL_inv_met}
(g^{ab})_{\text{BL}}=
\left[ \begin{array}{c|cc|c}
-1+\frac{2mr(r^2+a^2)}{\rho^2\Delta} & 0 & 0 & -\frac{2mar}{\rho^2\Delta}\\
\hline
0 & \frac{\Delta}{\rho^2} & 0 & 0\\
0 & 0 & \frac{1}{\rho^2} & 0\\
\hline
-\frac{2mar}{\rho^2\Delta} & 0 & 0 & \frac{1-2mr/\rho^2}{\Delta\sin^2(\theta)}
\end{array}\right]^{ab} \quad .
\ee
Hence we see that the Kerr metric in this form is not unit-lapse since $(g^{tt})_{\text{BL}}\neq -1$. Recall, to put this metric into unit-lapse form we solve  
\be
-1=-N^{-2}(1+v^iT_i)^2+h^{ij}T_iT_j \quad .
\ee
For this metric we have $v^iT_i=0$. Hence, this equation reduces to
\be
N^{-2}-1=h^{ij}T_iT_j \quad .
\ee
More explicitly
\be
\frac{2mr(r^2+a^2)}{\rho^2\Delta}=\frac{\Delta}{\rho^2}T^2_r+\frac{1}{\rho^2}T^2_{\theta} \quad .
\ee
Thence
\be
\frac{2mr(r^2+a^2)}{\Delta}=\Delta T^2_r+T^2_{\theta} \quad .
\ee
This equation has the following solution
\be
T_{\theta}=0; \qquad T_r=\pm \frac{\sqrt{2mr(r^2+a^2)}}{\Delta} \quad .
\ee
Hence, we see that $T(r, \theta)$ is independent of $\theta$ and therefore
\be
T(r)=\pm \int \frac{\sqrt{2mr(r^2+a^2)}}{\Delta}\; dr \quad .
\ee
Thence
\be
\bar{t}=t+T(r); \qquad d\bar{t}=dt+T_r dr; \qquad dt=d\bar{t}-T_r dr \quad .
\ee
If we suppress the overbar and take the Boyer-Lindquist form of the Kerr metric and make the replacement
\be
dt \rightarrow dt \mp \frac{\sqrt{2mr(r^2+a^2)}}{\Delta}\; dr
\ee
our resulting metric will then be in unit-lapse form. Notice that here we have two roots, hence we have to be careful as to which root we choose to take as one corresponds to a black hole and the other a white hole. If we retrospectively check, then we find that the negative root corresponds to the black hole, which we desire. The resulting metric is then
\be \label{Kerr_BL-rain}
\begin{split}
(ds^2)_{\text{BL-rain}} = & -\left(1-\frac{2mr}{\rho^2}\right)\left(dt-\frac{\sqrt{2mr(r^2+a^2)}}{\Delta} dr\right)^2\\
& -\frac{4mar\sin^2(\theta)}{\rho^2} d\phi \left(1-\frac{2mr}{\rho^2}\right)\left(dt-\frac{\sqrt{2mr(r^2+a^2)}}{\Delta} dr\right) \\
& + \frac{\rho^2}{\Delta} dr^2 + \rho^2 d\theta^2 +\Sigma\sin^2(\theta) d\phi^2 \quad .
\end{split}
\ee
Therefore, our covariant metric $(g_{ab})_{\text{BL-rain}}$ is given by
\be
(g_{ab})_{\text{BL-rain}}=\left[
\begin{array}{c|cc|c}
-1+\frac{2mr}{\rho^2} & g_{tr} & 0 & -\frac{2mar\sin^2(\theta)}{\rho^2} \\
\hline
g_{tr} & g_{rr} & 0 & g_{r\phi} \\
0 & 0 & \rho^2 & 0 \\
\hline
-\frac{2mar\sin^2(\theta)}{\rho^2} & g_{r\phi} & 0 & \Sigma\sin^2(\theta)
\end{array}
\right]_{ab}
\ee
where
\be
g_{tr}=\left(1-\frac{2mr}{\rho^2}\right)\frac{\sqrt{2mr(r^2+a^2)}}{\Delta}
\ee
\be
g_{rr}=\frac{\rho^2}{\Delta}-\left(1-\frac{2mr}{\rho^2}\right)\frac{2mr(r^2+a^2)}{\Delta^2}
\ee
\be
g_{r\phi}=\frac{2mar\sin^2(\theta)}{\rho^2}\frac{\sqrt{2mr(r^2+a^2)}}{\Delta}
\ee
and where we still have 
\be
\text{det}[(g_{ab})_{\text{BL-rain}}]=-\rho^4\sin^2(\theta) \quad .
\ee
The inverse metric is now relatively simple and is given by
\be
(g^{ab})_{\text{BL-rain}} = \left[
\begin{array}{c|cc|c}
-1 & \frac{\sqrt{2mr(r^2+a^2)}}{\rho^2} & 0 & -\frac{2mar}{\rho^2\Delta} \\
\hline
\frac{\sqrt{2mr(r^2+a^2)}}{\rho^2} & \frac{\Delta}{\rho^2} & 0 & 0 \\ 
0 & 0 & \frac{1}{\rho^2} & 0 \\
\hline
-\frac{2mar}{\rho^2\Delta} & 0 & 0 & \frac{1-2mr/\rho^2}{\Delta\sin^2(\theta)}
\end{array}
\right]^{ab} \quad .
\ee
Hence, we see that we have succeeded in putting our metric into unit-lapse form. However, this has come at a cost. The flow vector is now given by
\be
(v^i)_{\text{BL-rain}}=\left(-\frac{\sqrt{2mr(r^2+a^2)}}{\rho^2}, 0, \frac{2mar}{\rho^2\Delta}\right)
\ee
which by comparison to equation \ref{Kerr_BL_inv_met} is more complex than the flow vector in the usual Boyer-Lindquist coordinate system. Also note that for the rain geodesics $d\theta/dt=0$, hence $\theta(t)=\theta_{\infty}$ is constant along the geodesics. Furthermore
\be
\left(\frac{d\phi}{dr}\right)_{\text{BL-rain}} = \frac{d\phi/dt}{dr/dt} = -\frac{a\sqrt{2mr}}{\Delta\sqrt{r^2+a^2}} \quad .
\ee
Thence, for these geodesics
\be
\phi(r)=\phi_{\infty}+\int^{\infty}_r \frac{a\sqrt{2m\bar{r}}}{\Delta\sqrt{\bar{r}^2+a^2}}\; d\bar{r} \quad .
\ee

\subsection{Eddington–Finkelstein-rain metric}

The Eddington–Finkelstein coordinate system is the original coordinate system that Kerr wrote his solution in, which is given as 
\be 
\begin{split}
ds^2 = & (r^2+a^2\cos^2(\theta))(d\theta^2+\sin^2(\theta)d\phi^2)\\
& +2(du-a\sin^2(\theta)d\phi)(dr-a\sin^2(\theta)d\phi)\\
& -\left(1-\frac{2mr}{r^2+a^2\cos^2(\theta)}\right)(du-a\sin^2(\theta)d\phi)^2 \quad .
\end{split}
\ee 
However, as previously stated, this coordinate system is rather hard to physically interpret. We can make a slight change to these coordinates to transform them to the ``advanced Eddington–Finkelstein" coordinate system via the transformation
\be
u=t+r, \qquad du=dt+dr,
\ee
hence our metric becomes
\be
\begin{split}
(ds^2)_{\text{ad-EF}}= & -dt^2+dr^2-2a\sin^2(\theta) dr d\phi + \rho^2 d\theta^2 + (r^2+a^2)\sin^2(\theta) d\phi^2\\
& +\frac{2mr}{\rho^2}(dt+dr-a\sin^2(\theta) d\phi)^2 \quad .
\end{split}
\ee
In these coordinates, our covariant metric $(g_{ab})_{\text{ad-EF}}$ is given as
\be
\left[
\begin{array}{c|cc|c}
-1+\frac{2mr}{\rho^2} & \frac{2mr}{\rho^2} & 0 & -\frac{2mar}{\rho^2}\sin^2(\theta) \\
\hline
\frac{2mr}{\rho^2} & 1+\frac{2mr}{\rho^2} & 0 & -a(1+\frac{2mr}{\rho^2})\sin^2(\theta) \\
0 & 0 & \rho^2 & 0 \\
\hline
-\frac{2mar}{\rho^2}\sin^2(\theta) & -a(1+\frac{2mr}{\rho^2})\sin^2(\theta) & 0 & \Sigma\sin^2(\theta)
\end{array}
\right]_{ab} \quad .
\ee
While the inverse metric is given by
\be
(g^{ab})_{\text{ad-EF}}= \left[
\begin{array}{c|cc|c}
-1-\frac{2mr}{\rho^2} & \frac{2mr}{\rho^2} & 0 & 0 \\
\hline 
\frac{2mr}{\rho^2} & \frac{\Delta}{\rho^2} & 0 & \frac{a}{\rho^2} \\
0 & 0 & \frac{1}{\rho^2} & 0 \\
\hline
0 & \frac{a}{\rho^2} & 0 & \frac{1}{\rho^2\sin^2(\theta)}
\end{array}
\right]^{ab} \quad .
\ee
We can see, this metric is not in unit-lapse form. To force this condition, we have to solve 
\be
-1=g^{tt}+2T_ig^{ti}+T_iT_jg^{ij} \quad .
\ee
More specifically
\be
-1=-\left(1+\frac{2mr}{\rho^2}\right)+2T_r\frac{2mr}{\rho^2}+\frac{\Delta}{\rho^2}T^2_r+\frac{T^2_{\theta}}{\rho^2} \quad
\ee
Which simplifies to
\be
0=-2mr+4mrT_r+\Delta T^2_r+T^2_{\theta}
\ee
which has the solution
\be
T_{\theta}=0; \qquad T_r=\frac{-2mr\pm \sqrt{(2mr)^2+2mr\Delta}}{\Delta}=\frac{-2mr\pm \sqrt{2mr(r^2+a^2)}}{\Delta} \quad .
\ee
As before, we have to be careful to choose the root which corresponds to a black hole rather than a white hole. Now, $T_r$ can be recast in a slightly different form
\be
\begin{split}
\frac{-2mr\pm \sqrt{2mr(r^2+a^2)}}{\Delta} & =\frac{-2mr\pm \sqrt{2mr(r^2+a^2)}}{\Delta}\frac{-2mr\mp \sqrt{2mr(r^2+a^2)}}{-2mr\mp \sqrt{2mr(r^2+a^2)}}\\
& =\frac{-2mr}{-2mr\mp \sqrt{2mr(r^2+a^2)}}\\
& = \frac{2mr/(r^2+a^2)}{2mr/(r^2+a^2) \pm \sqrt{2mr/(r^2+a^2)}}\\
& = \frac{\sqrt{2mr/(r^2+a^2)}}{\sqrt{2mr/(r^2+a^2)}\pm 1}\\
& = \pm \frac{\sqrt{2mr/(r^2+a^2)}}{1\pm \sqrt{2mr/(r^2+a^2)}} \quad .
\end{split}
\ee
Therefore, $T(r, \theta)$ is independent of $\theta$ and we have
\be
T(r)=\pm \int \frac{\sqrt{2mr/(r^2+a^2)}}{1\pm \sqrt{2mr/(r^2+a^2)}} \; dr
\ee
Hence to put the metric into unit-lapse form we make the coordinate transformation (here we suppress the overbar)
\be
dt\rightarrow dt \mp \frac{\sqrt{2mr/(r^2+a^2)}}{1\pm \sqrt{2mr/(r^2+a^2)}} \; dr \quad .
\ee
Therefore, our metric, which we shall call the Eddington-Finkelstein-rain metric (or EF-rain), becomes
\be \label{Kerr_EF-rain}
\begin{split}
(ds^2)_{\text{EF-rain}}= & -\left(dt \mp \frac{\sqrt{2mr/(r^2+a^2)}}{1\pm \sqrt{2mr/(r^2+a^2)}} \; dr\right)^2\\
& + dr^2 -2a\sin^2(\theta) dr d\theta + \rho^2 d\theta^2 + (r^2+a^2)\sin^2(\theta) d\phi^2\\
& + \frac{2mr}{\rho^2} \left(dt + \left[ 1 \mp \frac{\sqrt{2mr/(r^2+a^2)}}{1\pm \sqrt{2mr/(r^2+a^2)}}\right] dr - a\sin^2(\theta) d\phi^2 \right)^2 .
\end{split}
\ee
Which simplifies to
\be
\begin{split}
(ds^2)_{\text{EF-rain}}= & -\left( dt\mp \frac{\sqrt{2mr/(r^2+a^2)}}{1\pm \sqrt{2mr/(r^2+a^2)}}\; dr\right)^2\\
& +dr^2-2a\sin^2(\theta) dr d\theta + \rho^2d\theta^2 + (r^2+a^2)\sin^2(\theta) d\theta^2\\
& +\frac{2mr}{\rho^2}\left( dt + \frac{dr}{1\pm \sqrt{2mr/(r^2+a^2)}} -a\sin^2(\theta) d\phi \right)^2 \quad .
\end{split}
\ee
Retrospectively, we can check that the upper sign corresponds to a black hole, hence we have
\be
\begin{split}
(ds^2)_{\text{EF-rain}}= & -\left( dt - \frac{\sqrt{2mr/(r^2+a^2)}}{1+ \sqrt{2mr/(r^2+a^2)}}\; dr\right)^2\\
& +dr^2-2a\sin^2(\theta) dr d\theta + \rho^2d\theta^2 + (r^2+a^2)\sin^2(\theta) d\theta^2\\
& +\frac{2mr}{\rho^2}\left( dt + \frac{dr}{1 + \sqrt{2mr/(r^2+a^2)}} -a\sin^2(\theta) d\phi \right)^2 \quad .
\end{split}
\ee
The covariant metric in these coordinates is given by
\be
(g_{ab})_{\text{EF-rain}}=\left[ 
\begin{array}{c|cc|c}
-1+\frac{2mr}{\rho^2} & g_{tr} & 0 & -\frac{2mar}{\rho^2}\sin^2(\theta) \\
\hline
g_{tr} & g_{rr} & 0 & g_{r\phi} \\
0 & 0 & \rho^2 & 0 \\
\hline
-\frac{2mar}{\rho^2}\sin^2(\theta) & g_{r\phi} & 0 & \Sigma\sin^2(\theta)
\end{array} \right]_{ab}
\ee
where
\be
g_{rr}=1 + \frac{a^2\sin^2(\theta)(2mr/\rho^2)}{(r^2+a^2)\left( 1+\sqrt{2mr/(r^2+a^2)}\right)^2} \quad ;
\ee
\be
g_{tr}=\frac{2mr/\rho^2+\sqrt{2mr/(r^2+a^2)}}{1+\sqrt{2mr/(r^2+a^2)}} \quad ;
\ee
\be
g_{r\phi}=-a\sin^2(\theta)\left(\frac{1+2mr/\rho^2+\sqrt{2mr/(r^2+a^2)}}{1+\sqrt{2mr/(r^2+a^2)}}\right) \quad .
\ee
The inverse metric is
\be
(g^{ab})_{\text{EF-rain}}=\left[ 
\begin{array}{c|cc|c}
-1 & \frac{\sqrt{2mr/(r^2+a^2)}}{\rho^2} & 0 & \frac{\sqrt{2mra^2/(r^2+a^2)}}{\rho^2\left(1+\sqrt{2mr/(r^2+a^2)}\right)} \\
\hline
\frac{\sqrt{2mr/(r^2+a^2)}}{\rho^2} & \frac{\Delta}{\rho^2} & 0 & \frac{a}{\rho^2} \\
0 & 0 & \frac{1}{\rho^2} & 0 \\
\hline
\frac{\sqrt{2mra^2/(r^2+a^2)}}{\rho^2\left(1+\sqrt{2mr/(r^2+a^2)}\right)} & \frac{a}{\rho^2} & 0 & \frac{1}{\rho^2\sin^2(\theta)}
\end{array} \right]^{ab} .
\ee
Hence, this metric has been put into unit-lapse form. However, there is a cost to this, the flow vector has now been complicated
\be
(v^i)_{\text{EF-rain}}=-\left( \frac{\sqrt{2mr(r^2+a^2)}}{\rho^2}, 0, \frac{\sqrt{2mra^2/(r^2+a^2)}}{\rho^2\left(1+\sqrt{2mr/(r^2+a^2)}\right)} \right) \quad .
\ee
Furthermore, for rain geodesics, we have $d\theta/dt=0$ hence $\theta(t)=\theta_{\infty}$ is conserved along the geodesic. Also, we have
\be
\left(\frac{d\phi}{dr}\right)_{\text{EF-rain}}=\frac{d\phi/dt}{dr/dt}=\frac{a}{(r^2+a^2)\left(1+\sqrt{2mr/(r^2+a^2)}\right)} \quad .
\ee
Hence
\be
\phi(r)=\phi_{\infty}-\int_r^{\infty} \frac{a}{(\bar{r}^2+a^2)\left(1+\sqrt{2m\bar{r}/(\bar{r}^2+a^2)}\right)} \; d\bar{r} \quad .
\ee

\subsection{Adjusting the flow vector}

We have used our freedom in our choice of time coordinate to put our metrics into unit-lapse form, we now use the freedom in our choice of azimuthal coordinate to simplify the flow vector $v^i$. In general, under a change in in azimuthal coordinate our flow vector transforms as $v^{\phi}\rightarrow \bar{v}^{\phi}=v^{\phi}+\Phi_rv^r+\Phi_{\theta}v^{\theta}$. However, for both the BL-rain and EL-rain metrics we have $v^{\theta}=0$. Hence in this case, we have $v^{\phi}\rightarrow \bar{v}^{\phi}=v^{\phi}+\Phi_rv^r$. Furthermore, for the BL-rain and EL-rain metrics, the only angular dependence in the $v^r$ and $v^{\theta}$ terms arises from the common factor $1/\rho^2$. Therefore, this indicates that we can eliminate $v^{\phi}$ entirely by choosing a suitable transformation of the $\phi$ coordinate $\bar{\phi}=\phi+\Phi(r)$. Hence, we will use this freedom in our choice of $\phi$ coordinate to derive the Doran form of the Kerr metric from the BL-rain, EF-rain and the EF metrics.

The Doran form of the Kerr metric is one of the most common unit-lapse forms of the Kerr metric and is given as
\be \label{Kerr_Doran}
\begin{split}
(ds^2)_{\text{Doran}} = & -dt^2 + \rho^2 d\theta^2 + (r^2+a^2)\sin^2(\theta) d\phi^2\\
& + \left\{ \frac{\rho}{\sqrt{r^2+a^2}}dr + \frac{\sqrt{2mr}}{\rho}(dt-a\sin^2(\theta)d\phi)\right\}^2 \quad .
\end{split}
\ee
The covariant metric is given by
\be
(g_{ab})_{\text{Doran}}=\left[
\begin{array}{c|cc|c}
-1+\frac{2mr}{\rho^2} & \sqrt{\frac{2mr}{a^2+r^2}} & 0 & -\frac{2mar\sin^2(\theta)}{\rho^2} \\
\hline
\sqrt{\frac{2mr}{a^2+r^2}} & \frac{\rho^2}{r^2+a^2} & 0 & -a\sqrt{\frac{2mr}{a^2+r^2}}\sin^2(\theta) \\
0 & 0 & \rho^2 & 0 \\
-\frac{2mar\sin^2(\theta)}{\rho^2} & -a\sqrt{\frac{2mr}{a^2+r^2}}\sin^2(\theta) & 0 & \Sigma\sin^2(\theta)
\end{array} \right]_{ab}
\ee
with inverse metric
\be
(g^{ab})_{\text{Doran}}=\left[
\begin{array}{c|cc|c}
-1 & \frac{\sqrt{2mr(a^2+r^2)}}{\rho^2} & 0 & 0 \\
\hline
\frac{\sqrt{2mr(a^2+r^2)}}{\rho^2} & \frac{\Delta}{\rho^2} & 0 & \frac{a}{\rho^2}\sqrt{\frac{2mr}{a^2+r^2}}\\
0 & 0 & \frac{1}{\rho^2} & 0 \\
\hline
0 & \frac{a}{\rho^2}\sqrt{\frac{2mr}{a^2+r^2}} & 0 & \frac{1}{(a^2+r^2)\sin^2(\theta)}
\end{array} \right]^{ab} \quad .
\ee
Hence, we see that this metric is indeed unit-lapse. 

Let us start from the BL-rain (inverse) metric, given by 
\be
(g^{ab})_{\text{BL-rain}} = \left[
\begin{array}{c|cc|c}
-1 & \frac{\sqrt{2mr(r^2+a^2)}}{\rho^2} & 0 & -\frac{2mar}{\rho^2\Delta} \\
\hline
\frac{\sqrt{2mr(r^2+a^2)}}{\rho^2} & \frac{\Delta}{\rho^2} & 0 & 0 \\ 
0 & 0 & \frac{1}{\rho^2} & 0 \\
\hline
-\frac{2mar}{\rho^2\Delta} & 0 & 0 & \frac{1-2mr/\rho^2}{\Delta\sin^2(\theta)}
\end{array}
\right]^{ab} 
\ee
where the flow vector is given by
\be
(v^i)_{\text{BL-rain}}=\left(-\frac{\sqrt{2mr(r^2+a^2)}}{\rho^2}, 0, \frac{2mar}{\rho^2\Delta}\right) \quad .
\ee
If we choose
\be
\Phi_r=-\left(\frac{v^{\phi}}{v^r}\right)_{\text{BL-rain}}=\frac{a}{\Delta}\sqrt{\frac{2mr}{r^2+a^2}}; \qquad \Phi(r)=\int \frac{a}{\Delta}\sqrt{\frac{2mr}{r^2+a^2}} \; dr
\ee
then $\bar{v}^{\phi}\rightarrow 0$. However, by equations \eqref{Kerr_UL_Phi_3_Met_1} - \eqref{Kerr_UL_Phi_3_Met_3}, we see that the 3-metric will now be complicated via this transformation. But by conducting this transformation, the BL-rain (inverse) metric now becomes
\be
(g^{ab})_{\text{Doran}}=\left[
\begin{array}{c|cc|c}
-1 & \frac{\sqrt{2mr(a^2+r^2)}}{\rho^2} & 0 & 0 \\
\hline
\frac{\sqrt{2mr(a^2+r^2)}}{\rho^2} & \frac{\Delta}{\rho^2} & 0 & \frac{a}{\rho^2}\sqrt{\frac{2mr}{a^2+r^2}}\\
0 & 0 & \frac{1}{\rho^2} & 0 \\
\hline
0 & \frac{a}{\rho^2}\sqrt{\frac{2mr}{a^2+r^2}} & 0 & \frac{1}{(a^2+r^2)\sin^2(\theta)}
\end{array} \right]^{ab} \quad .
\ee
Hence, starting from the BL-rain metric, we can transform the metric into Doran form. This implies that starting from the Boyer-Lindquist form of the Kerr metric, we can transform the metric into Doran form via the following coordinate transformations:
\be
dt\rightarrow dt-\frac{\sqrt{2mr(r^2+a^2)}}{\Delta}\; dr \quad ,
\ee
\be
d\phi \rightarrow d\phi-\frac{a}{\Delta}\sqrt{\frac{2mr}{r^2+a^2}} \; dr \quad .
\ee
\\

\noindent If we now start from the EF-rain (inverse) metric, which is given as 
\be
(g^{ab})_{\text{EF-rain}}=\left[ 
\begin{array}{c|cc|c}
-1 & \frac{\sqrt{2mr/(r^2+a^2)}}{\rho^2} & 0 & \frac{\sqrt{2mra^2/(r^2+a^2)}}{\rho^2\left(1+\sqrt{2mr/(r^2+a^2)}\right)} \\
\hline
\frac{\sqrt{2mr/(r^2+a^2)}}{\rho^2} & \frac{\Delta}{\rho^2} & 0 & \frac{a}{\rho^2} \\
0 & 0 & \frac{1}{\rho^2} & 0 \\
\hline
\frac{\sqrt{2mra^2/(r^2+a^2)}}{\rho^2\left(1+\sqrt{2mr/(r^2+a^2)}\right)} & \frac{a}{\rho^2} & 0 & \frac{1}{\rho^2\sin^2(\theta)}
\end{array} \right]^{ab}
\ee
where the flow vector is given by
\be
(v^i)_{\text{EF-rain}}=-\left( \frac{\sqrt{2mr(r^2+a^2)}}{\rho^2}, 0, \frac{\sqrt{2mra^2/(r^2+a^2)}}{\rho^2\left(1+\sqrt{2mr/(r^2+a^2)}\right)} \right) \quad .
\ee
If we choose
\be
\Phi_r=-\left( \frac{v^{\phi}}{v^r}\right)_{\text{EF-rain}}=-\frac{a/(r^2+a^2)}{1+\sqrt{2mr/(r^2+a^2)}}
\ee
such that 
\be
\Phi(r)=-\int \frac{a/(r^2+a^2)}{1+\sqrt{2mr/(r^2+a^2)}} \; dr \quad ,
\ee
then $\bar{v}^{\phi}\rightarrow 0$. However, as we saw for the BL-rain metric, this transformation complicates the 3-metric. After conducting this transformation, our EF-rain inverse metric now becomes
\be
(g^{ab})_{\text{Doran}}=\left[
\begin{array}{c|cc|c}
-1 & \frac{\sqrt{2mr(a^2+r^2)}}{\rho^2} & 0 & 0 \\
\hline
\frac{\sqrt{2mr(a^2+r^2)}}{\rho^2} & \frac{\Delta}{\rho^2} & 0 & \frac{a}{\rho^2}\sqrt{\frac{2mr}{a^2+r^2}}\\
0 & 0 & \frac{1}{\rho^2} & 0 \\
\hline
0 & \frac{a}{\rho^2}\sqrt{\frac{2mr}{a^2+r^2}} & 0 & \frac{1}{(a^2+r^2)\sin^2(\theta)}
\end{array} \right]^{ab} \quad .
\ee
Hence, starting from the EF-rain metric, we can transform the metric into Doran form. This implies that starting from the advanced Eddington-Finkelstein form of the Kerr metric, we can transform the metric into Doran form via the following coordinate transformations:
\be
dt\rightarrow dt-\frac{\sqrt{2mr/(r^2+a^2)}}{1+\sqrt{2mr/(r^2+a^2)}}\; dr \quad ,
\ee
\be
d\phi \rightarrow d\phi-\frac{a/(r^2+a^2)}{1+\sqrt{2mr/(r^2+a^2)}} \; dr \quad .
\ee
However, the Kerr metric was originally written in Eddington-Finkelstein coordinates as 
\be 
\begin{split}
ds^2 = & (r^2+a^2\cos^2(\theta))(d\theta^2+\sin^2(\theta)d\phi^2)\\
& +2(du-a\sin^2(\theta)d\phi)(dr-a\sin^2(\theta)d\phi)\\
& -\left(1-\frac{2mr}{r^2+a^2\cos^2(\theta)}\right)(du-a\sin^2(\theta)d\phi)^2 \quad .
\end{split}
\ee 
This metric can be transformed into Doran form by firstly transforming to advanced Eddington-Finkelstein coordinates, then to Doran coordinates via the following transformation
\be
du\rightarrow dt + \frac{dr}{1+\sqrt{2mr(r^2+a^2)}} \quad ,
\ee
\be
d\phi \rightarrow d\phi + \frac{a \; dr}{r^2+a^2+\sqrt{2mr(r^2+a^2)}} \quad .
\ee

\subsection{Nat\'{a}rio metric}

There exists yet another unit-lapse form of the Kerr metric, the Nat\'{a}rio form. If we start from Boyer-Lindquist coordinates and make the following coordinate transformation
\be
dt\rightarrow dt-\frac{2mr(r^2+a^2)}{\Delta} \quad ,
\ee
\be
d\phi\rightarrow d\phi + \Phi_r dr + \Phi_{\theta} d\theta \quad , 
\ee
then the resulting metric is
\be \label{Kerr_Natario}
(ds^2)_{\text{Nat\'{a}rio}}= -dt^2+\frac{\rho^2}{\Sigma}(dr-v \, dt)^2 + \rho^2 d\theta^2 + \Sigma\sin^2(\theta)(d\phi+\delta d\theta-\Omega dt)^2 \quad .
\ee
Notice that the temporal coordinate transformation above just takes the Kerr metric in Boyer-Lindquist form to the Boyer-Lindquist-rain metric, the azimuthal transformation however, is different from what we have seen for the Doran metric. Here Nat\'{a}rio chose to set
\be
(v^{\phi})_{\text{Nat\'{a}rio}}=\Omega=\frac{2mra}{\rho^2\Sigma} \quad .
\ee
This choice then causes the form of $\Phi(r, \theta)$ to be rather complicated. More specifically
\be
\Phi(r, \theta)=\frac{(v^{\phi})_{\text{Nat\'{a}rio}}-(v^{\phi})_{\text{BL-rain}}}{(v^{r})_{\text{BL-rain}}}=\frac{(v^{\phi})_{\text{Nat\'{a}rio}}-(v^{\phi})_{\text{BL-rain}}}{(v^{r})_{\text{Nat\'{a}rio}}} \quad ,
\ee
we could then substitute these known variables and explicitly integrate, but this exercise doesn't give too much relevant information. The quantity 
\be
v=-\frac{2mr(r^2+a^2)}{\rho^2} \quad ,
\ee
is tractable, however the quantity
\be
\delta(r, \theta)=-a^2\sin(2\theta)\int^{\infty}_{r} \frac{v\Omega}{\Sigma}\, d\bar{r} \quad ,
\ee
is not. More explicitly 
\be
\delta(r, \theta)=-a^2\sin(2\theta) \int^{\infty}_r \frac{2ma\bar{r}\sqrt{2m\bar{r}(\bar{r}^2+a^2)}}{\left[ (\bar{r}^2+a^2)(\bar{r}^2+a^2\cos^2(\theta))+2\sin^2(\theta)ma^2\bar{r}\right]^2} \, d\bar{r} \quad ,
\ee
this integration then gives incomplete Elliptic integrals. However, we notice that the quantity $\delta(r, \theta)$ is present in the metric, hence this implies the existence of incomplete Elliptic integrals \emph{in the metric components}. This makes the Nat\'{a}rio metric rather troublesome to work with since we are then working with implicit metric components rather than explicit components. 

The covariant metric is given as
\be
(g_{ab})_{\text{Nat\'{a}rio}}=\left[ 
\begin{array}{c|cc|c}
-1+\frac{\rho^2v^2}{\Sigma} & -\frac{\rho^2v}{\Sigma} & -\delta\Sigma\sin^2(\theta)\Omega & -\Sigma\sin^2(\theta)\Omega \\
\hline
-\frac{\rho^2v}{\Sigma} & \frac{\rho^2}{\Sigma} & 0 & 0 \\
-\delta\Sigma\sin^2(\theta)\Omega & 0 & \rho^2+\delta^2\Sigma\sin^2(\theta) & \delta\Sigma\sin^2(\theta) \\
\hline
-\Sigma\sin^2(\theta)\Omega & 0 & \delta\Sigma\sin^2(\theta) & \Sigma\sin^2(\theta)
\end{array}
\right]_{ab}
\ee
with determinant 
\be
\text{det}[(g_{ab})_{\text{Nat\'{a}rio}}]=-\rho^4\sin^2(\theta)
\ee
as we expect. The inverse metric is 
\be
(g^{ab})_{\text{Nat\'{a}rio}}=\left[ 
\begin{array}{c|cc|c}
-1 & -v & 0 & -\Omega \\
\hline
-v & \frac{\Sigma}{\rho^2}-v^2 & 0 & -\Omega v \\
0 & 0 & \frac{1}{\rho^2} & -\frac{\delta}{\rho^2} \\
\hline
-\Omega & -\Omega v & -\frac{\delta}{\rho^2} & \frac{1}{\Sigma\sin^2(\theta)}+\frac{\delta^2}{\rho^2}-\Omega^2
\end{array}
\right]^{ab}
\ee
hence the metric is unit-lapse, as advertised. 

\subsection{General unit-lapse form of the Kerr metric}

We can now see that we can quite easily create an infinite number of unit-lapse forms of the Kerr metric by giving a general representation of the Kerr metric in unit-lapse form. To do this, we just have to start with any of the pre-existing unit-lapse forms of the Kerr metric (be it BL-rain, EF-rain, Doran, or Nat\'{a}rio) then make the coordinate transformation $\phi\rightarrow \phi-\Phi(r, \theta)$ for some arbitrary function $\Phi(r, \theta)$, while leaving the $t$ coordinate unchanged. That is to say, in the line element, we make the following replacement
\be
d\phi\rightarrow d\phi-\Phi_r dr-\Phi_{\theta} d\theta \quad .
\ee
As an example, we shall do this replacement for the Doran form of the Kerr metric. We get
\be
\begin{split}
(ds^2)_{\text{General}}= & -dt^2 + \rho^2 d\theta^2 + (r^2+a^2)\sin^2(\theta)(d\phi - \Phi_r dr - \Phi_{\theta} d\theta)^2\\
& \left[ \frac{\rho dr}{\sqrt{r^2+a^2}} + \frac{\sqrt{2mr}}{\rho} (dt - a\sin^2(\theta)(d\phi - \Phi_r dr - \Phi_{\theta} d\theta))\right]^2 \quad .
\end{split}
\ee
We shall write 
\be
(g_{ab})_{\text{General}}=(g_{ab})_{\text{Doran}} + \Delta_1(g_{ab}) + \Delta_2(g_{ab})
\ee
where the first and second order shifts (linear and quadratic in gradients of $\Phi$ respectively) are given by
\be
\Delta_1(g_{ab})=\sin^2(\theta)\left[
\begin{array}{c|cc|c}
0 & \frac{2mar}{\rho^2} \Phi_r & \frac{2mar}{\rho^2} \Phi_{\theta} & 0 \\
\hline
\frac{2mar}{\rho^2} \Phi_r & 2a\sqrt{\frac{2mar}{r^2+a^2}} \Phi_r & a \sqrt{\frac{2mar}{r^2+a^2}} \Phi_{\theta} & -\Sigma \Phi_r \\
\frac{2mar}{\rho^2} \Phi_{\theta} & a \sqrt{\frac{2mar}{r^2+a^2}} \Phi_{\theta} & 0 & -\Sigma \Phi_{\theta} \\
\hline
0 & -\Sigma \Phi_r & -\Sigma \Phi_{\theta} & 0
\end{array}
\right]_{ab}
\ee
and
\be
\Delta_2(g_{ab})=\Sigma \sin^2(\theta) \left[ 
\begin{array}{c|cc|c}
0 & 0 & 0 & 0 \\
\hline
0 & \Phi^2_r & \Phi_r\Phi_{\theta} & 0 \\
0 & \Phi_r\Phi_{\theta} & \Phi^2_{\theta} & 0 \\
\hline
0 & 0 & 0 & 0
\end{array}
\right]_{ab} = \Sigma \sin^2(\theta) \Phi_a \Phi_b \quad .
\ee
We can now simply invert $(g_{ab})_{\text{General}}$ to find $(g^{ab})_{\text{General}}$. We write 
\be
(g^{ab})_{\text{General}}=(g^{ab})_{\text{Doran}} + \Delta_1(g^{ab}) + \Delta_2(g^{ab})
\ee
where the first and second order shifts are given by
\be
\Delta_1(g^{ab})=\frac{1}{\rho} \left[ 
\begin{array}{c|cc|c}
0 & 0 & 0 & \sqrt{2mr(r^2+a^2)} \Phi_r \\
\hline
0 & 0 & 0 & \Delta \Phi_r \\
0 & 0 & 0 & \Phi_{\theta} \\
\hline
\sqrt{2mr(r^2+a^2)} \Phi_r & \Delta \Phi_r & \Phi_{\theta} & 2a\sqrt{\frac{2mr}{r^2+a^2}} \Phi_r
\end{array}
\right]^{ab}
\ee
and 
\be
\Delta_2(g^{ab})=\frac{\Delta \Phi^2_r + \Phi^2_{\theta}}{\rho^2} \left[
\begin{array}{c|cc|c}
0 & 0 & 0 & 0 \\
\hline
0 & 0 & 0 & 0 \\
0 & 0 & 0 & 0 \\
\hline
0 & 0 & 0 & 1 \\
\end{array}
\right]^{ab} \quad .
\ee
A coordinate transformation of this form is the most general coordinate transformation we can perform while: retaining unit-lapse, preserving axial-symmetry, and retaining the usual oblate spheroidal coordinates $(r, \theta,$ $\phi)$.

\section{Painlev\'{e}-Gullstrand form of the Kerr metric}

We now look at the possibility of the existence of a Painlev\'{e}-Gullstrand form of the Kerr metric. Recall, a metric is in Painlev\'{e}-Gullstrand form if it has unit-lapse and if the 3-metric is diagonal. In the last section we analysed many unit-lapse forms of the Kerr metric. However, they all had one thing in common, the 3-metrics of all the above mentioned unit-lapse forms of the Kerr metric are not diagonal. Hence, the question is: can we diagonalise the 3-metric of one of these unit-lapse forms of the Kerr metric to yield the Painlev\'{e}-Gullstrand form of the Kerr metric? A classical mathematical result due to Darboux is that the 3-metric of a manifold can be diagonalised under some mild conditions. This seems to suggest that this may be applicable to the Kerr metric, yielding the desired result. A metric has 10 independent components, by enforcing unit-lapse and requiring the 3-metric to be diagonal, we enforce 4 conditions on the metric while leaving 6 degrees of freedom in the metric components. This naively seems plausible to enforce such conditions on the Kerr metric.

In section \ref{Sec:Azi_Trans} we saw that azimuthal-only transformations leave the $g^{tt}$ component of the inverse metric invariant. If fact, if we were to make any general spatial coordinate transformation (that is leaving the time coordinate unchanged), then the $g^{tt}$ component of the inverse metric will be invariant under this transformation. So our tactic in attempting to construct a Painlev\'{e}-Gullstrand form of the Kerr metric is to assume our metric is already in unit-lapse form, then make a general spatial coordinate transformation and demand the off-diagonal components of the 3-metric to be zero, then solve for the analytic functions which gives this coordinate transformation. We start by looking at a general azimuthal only coordinate transformation.

\subsection{Azimuthal-only coordinate transformation}

We shall make a general azimuthal only coordinate transformation, while leaving the $r$ and $\theta$ coordinates unchanged. So we have
\be
r=\bar{r}; \qquad \theta=\bar{\theta}; \qquad \phi=\bar{\phi}+\Phi(\bar{r},\bar{\theta})
\ee
therefore
\be
dr=d\bar{r}; \qquad d\theta=d\bar{\theta}; \qquad d\phi=d\bar{\phi}+\Phi_{\bar{r}} d\bar{r}+\Phi_{\bar{\theta}} d\bar{\theta} \quad .
\ee
Hence, we can write the 3-metric as
\be
\begin{split}
ds^2= & g_{rr}(\bar{r},\bar{\theta}) d\bar{r}^2+g_{\theta \theta}(\bar{r},\bar{\theta}) d\bar{\theta}^2 + g_{\phi \phi}(\bar{r},\bar{\theta}) (d\bar{\phi}+\Phi_{\bar{r}} d\bar{r}+\Phi_{\bar{\theta}}d\bar{\theta})^2\\
& + 2g_{r \phi}(\bar{r},\bar{\theta}) d\bar{r} (d\bar{\phi}+\Phi_{\bar{r}} d\bar{r}+\Phi_{\bar{\theta}}d\bar{\theta}) \quad .
\end{split}
\ee
Hence, the off-diagonal components are
\begin{align}
g_{\bar{r}\bar{\phi}}= & g_{r\phi}(\bar{r},\bar{\theta})+g_{\phi\phi}(\bar{r},\bar{\theta}) \Phi_{\bar{r}}\\
g_{\bar{\theta}\bar{\phi}}= & g_{\phi\phi}(\bar{r},\bar{\theta}) \Phi_{\bar{\theta}}\\
g_{\bar{r}\bar{\theta}}= & \Phi_{\bar{\theta}}(g_{\phi\phi}(\bar{r},\bar{\theta}) \Phi_{\bar{r}}+g_{r\phi}(\bar{r},\bar{\theta})) \quad .
\end{align}
We now demand all of these components to vanish. That is, we demand the following equations to hold
\begin{align}
E1: & \qquad \qquad g_{r\phi}(\bar{r},\bar{\theta})+g_{\phi\phi}(\bar{r},\bar{\theta}) \Phi_{\bar{r}}=0\\
E2: & \qquad \qquad g_{\phi\phi}(\bar{r},\bar{\theta}) \Phi_{\bar{\theta}}=0\\
E3: & \qquad \qquad \Phi_{\bar{\theta}}(g_{\phi\phi}(\bar{r},\bar{\theta}) \Phi_{\bar{r}}+g_{r\phi}(\bar{r},\bar{\theta}))=0 \quad .
\end{align}
Now, we know that $g_{\phi\phi}\neq 0$, hence, from $E2$ we have $\Phi_{\bar{\theta}}=0$. This guarantees that $E3$ is satisfied. However, this implies that $\Phi(\bar{r},\bar{\theta})\rightarrow \Phi(\bar{r})$, so from $E1$ we have
\be
\frac{d\Phi(\bar{r})}{d\bar{r}}=-\frac{g_{r\phi}(\bar{r},\bar{\theta})}{g_{\phi\phi}(\bar{r},\bar{\theta})}
\ee
This equation is consistent iff the RHS is a purely a function of $\bar{r}$, that is if
\be
\pd_{\bar{\theta}}\left(\frac{g_{r\phi}(\bar{r},\bar{\theta})}{g_{\phi\phi}(\bar{r},\bar{\theta})}\right)=0 \quad .
\ee
However, note that we left the $r$ and $\theta$ coordinates unchanged, hence this condition is equivalent to
\be \label{PG_Kerr_Const_Con_1}
\pd_{\theta}\left(\frac{g_{r\phi}(r,\theta)}{g_{\phi\phi}(r,\theta)}\right)=0 \quad .
\ee
However, via inspection of the BL-rain, EF-rain and Doran metrics, we find
\begin{align}
\left(\frac{g_{r\phi}(r,\theta)}{g_{\phi\phi}(r,\theta)}\right)_{\text{BL-rain}}= & \frac{2mar\sqrt{2mr(r^2+a^2)}}{\Delta\rho^2\Sigma}\\
\left(\frac{g_{r\phi}(r,\theta)}{g_{\phi\phi}(r,\theta)}\right)_{\text{EF-rain}}= & -\frac{a}{\Sigma}\left(\frac{1+2mr/\rho^2+\sqrt{2mr/(r^2+a^2)}}{1+\sqrt{2mr/(r^2+a^2)}}\right)\\
\left(\frac{g_{r\phi}(r,\theta)}{g_{\phi\phi}(r,\theta)}\right)_{\text{Doran}}= & -\frac{a}{\Sigma}\sqrt{\frac{2mr}{r^2+a^2}} \quad .
\end{align}
Recall that $\rho=\sqrt{r^2+a^2\cos^2(\theta)}$, $\Delta=r^2+a^2-2mr$ and $\Sigma=r^2+a^2+2mra^2\sin^2(\theta)/\rho^2$. Hence there is $\theta$ dependence in both $\rho$ and $\Sigma$. Therefore, the consistency condition in equation \ref{PG_Kerr_Const_Con_1} is not satisfied. Hence, we cannot construct a Painlev\'{e}-Gullstrand form of the Kerr metric via a azimuthal-only coordinate transformation.

\subsection{Polar-only $(r,\theta)$ coordinate transformation}

We shall now conduct a polar-only coordinate transformation, that is transform the $(r, \theta)$ while keeping the $\phi$ coordinate unchanged. So we have
\be
r=G(\bar{r},\bar{\theta}); \qquad \theta=H(\bar{r},\bar{\theta}); \qquad \phi=\bar{\phi}
\ee
therefore
\be
dr=G_{\bar{r}} d\bar{r} + G_{\bar{\theta}} d\bar{\theta}; \qquad d\theta=H_{\bar{r}} d\bar{r} + H_{\bar{\theta}} d\bar{\theta}; \qquad d\phi=d\bar{\phi} \quad .
\ee
Hence, we can write the 3-metric as
\be
\begin{split}
ds^2= & g_{rr}(G(\bar{r},\bar{\theta}),H(\bar{r},\bar{\theta}))(G_{\bar{r}} d\bar{r}+G_{\bar{\theta}} d\bar{\theta})^2\\
& +g_{\theta\theta}(G(\bar{r},\bar{\theta}),H(\bar{r},\bar{\theta}))(H_{\bar{r}} d\bar{r}+H_{\bar{\theta}} d\bar{\theta})^2\\
& +g_{\phi\phi}(G(\bar{r},\bar{\theta}),H(\bar{r},\bar{\theta})) d\bar{\phi}^2\\
& +2g_{r\phi}(G(\bar{r},\bar{\theta}),H(\bar{r},\bar{\theta}))(G_{\bar{r}} d\bar{r}+G_{\bar{\theta}} d\bar{\theta})d\bar{\phi} \quad .
\end{split}
\ee
To simplify our notation, we shall write $g_{ij}(G(\bar{r},\bar{\theta}),H(\bar{r},\bar{\theta}))\rightarrow g_{ij}(\bar{r},\bar{\theta})$. The off-diagonal components of the 3-metric are
\begin{align}
g_{\bar{r}\bar{\phi}}= & g_{r\phi}(\bar{r},\bar{\theta})G_{\bar{r}}\\
g_{\bar{\theta}\bar{\phi}}= & g_{r\phi}(\bar{r},\bar{\theta})G_{\bar{\theta}}\\
g_{\bar{r}\bar{\theta}}= & g_{rr}(\bar{r},\bar{\theta})G_{\bar{r}}G_{\bar{\theta}}+g_{\theta\theta}(\bar{r},\bar{\theta})H_{\bar{r}}H_{\bar{\theta}} \quad .
\end{align}
We now demand all of these components to vanish. That is, we demand the following equations to hold
\begin{align}
E1: \qquad \qquad & g_{r\phi}(\bar{r},\bar{\theta})G_{\bar{r}}=0\\
E2:\qquad \qquad & g_{r\phi}(\bar{r},\bar{\theta})G_{\bar{\theta}}=0\\
E3: \qquad \qquad & g_{rr}(\bar{r},\bar{\theta})G_{\bar{r}}G_{\bar{\theta}}+g_{\theta\theta}(\bar{r},\bar{\theta})H_{\bar{r}}H_{\bar{\theta}}=0 \quad .
\end{align}
Now, we know that $g_{r\phi}\neq 0$, hence, from $E1$ we have $G_{\bar{r}}=0$ and from $E2$ we have $G_{\bar{\theta}}=0$. This implies that $G(\bar{r},\bar{\theta})$ is constant. Therefore, it not a good choice of coordinate, so we have an inconsistency. Hence, we cannot construct a Painlev\'{e}-Gullstrand form of the Kerr metric via a polar-only coordinate transformation.

\subsection{Axisymmetry preserving coordinate transformation}

We now look at the most general coordinate transformation we can perform while retaining the axisymmetry present in the Kerr spacetime. This coordinate transformation is of the form 
\be
r=G(\bar{r},\bar{\theta}); \qquad \theta=H(\bar{r},\bar{\theta}); \qquad \phi=\bar{\phi}+\Phi(\bar{r}, \bar{\theta})
\ee
therefore
\be
dr=G_{\bar{r}} d\bar{r} + G_{\bar{\theta}} d\bar{\theta}; \qquad d\theta=H_{\bar{r}} d\bar{r} + H_{\bar{\theta}} d\bar{\theta}; \qquad d\phi=d\bar{\phi}+\Phi_{\bar{r}} d\bar{r}+\Phi_{\bar{\theta}} d\bar{\theta} \quad .
\ee
Hence, we can write the 3-metric as
\be
\begin{split}
ds^2= & g_{rr}(G(\bar{r},\bar{\theta}),H(\bar{r},\bar{\theta}))(G_{\bar{r}} d\bar{r}+G_{\bar{\theta}} d\bar{\theta})^2\\
& +g_{\theta\theta}(G(\bar{r},\bar{\theta}),H(\bar{r},\bar{\theta}))(H_{\bar{r}} d\bar{r}+H_{\bar{\theta}} d\bar{\theta})^2\\
& +g_{\phi\phi}(G(\bar{r},\bar{\theta}),H(\bar{r},\bar{\theta}))(d\bar{\phi}+\Phi_{\bar{r}} d\bar{r}+\Phi_{\bar{\theta}} d\bar{\theta})^2\\
& +2g_{r\phi}(G(\bar{r},\bar{\theta}),H(\bar{r},\bar{\theta}))(G_{\bar{r}} d\bar{r}+G_{\bar{\theta}} d\bar{\theta})(d\bar{\phi}+\Phi_{\bar{r}} d\bar{r}+\Phi_{\bar{\theta}} d\bar{\theta}) \quad .
\end{split}
\ee
To simplify our notation, we shall write $g_{ij}(G(\bar{r},\bar{\theta}),H(\bar{r},\bar{\theta}))\rightarrow g_{ij}(\bar{r},\bar{\theta})$. The off-diagonal components of the 3-metric are
\begin{align}
g_{\bar{r}\bar{\phi}}= & g_{r\phi}(\bar{r},\bar{\theta}) G_{\bar{r}} + g_{\phi\phi}(\bar{r},\bar{\theta}) \Phi_{\bar{r}}\\
g_{\bar{\theta}\bar{\phi}}= & g_{r\phi}(\bar{r},\bar{\theta}) G_{\bar{\theta}} + g_{\phi\phi}(\bar{r},\bar{\theta}) \Phi_{\bar{\theta}}\\
\begin{split}
g_{\bar{r}\bar{\theta}}= & g_{rr}(\bar{r},\bar{\theta})G_{\bar{r}}G_{\bar{\theta}}+g_{\theta\theta}(\bar{r},\bar{\theta})H_{\bar{r}}H_{\bar{\theta}}+g_{\phi\phi}(\bar{r},\bar{\theta})\Phi_{\bar{r}}\Phi_{\bar{\theta}}\\
& +g_{r\phi}(\bar{r},\bar{\theta})[\Phi_{\bar{r}}G_{\bar{\theta}}+G_{\bar{r}}\Phi_{\bar{\theta}}] \quad .
\end{split}
\end{align}
We now demand all of these components to vanish. That is, we demand the following equations to hold
\begin{align}
E1: \qquad \qquad & g_{r\phi}(\bar{r},\bar{\theta}) G_{\bar{r}} + g_{\phi\phi}(\bar{r},\bar{\theta}) \Phi_{\bar{r}}=0\\
E2:\qquad \qquad & g_{r\phi}(\bar{r},\bar{\theta}) G_{\bar{\theta}} + g_{\phi\phi}(\bar{r},\bar{\theta}) \Phi_{\bar{\theta}}=0\\
\begin{split}
E3: \qquad \qquad & g_{rr}(\bar{r},\bar{\theta})G_{\bar{r}}G_{\bar{\theta}}+g_{\theta\theta}(\bar{r},\bar{\theta})H_{\bar{r}}H_{\bar{\theta}}+g_{\phi\phi}(\bar{r},\bar{\theta})\Phi_{\bar{r}}\Phi_{\bar{\theta}}\\
&+g_{r\phi}(\bar{r},\bar{\theta})[\Phi_{\bar{r}}G_{\bar{\theta}}+G_{\bar{r}}\Phi_{\bar{\theta}}]=0 \quad .
\end{split}
\end{align}
So we have 3 partial-differential equations (PDEs) involving 3 unknown functions, hence (assuming these PDEs are solvable), we should be able to analytically solve for these functions. But let us consider the following linear combination of equations 
\be
\frac{1}{g_{\phi\phi}(\bar{r},\bar{\theta})}(G_{\bar{\theta}}\, E1-G_{\bar{r}}\, E2) \quad .
\ee
That is 
\be
G_{\theta}\Phi_{\bar{r}}-G_{\bar{r}}\Phi_{\theta}=0 \quad .
\ee
We notice that this looks like the cross product between two vectors and moreover this cross product vanishes. More explicitly, the following cross product vanishes
\be
(\Phi_{\bar{r}},\Phi_{\bar{\theta}}) \times (G_{\bar{r}},G_{\theta}) =0 \quad .
\ee
This means that 
\be
(\Phi_{\bar{r}},\Phi_{\bar{\theta}}) \propto (G_{\bar{r}},G_{\theta}) \quad .
\ee
Which has the solution
\be
\Phi(\bar{r},\bar{\theta})=W(G(\bar{r},\bar{\theta}))
\ee
where $W(G)$ is some arbitrary function of $G(\bar{r},\bar{\theta})$. We now substitute this solution back into $E1$ and $E2$
\begin{align}
E1': \qquad \qquad & G_{\bar{r}}(\bar{r},\bar{\theta}) [g_{r\phi}(\bar{r},\bar{\theta}) + g_{\phi\phi}(\bar{r},\bar{\theta}) W'(G(\bar{r},\bar{\theta}))]=0\\
E2': \qquad \qquad & G_{\bar{\theta}}(\bar{r},\bar{\theta}) [g_{r\phi}(\bar{r},\bar{\theta}) + g_{\phi\phi}(\bar{r},\bar{\theta}) W'(G(\bar{r},\bar{\theta}))]=0 \quad .
\end{align}
There are two sets of solutions for this system of equations. The first is that both $G_{\bar{r}}=0$ and $G_{\bar{\theta}}=0$, however, this would imply that $G$ is constant, which not be a good choice of coordinate. Hence, we must enforce the second solution. That is, we enforce
\be
g_{r\phi}(\bar{r},\bar{\theta}) + g_{\phi\phi}(\bar{r},\bar{\theta}) W'(G(\bar{r},\bar{\theta}))=0 \quad .
\ee
That is
\be
W'(G(\bar{r},\bar{\theta}))=-\frac{g_{r\phi}(\bar{r},\bar{\theta})}{g_{\phi\phi}(\bar{r},\bar{\theta})} \quad .
\ee
Rewriting this condition in our original coordinate system, this condition reads
\be \label{PG_Kerr_Const_Con_2}
W'(r)=-\frac{g_{r\phi}(r,\theta)}{g_{\phi\phi}(r,\theta)} \quad .
\ee
Now recall that we have 
\begin{align}
\left(\frac{g_{r\phi}(r,\theta)}{g_{\phi\phi}(r,\theta)}\right)_{\text{BL-rain}}= & \frac{2mar\sqrt{2mr(r^2+a^2)}}{\Delta\rho^2\Sigma}\\
\left(\frac{g_{r\phi}(r,\theta)}{g_{\phi\phi}(r,\theta)}\right)_{\text{EF-rain}}= & -\frac{a}{\Sigma}\left(\frac{1+2mr/\rho^2+\sqrt{2mr/(r^2+a^2)}}{1+\sqrt{2mr/(r^2+a^2)}}\right)\\
\left(\frac{g_{r\phi}(r,\theta)}{g_{\phi\phi}(r,\theta)}\right)_{\text{Doran}}= & -\frac{a}{\Sigma}\sqrt{\frac{2mr}{r^2+a^2}} \quad .
\end{align}
Where there is $\theta$ dependence in both $\rho$ and $\Sigma$. Hence, equation \ref{PG_Kerr_Const_Con_2} cannot be satisfied, we cannot have a function purely of $r$ be equal to some function of both $r$ and $\theta$. Since this condition cannot be satisfied, there exists no coordinate transformation which diagonalises the 3-metric of a unit-lapse form of the Kerr metric while retaining both unit-lapse and axisymmetry. Therefore, there does not exist a Painlev\'{e}-Gullstrand form of the Kerr metric.

\chapter{The Lense-Thirring spacetime}

\section{Introduction to the Lense-Thirring spacetime}

It took 47 years after the discovery of the Schwarzschild solution to discover the Kerr solution, the exact solution to Einstein's equation that describes a rotating central mass in a vacuum. However, it was not the first \emph{spacetime} to model this physical system. Just 2 years after the discovery of the Schwarzschild solution Josef Lense and Hans Thirring discovered an \emph{approximate} solution to the vacuum Einstein equation that describes a rotating central mass in a vacuum at large distances from the central mass. The Lense-Thirring spacetime is a slow rotation approximation to the Kerr solution, however even for rapid rotation, Lense-Thirring approximates Kerr at large distances (large $r$). The Lense-Thirring metric is most commonly written as follows
\be \label{LT_Met}
\begin{split}
ds^2= & -\left[ 1-\frac{2m}{r}+\OO\left(\frac{1}{r^2}\right)\right] dt^2 - \left[\frac{4J\sin^2(\theta)}{r}+\OO\left(\frac{1}{r^2}\right)\right] dt d\phi\\
& +\left[ 1+\frac{2m}{r}+\OO\left(\frac{1}{r^2}\right)\right] \left[ dr^2+r^2(d\theta^2+\sin^2(\theta) d\phi^2)\right]
\end{split}
\ee 
where $J$ is the angular momentum of the rotating object. This metric is rather useful for a few reasons. Firstly, it is much easier to use this metric that the rather complex Kerr solution (we can already notice that the metric components of the Lense-Thirring metric are much simpler than the metric components of the Kerr metric). Secondly, since there exists no Birkhoff theorem for axisymmetric solutions in 3+1 dimensions, the Kerr solution will not \emph{perfectly} describe any physical rotating star or planet in the universe considering these objects will typically possess non-trivial mass multipole moments. Hence, the Kerr solution will only be a good model in the asymptotic regime, which is where the Lense-Thirring spacetime approximates Kerr. 

Unlike the Kerr solution, the Lense-Thirring spacetime does admit a Painlev\'{e}-Gullstrand form of its metric. We now modify the metric given in equation \eqref{LT_Met} to generate a Painlev\'{e}-Gullstrand form of the metric. Consider the following modification
\be
\begin{split}
ds^2= & -\left( 1-\frac{2m}{r}\right) dt^2 -\left[ \frac{4J\sin^2(\theta)}{r}+\OO\left(\frac{1}{r^2}\right)\right] dt d\phi \\
& +\frac{dr^2}{1-2m/r}+r^2(d\theta^2+\sin^2(\theta) d\phi^2) \quad .
\end{split}
\ee
Notice that this metric approaches equation \eqref{LT_Met} for large $r$ and furthermore notice that for $J=0$, this modified version of the Lense-Thirring metric reduces to Schwarzschild for large $r$. We now ``complete the square" (foreshadowing the tetrad discussion below)
\be
\begin{split}
ds^2= &  -\left( 1-\frac{2m}{r}\right) dt^2+\frac{dr^2}{1-2m/r}\\
& + r^2\left( d\theta^2+\sin^2(\theta)\left( d\phi -\left[ \frac{2J}{r^3}+\OO\left(\frac{1}{r^4}\right)\right] dt\right)^2\right) \quad .
\end{split}
\ee
Here the azimuthal components have been put into partial Painlev\'{e}-Gull- strand form, that is: $g_{\phi\phi}(d\phi-v^{\phi} dt)^2=g_{\phi\phi}(d\phi-\omega dt)^2$. Now to put the $t-r$ plane into Painlev\'{e}-Gullstrand form we make the following coordinate transformation 
\be 
dt\rightarrow dt+\frac{1}{1-2m/r}\sqrt{\frac{2m}{r}} dr \quad .
\ee
Doing so then yields
\be \label{LT_Met_2}
\begin{split}
ds^2= &  -dt^2 + \left( dr + \sqrt{\frac{2m}{r}} dt \right)^2\\
& + r^2\left( d\theta^2+\sin^2(\theta)\left( d\phi -\left[ \frac{2J}{r^3}+\OO\left(\frac{1}{r^4}\right)\right] dt\right)^2\right) \quad .
\end{split}
\ee
Notice that now the 3-metric of equation \eqref{LT_Met_2} is flat. We now discard the $\OO(1/r^4)$ terms, that is, we now have the explicit metric
\be \label{PGLT_Met}
ds^2=-dt^2+\left( dr+\sqrt{\frac{2m}{r}} dt \right)^2+r^2\left( d\theta^2+\sin^2(\theta)\left( d\phi -\frac{2J}{r^3} dt\right)^2 \right) \quad .
\ee
Notice that for $J=0$, this metric reduces to the Schwarzschild solution in Painlev\'{e}-Gullstrand coordinates exactly and for large $r$ this metric approaches equation \eqref{LT_Met}. 

From equation \eqref{PGLT_Met}, the covariant metric is given by
\be
g_{ab}=\left[ 
\begin{array}{c|ccc}
-1+\frac{2m}{r}+\frac{4J^2\sin^2(\theta)}{r^4} & \sqrt{\frac{2m}{r}} & 0 & -\frac{2J\sin^2(\theta)}{r} \\
\hline
\sqrt{\frac{2m}{r}} & 1 & 0 & 0 \\
0 & 0 & r^2 & 0 \\
-\frac{2J\sin^2(\theta)}{r} & 0 & 0 & r^2\sin^2(\theta)
\end{array}
\right]_{ab} 
\ee
while the inverse metric is given by
\be
g^{ab}=\left[ 
\begin{array}{c|ccc}
-1 & \sqrt{\frac{2m}{r}} & 0 & -\frac{2J}{r^3} \\
\hline
\sqrt{\frac{2m}{r}} & 1-\frac{2m}{r} & 0 & \sqrt{\frac{2m}{r}}\frac{2J}{r^3} \\
0 & 0 & \frac{1}{r^2} & 0 \\
-\frac{2J}{r^3} & \sqrt{\frac{2m}{r}}\frac{2J}{r^3} & 0 & \frac{1}{r^2\sin^2(\theta)}-\frac{4J^2}{r^6}
\end{array}
\right]^{ab} \quad .
\ee
Hence, the metric is in unit-lapse form, as advertised. 

\subsection{Tetrad}

A tetrad (or vierbein) is essentially an orthonormal basis defined on our manifold. Therefore, most tensors defined on our manifold become more simplified when written in a tetrad basis. When we write a tensor using a tetrad basis, we will write the corresponding labels of the indices with overhats, for example $\hat{a}, \hat{b} \in \{\hat{t}, \hat{r}, \hat{\theta}, \hat{\phi}\}$. Let $\eta_{\hat{a}\hat{b}}=\text{diag}(-1,1,1,1)$. Then a covariant tetrad (or co-tetrad) $e^{\hat{a}}{}_a$, will satisfy the following condition $g_{ab}=\eta_{\hat{a}\hat{b}}e^{\hat{a}}{}_a e^{\hat{b}}{}_b$. By examining equation \eqref{PGLT_Met}, an obvious choice for a suitable co-tetrad is
\begin{align}
e^{\hat{t}}{}_a & = (1;0,0,0); & e^{\hat{r}}{}_a & = \left( \sqrt{\frac{2m}{r}};1,0,0\right)\\
e^{\hat{\theta}}{}_a & = r(0;0,1,0); & e^{\hat{\phi}}{}_a & = r\sin(\theta) \left( -\frac{2J}{r^3};0,0,1\right) \quad .
\end{align}
We note that a tetrad basis is not unique. This is because the metric is invariant under a local Lorentz transformations $L^{\hat{a}}{}_{\hat{b}}$ on the co-tetrad/tetrad indices. However, we have picked this particular tetrad since it is well adapted to the coordinate system we have chosen. Once we have a co-tetrad, we can solve for the contravariant tetrad (or just tetrad) by using the following condition $e_{\hat{a}}{}^a=\eta_{\hat{a}\hat{b}}e^{\hat{b}}{}_bg^{ba}$. Therefore, our tetrad is given by:
\begin{align}
e_{\hat{t}}{}^a & = \left( 1;-\sqrt{\frac{2m}{r}},0,\frac{2J}{r^3}\right) ; & e_{\hat{r}}{}^a & = (0;1,0,0)\\
e_{\hat{\theta}}{}^a & = \frac{1}{r}(0;0,1,0); & e_{\hat{\phi}}{}^a & = \frac{1}{r\sin(\theta)} (0;0,0,1) \quad .
\end{align}
We note the last three tetrad vectors given above are exactly the tetrad vectors we would get for flat Euclidean 3-space written in spherical coordinates. Therefore, for this tetrad, all of the non-trivial physics lies within the first tetrad vector $e_{\hat{t}}{}^a$. This tetrad is rather simple in form, this is what motivated us to ``complete the square" above, not taking that step complicates the tetrad significantly. As stated before, tensors defined on our manifold become more simplified when written in a tetrad basis. Hence, we shall give some tensors and curvature invariants written in our tetrad basis. 

\subsection{Curvature tensors}

As previously stated, the Lense-Thirring metric is not an exact solution to the vacuum Einstein equation, it is an \emph{approximate} solution. This means that the Lense-Thirring spacetime is not Ricci flat, that is $R_{ab} \neq 0$. The Ricci scalar in the Lense-Thirring spacetime is given as
\be
R=\frac{18 J^2\sin^2(\theta)}{r^6} \quad .
\ee
Hence, we see as $r \rightarrow \infty$, $R \rightarrow 0$. The Ricci tensor in the tetrad basis is given by $R_{\hat{a}\hat{b}}=e_{\hat{a}}{}^a e_{\hat{b}}{}^b R_{ab}$, and is explicitly given as
\be
R_{\hat{a}\hat{b}}=R \left[
\begin{array}{c|ccc}
-1 & 0 & 0 & 0 \\
\hline
0 & 1 & 0 & 0 \\
0 & 0 & 0 & 0 \\
0 & 0 & 0 & -1
\end{array}
\right]_{\hat{a}\hat{b}} \quad .
\ee
Therefore, as $r \rightarrow \infty$, $R_{\hat{a}\hat{b}} \rightarrow 0$. This shows us that the Lense-Thirring metric approaches an exact solution of the vacuum Einstein equation at large distances. To show that this approximates Kerr, we would have to conduct a Taylor series expansion of the metric components of Kerr. We would then see at large $r$, the components of the Lense-Thirring metric and the components of the Kerr metric are approximately equal (however, we shall not explicitly conduct this calculation in this text).

Now, the Einstein tensor in the tetrad basis is given by $G_{\hat{a}\hat{b}}=e_{\hat{a}}{}^a e_{\hat{b}}{}^b R_{ab}$, and is explicitly given as
\be
G_{\hat{a}\hat{b}}=\frac{R}{2} \left[
\begin{array}{c|ccc}
-1 & 0 & 0 & 0 \\
\hline
0 & 1 & 0 & 0 \\
0 & 0 & -1 & 0 \\
0 & 0 & 0 & -3
\end{array}
\right]_{\hat{a}\hat{b}} \quad .
\ee
Notice the interesting pattern of minus signs and zeros in both the Ricci and Einstein tensors. However, these tensors are rather simple. The Weyl and Riemann tensors on the other hand, are rather complex and tedious to calculate. 

The Weyl tensor in the tetrad basis is given by $C_{\hat{a}\hat{b}\hat{c}\hat{d}}=e_{\hat{a}}{}^a e_{\hat{b}}{}^b e_{\hat{c}}{}^c e_{\hat{d}}{}^d C_{abcd}$. The terms quadratic in $J$ are given by
\be
\begin{split}
C_{\hat{t}\hat{r}\hat{t}\hat{r}} & = -2C_{\hat{t}\hat{\theta}\hat{t}\hat{\theta}} = -2C_{\hat{t}\hat{\phi}\hat{t}\hat{\phi}} = 2C_{\hat{r}\hat{\theta}\hat{r}\hat{\theta}} = 2C_{\hat{r}\hat{\phi}\hat{r}\hat{\phi}} = -C_{\hat{\theta}\hat{\phi}\hat{\theta}\hat{\phi}} \\
& = -\frac{2m}{r^3}-\frac{12 J^2\sin^2(\theta)}{r^6} = -\frac{2m}{r^3}-\frac{2}{3}R \quad .
\end{split}
\ee 
The terms linear in $J$ are given by
\begin{align}
\frac{1}{2}C_{\hat{t}\hat{r}\hat{\theta}\hat{\phi}} & = C_{\hat{t}\hat{\theta}\hat{r}\hat{\phi}} = -C_{\hat{t}\hat{\phi}\hat{r}\hat{\theta}} = \frac{3 J\cos(\theta)}{r^4} \\
C_{\hat{t}\hat{r}\hat{r}\hat{\phi}} & = -C_{\hat{t}\hat{\theta}\hat{\theta}\hat{\phi}} = -\frac{3 J\sin(\theta)}{r^4} \\
C_{\hat{t}\hat{r}\hat{t}\hat{\phi}} & = -C_{\hat{r}\hat{\theta}\hat{\theta}\hat{\phi}} = \frac{3 J\sin(\theta)\sqrt{2m/r}}{r^4} \quad .
\end{align}
The Riemann tensor in the tetrad basis is given by $R_{\hat{a}\hat{b}\hat{c}\hat{}}=e_{\hat{a}}{}^a e_{\hat{b}}{}^b e_{\hat{c}}{}^c e_{\hat{d}}{}^d R_{abcd}$. The terms quadratic in $J$ are given by
\begin{align}
R_{\hat{t}\hat{r}\hat{t}\hat{r}} & = -\frac{2m}{r^3}-\frac{27 J^2\sin^2(\theta)}{r^6} = -\frac{2m}{r^3}-\frac{3}{2}R \\
R_{\hat{t}\hat{\phi}\hat{t}\hat{\phi}} & = -R_{\hat{r}\hat{\phi}\hat{r}\hat{\phi}} = \frac{m}{r^3}+\frac{9 J^2\sin^2(\theta)}{r^6} = \frac{m}{r^3}+\frac{1}{2}R \quad .
\end{align}
The terms linear in $J$ are given by
\begin{align}
R_{\hat{t}\hat{r}\hat{\theta}\hat{\phi}} & = 2R_{\hat{t}\hat{\theta}\hat{r}\hat{\phi}} = -2R_{\hat{t}\hat{\phi}\hat{r}\hat{\theta}} = \frac{6 J\cos(\theta)}{r^4} \\
R_{\hat{t}\hat{r}\hat{r}\hat{\phi}} & = -R_{\hat{t}\hat{\theta}\hat{\theta}\hat{\phi}} = -\frac{3 J\sin(\theta)}{r^4} \\
R_{\hat{t}\hat{r}\hat{t}\hat{\phi}} & = -R_{\hat{r}\hat{\theta}\hat{\theta}\hat{\phi}} = \frac{3 J\sin(\theta)\sqrt{2m/r}}{r^4} \quad .
\end{align}
And the terms independent of $J$ are given by 
\be
R_{\hat{t}\hat{\theta}\hat{t}\hat{\theta}} = -R_{\hat{r}\hat{\theta}\hat{r}\hat{\theta}} = \frac{1}{2}R_{\hat{\theta}\hat{\phi}\hat{\theta}\hat{\phi}} = \frac{m}{r^3} \quad .
\ee
While these tensors are rather complex, they are much simpler when written in the tetrad basis as written here.

\subsection{Curvature invariants}

As previously mentioned, the Ricci scalar of the Lense-Thirring spacetime is given as 
\be
R=\frac{18 J^2\sin^2(\theta)}{r^6}
\ee
also the Ricci invariant is
\be
R_{ab}R^{ab} = 3R^2 \quad .
\ee
Notice that for $J \rightarrow 0$, both of these quantities vanish. This is exactly what we expect since for $J \rightarrow 0$, the Lense-Thirring metric, equation \eqref{PGLT_Met}, reduces to the Schwarzschild metric written in Painlev\'{e}-Gullstrand coordinates.

For the Weyl invariant we find
\be
\begin{split}
C_{abcd}C^{abcd} & = \frac{48m^2}{r^6}-\frac{144 J^2(2\cos^2(\theta)+1)}{r^8}+\frac{864 J^2\sin^2(\theta)}{r^9}+\frac{1728 J^4\sin^4(\theta)}{r^{12}} \\
& = \frac{48m^2}{r^6}-\frac{144 J^2(3-2\sin^2(\theta))}{r^8}+\frac{48m}{r^3} R+\frac{16}{3} R^2 \\
& = \frac{48m^2}{r^6}-\frac{432 J^2}{r^8}+\frac{16}{3}\left( 1+\frac{3m}{r}\right) R+\frac{16}{3} R^2 \quad .
\end{split}
\ee
For the Kretschmann scalar we find
\be
R_{abcd}R^{abcd}=C_{abcd}C^{abcd}+\frac{1728 J^4\sin^4(\theta)}{r^{12}}=C_{abcd}C^{abcd}+\frac{17}{3} R^2 \quad .
\ee
Notice that for both of these scalars, for $J \rightarrow 0$, these both reduce to their known values in Schwarzschild, as we expect. 

\section{Petrov type}

Finding solutions to Einstein's equation is considerably difficult in the abstract. Einstein's equation is equivalent to a system of 10 coupled, non-linear, partial differential equations (there are 10 independent equations because both $R_{ab}$ and $T_{ab}$ are symmetric). Such a system is considerably complicated and generally does not admit explicit solutions. However, there are various methods one can use to generate solutions to Einstein's equation. If the spacetime we are trying to solve for has a considerable degree of symmetry, then some of the PDEs in Einstein's equation degenerate, thus simplifying the system. The resulting system of PDEs may then admit a solution. One measure of the degree of symmetry in a spacetime is the Petrov type (or Petrov classification) of the spacetime. The Petrov type of spacetimes classifies the degree of symmetry in a spacetime by the algebraic symmetry present in the Weyl tensor of that spacetime. We shall not go into the details of the analysis of the algebraic symmetry of the Weyl tensor, but just simply state an important conclusion. Generally, there exists up to 4 distinct null vectors $k^a$ where the following condition holds:
\be \label{Petrov_Def}
k^b k^c k_{[e} C_{a]bc[d} k_{f]} = 0 \quad .
\ee
The null vectors which satisfy the above equation are called principal null directions. While these principal null directions are in general distinct, if there exists symmetry in the spacetime, some of these principal null directions can coincide, that is, they can become degenerate. Spacetimes with degenerate principal null directions are called algebraically special spacetimes. These spacetimes can use the degeneracy of the principal null directions to cause degeneracy in the system of PDEs present in Einstein's equation which can then allow a solution to be generated. Spacetimes are classified by the number of degenerate principal null directions as shown in table \ref{Tab:Petrov_Class}. 

So to find the Petrov type of a spacetime, one can solve for all principal null directions of that spacetime, which can be rather tedious. However, as shown in pages 49 and 50 of the “Exact solutions” book by Stephani \emph{et al} \cite{exact}, we can find the Petrov type by a much simpler method. Consider the mixed Weyl tensor $C^{ab}{}_{cd}$. This tensor is antisymmetric in the index pairs $[ab]$ and $[cd]$. Hence, we can think of this tensor as a $6 \times 6$ real matrix $C^A{}_B$ by making the correspondence $A \longleftrightarrow [ab]$ and $B \longleftrightarrow [cd]$. This matrix $C^A{}_B$ will be asymmetric (i.e. not symmetric). Then by finding the eigenvalues of this matrix, we can find the Petrov type of the spacetime in question.

\begin{table}[h!]
\centering
\begin{tabular}{||c|c|c||} 
 \hline
 \specialcell{Petrov \\ Type} & Description & Mathematical Condition \\  
 \hline\hline
 \Ronumb{1} & \specialcell{Not algebraically special; 4 distinct \\ principal null directions} & $k^b k^c k_{[e} C_{a]bc[d} k_{f]} = 0$ \\ 
 \hline
 \Ronumb{2} & \specialcell{One pair of degenerate \\ principal null directions} & $k^b k^c C_{abc[d} k_{e]} = 0$ \\
 \hline
 D & \specialcell{Two pairs of degenerate \\ principal null directions} & \specialcell{$k^b k^c C_{abc[d} k_{e]} = 0$ \\ (two solutions)} \\
 \hline
 \Ronumb{3} & \specialcell{Set of three degenerate \\ principal null directions} & $k^c C_{abc[d} k_{e]} = 0$ \\
 \hline
 N & \specialcell{All four principal null directions \\ are degenerate} & $k^c C_{abcd} = 0$ \\
 \hline
 O & The Weyl tensor vanishes & $C_{abcd} = 0$ \\
 \hline
\end{tabular}
\caption{Petrov classification of various spacetimes.}
\label{Tab:Petrov_Class}
\end{table}

\noindent For our Lense-Thirring spacetime, we calculate the mixed Weyl tensor in the tetrad basis, $C^{\hat{a}\hat{b}}{}_{\hat{c}\hat{d}}$ , and then make the correspondence $A \longleftrightarrow [\hat{a}\hat{b}]$ and $B \longleftrightarrow [\hat{c}\hat{d}]$ via the following scheme:
\be
1\leftrightarrow [\hat{1}\hat{2}]; \quad 2\leftrightarrow [\hat{1}\hat{3}]; \quad 3\leftrightarrow [\hat{1}\hat{4}]; \quad 4\leftrightarrow [\hat{3}\hat{4}]; \quad 5\leftrightarrow [\hat{4}\hat{2}]; \quad 6\leftrightarrow [\hat{2}\hat{3}] \quad .
\ee
We now define some useful quantities
\be
\Xi_1=\frac{3 J\sin(\theta)}{r^4}; \quad \Xi_2=-\frac{m}{r^3}-\frac{6 J^2\sin^2(\theta)}{r^6}; \quad \Xi_3=\frac{3 J\cos(\theta)}{r^4} \quad .
\ee 
For our Lense-Thirring spacetime, the $6\times 6$ matrix $C^A{}_B$ is given by
\be
C^A{}_B=\left[ 
\begin{array}{ccc|ccc}
-2\, \Xi_2 & 0 & -\Xi_1 \sqrt{\frac{2m}{r}} & -2\, \Xi_3 & -\Xi_1 & 0 \\
0 & \Xi_2 & 0 & -\Xi_1 & \Xi_3 & 0 \\
-\Xi_1 \sqrt{\frac{2m}{r}} & 0 & \Xi_2 & 0 & 0 & \Xi_3 \\
\hline
2\, \Xi_3 & \Xi_1 & 0 & -2\, \Xi_2 & 0 & -\Xi_1 \sqrt{\frac{2m}{r}} \\
\Xi_1 & -\Xi_3 & 0 & 0 & \Xi_2 & 0 \\
0 & 0 & -\Xi_3 & -\Xi_1 \sqrt{\frac{2m}{r}} & 0 & \Xi_2
\end{array}
\right] \quad .
\ee
Notice that the trace of this matrix vanishes, as we should expect since the Weyl tensor has zero trace. Also notice that this matrix has the partial symmetry 
\be
C^A{}_B=\left[ 
\begin{array}{c|c}
S_R & S_I \\
\hline
-S_I & S_R
\end{array}
\right] 
\ee
where $S_R$ and $S_I$ are symmetric $3\times 3$ matrices. 

The matrix $C^A{}_B$ has 6 distinct eigenvalues, more specifically 3 complex conjugate pairs. If we define the fourth quantity
\be
\Xi_4=\left(1-\frac{2m}{r}\right)\frac{9 J^2\sin^2(\theta)}{r^8}=\left(1-\frac{2m}{r}\right)\Xi_1{}^2
\ee
then we can write the 6 distinct eigenvalues of $C^A{}_B$ as
\be
\Xi_2+i\Xi_3; \quad -\frac{1}{2}(\Xi_2+i\Xi_3)\pm \sqrt{\frac{9}{4}(\Xi_2+i\Xi_3)^2-\Xi_4} \quad ;
\ee
and
\be
\Xi_2-i\Xi_3; \quad -\frac{1}{2}(\Xi_2-i\Xi_3)\pm \sqrt{\frac{9}{4}(\Xi_2-i\Xi_3)^2-\Xi_4} \quad .
\ee
Since there exists 6 distinct eigenvalues, this implies that the Jordan canonical form of $C^A{}_B$ is trivial, hence our form of the Lense-Thirring spacetime is Petrov type \Ronumb{1}. Note that while the Petrov type of a spacetime is independent of the coordinate system it is written in, we have made modifications to the usual form of the spacetime hence the distinction made above.\\

\noindent We now consider some special cases:
\begin{itemize}
    \item On the rotational axis, $\theta=0$. So we have $\Xi_1=\Xi_3=\frac{3 J}{r^4}$, $\Xi_2=-\frac{m}{r^3}$ and $\Xi_4=0$. Then the 6 distinct eigenvalues of $C^A{}_B$ reduce to
    \be
    \{ \lambda(\theta=0)\}=\left\{ -\frac{m}{r^3}\pm i\frac{3 J}{r^4},\;  -\frac{m}{r^3}\pm i\frac{3 J}{r^4}; \; -2\left(-\frac{m}{r^3}\pm i\frac{3 J}{r^4}\right) \right\} \quad.
    \ee
    Hence, the 6 on-axis eigenvalues of $C^A{}_B$ are degenerate, we have two sets of twice repeated eigenvalues.
    \item On the equator, $\theta=\frac{\pi}{2}$. So we have $\Xi_1=\frac{3 J}{r^4}$, $\Xi_2=-\frac{m}{r^3}-\frac{6 J^2}{r^6}$, $\Xi_3=0$ and $\Xi_4=\left( 1-\frac{2m}{r}\right)\frac{9 J^2}{r^8}$. Then the 6 distinct eigenvalues of $C^A{}_B$ reduce to
    \be
    \{ \lambda(\theta=\pi/2)\}=\left\{ \Xi_2, \; -\frac{1}{2}\Xi_2\pm \sqrt{\frac{9}{4}\Xi_2{}^2-\Xi_4} \right\} \quad .
    \ee
    Hence we have twice-repeated degenerate eigenvalues of $C^A{}_B$ on the equator.
    \item For $J=0$ we have $\Xi_1=\Xi_3=\Xi_4=0$ and $\Xi_2=-\frac{m}{r^3}$. Then the 6 distinct eigenvalues of $C^A{}_B$ reduce to
    \be
    \{ \lambda(J=0)\}=\left\{ -\frac{m}{r^3}, \; -\frac{m}{r^3}, \; -\frac{m}{r^3}, \; -\frac{m}{r^3}; \; \frac{2m}{r^3}, \; \frac{2m}{r^3} \right\} \quad .
    \ee
    This is exactly the repeated eigenvalue structure we expect for the Schwarzschild solution where we have a set of four times repeated eigenvalues and a set of twice repeated eigenvalues. 
\end{itemize}
While the eigenvalues of $C^A{}_B$ are degenerate on-axis, on the equator and in the non-rotating limit where $J\rightarrow 0$, in general, the 6 eigenvalues are distinct. Hence, the Lense-Thirring spacetime is of Petrov type \Ronumb{1}. 

\section{Rain geodesics}

As we saw in section \ref{S:Sch_Coord}, writing the Schwarzschild metric in Painlev\'{e}-Gullstrand form allowed for simple analysis of the rain geodesics. This effect carries over to the Lense-Thirring spacetime now that the metric has now been written in Painlev\'{e}-Gullstrand form. Consider the vector field
\be
V^a=-g^{ab}\n_b t=-g^{ta}=\left(1; -\sqrt{\frac{2m}{r}}, 0, \frac{2 J}{r^3}\right)
\ee
where the corresponding dual vector field is 
\be
V_a=-\n_a t=(-1;0,0,0) \quad .
\ee
Hence, $g_{ab}V^aV^b=V^aV_a=-1$, so $V^a$ is a timelike vector field with unit norm. This vector field has zero 4-acceleration
\be
A^a=V^b\n_bV^a=-V^b\n_b\n_at=-V^b\n_a\n_bt=V^b\n_aV^b=\frac{1}{2}\n_a(V^bV_b)=0
\ee
therefore the integral curves of $V^a$ are timelike geodesics. More specifically, the integral curves given by 
\be
\frac{dx^a}{d\tau}=\left(\frac{dt}{d\tau};\frac{dr}{d\tau},\frac{d\theta}{d\tau},\frac{d\phi}{d\tau}\right)=\left(1; -\sqrt{\frac{2m}{r}}, 0, \frac{2 J}{r^3}\right)
\ee
are timelike geodesics. We can trivially integrate 2 of these equations
\be
t(\tau)=\tau; \quad \theta(\tau)=\theta_{\infty}
\ee
thus $t$ is just the proper time of these geodesics, while $\theta_{\infty}$ is the original (and permanent) value of the $\theta$ coordinate for these geodesics. Furthermore, we have
\be
\frac{1}{2}\left(\frac{dr}{dt}\right)^2=\frac{m}{r} \quad .
\ee
Therefore, these geodesics mimic a particle with zero initial velocity falling towards a mass ($m$) from spatial infinity. As for the final equation, we have 
\be
\frac{d\phi}{dr}=\frac{d\phi/d\tau}{dr/d\tau}=-\frac{2 J/r^3}{\sqrt{2m/r}}=-\frac{2 J}{\sqrt{2m}} r^{-5/2}
\ee
which can be integrated to find
\be
\phi(r)=\phi_{\infty}+\frac{4 J}{3\sqrt{2m}} r^{-3/2}
\ee
where $\phi_{\infty}$ is the value of the $\phi$ coordinate evaluated at spatial infinity for these geodesics. Notice the relatively clean result that rotation of the central mass will cause particles traveling along these rain geodesics to be deflected. 

\section{On-axis geodesics}

On-axis we have either $\theta=0$ or $\theta=\pi$, in either case $d\theta/dt=0$. Also, since we are working on-axis, we can choose to conduct our analysis at any fixed value of $\phi$ and hence let $d\phi/dt=0$ without any loss of generality. Hence, we only consider the $t-r$ plane. In this case our Lense-Thirring metric reduces to
\be
(ds^2)_{\text{On-axis}}=-dt^2+\left( dr+\sqrt{\frac{2m}{r}} dt \right)^2 \quad .
\ee
Hence the contravariant metric and inverse metric (respectively) are given by
\be
(g_{ab})_{\text{On-axis}}=\left[ 
\begin{array}{c|c}
-1+\frac{2m}{r} & \sqrt{\frac{2m}{r}} \\
\hline
\sqrt{\frac{2m}{r}} & 1
\end{array}
\right]_{ab} ;
(g^{ab})_{\text{On-axis}}=\left[ 
\begin{array}{c|c}
-1 & \sqrt{\frac{2m}{r}} \\
\hline
\sqrt{\frac{2m}{r}} & 1-\frac{2m}{r}
\end{array}
\right]_{ab} \quad .
\ee
These metrics are exactly the same as the metrics of the Schwarzschild solution written in Painlev\'{e}-Gullstrand coordinates and limited to the $t-r$ plane. Hence, the on-axis geodesics of our Lense-Thirring spacetime will be the same as in the Schwarzschild solution. We will carry out the full calculation for completeness. For the on-axis null geodesics we have $g_{ab}(dx^a/dt)(dx^b/dt)=0$, hence
\be
-1+\left( \frac{dr}{dt}+\sqrt{\frac{2m}{r}}\right)^2=0 \quad .
\ee
Therefore we have 
\be \label{PGLT_OA_Null_drdt}
\frac{dr}{dt}=-\sqrt{\frac{2m}{r}}\pm 1 \quad .
\ee
For the on-axis timelike geodesics we have $g_{ab}(dx^a/dt)(dx^b/dt)=-1$, hence
\be
\left(\frac{dt}{d\tau}\right)^2 \left( -1+\left( \frac{dr}{dt}+\sqrt{\frac{2m}{r}}\right)^2\right)=-1 \quad .
\ee
From the results in appendix \ref{C:Killing}, we can construct a conserved quantity from the time translation Killing vector $\xi^a=(1;0,0,0)^a\rightarrow (1,0)^a$, this quantity is given by
\be
k=g_{ab}\xi^a\left(\frac{dx^b}{d\tau}\right)=\xi_a\left(\frac{dx^a}{d\tau}\right) \quad .
\ee
Therefore
\be
\left(\frac{dt}{d\tau}\right)\left(\left(-1+\frac{2m}{r}\right)+\sqrt{\frac{2m}{r}}\frac{dr}{dt}\right) = k \quad .
\ee
If we eliminate $dt/d\tau$, we find
\be
k^2\left(-1+\left(\frac{dr}{dt}+\sqrt{\frac{2m}{r}}\right)^2\right)=-\left(\left(-1+\frac{2m}{r}\right)+\sqrt{\frac{2m}{r}}\frac{dr}{dt}\right)^2 \quad .
\ee
This is a quadratic equation in $dr/dt$, which solving yields
\be
\frac{dr}{dt}=-\sqrt{\frac{2m}{r}}\frac{k^2-1+2m/r}{k^2+2m/r}\pm k\frac{\sqrt{k^2-1+2m/r}}{k^2+2m/r} \quad .
\ee
In the limit where $k\rightarrow \infty$, this reproduces the result for on-axis null geodesics, more specifically the result presented in equation \eqref{PGLT_OA_Null_drdt}. In the limit where $r\rightarrow \infty$, we have
\be
\lim_{r\rightarrow \infty} \left(\frac{dr}{dt}\right)=\pm\sqrt{1-\frac{1}{k^2}} \quad .
\ee
This gives us a physical interpretation of the quantity $k$ since now we have 
\be
k=\frac{1}{\sqrt{1-\left(\frac{dr}{dt}\right)^2_{\infty}}}
\ee
which is the asymptotic ``gamma factor" for the on-axis geodesic. 

\section{Generic equatorial non-circular geodesics}

On the equator we have $\theta=\pi/2$, hence $d\theta/dt=0$. Since we're on the equator we only consider the $t-r-\phi$ hypersurface, therefore our metric reduces to
\be
ds^2\rightarrow -dt^2+\left( dr+\sqrt{\frac{2m}{r}} dt^2\right)^2 + r^2\left( d\phi-\frac{2J}{r^2}\right)^2 \quad .
\ee
Therefore
\be
g_{ab} \rightarrow \left[ 
\begin{array}{c|cc}
-1+\frac{2m}{r}+\frac{4J^2}{r^4} & \sqrt{\frac{2m}{r}} & -\frac{2J}{r} \\
\hline
\sqrt{\frac{2m}{r}} & 1 & 0 \\
-\frac{2J}{r} & 0 & r^2
\end{array}
\right]_{ab}
\ee
so the inverse metric becomes 
\be
g^{ab}\rightarrow \left[ 
\begin{array}{c|cc}
-1 & \sqrt{\frac{2m}{r}} & -\frac{2J}{r^3} \\
\hline
\sqrt{\frac{2m}{r}} & 1-\frac{2m}{r} & \sqrt{\frac{2m}{r}}\frac{2J}{r^3} \\
-\frac{2J}{r^3} & \sqrt{\frac{2m}{r}}\frac{2J}{r^3} & \frac{1}{r^2}-\frac{4J^2}{r^6}
\end{array}
\right]^{ab} \quad .
\ee

\subsection{Non-circular equatorial null geodesics}

For these null geodesics we will parameterise the curve $x^a(\lambda)$ by some arbitrary affine parameter $\lambda$. The null condition $g_{ab}(dx^a/dt)(dx^b/dt)=0$ then reads
\be \label{PGLT_Null_Con}
-1+\left(\frac{dr}{dt}+\sqrt{\frac{2m}{r}}\right)^2+r^2\left(\frac{d\phi}{dt}-\frac{2J}{r^3}\right)^2=0 \quad .
\ee
We can construct two conserved quantities from the time translation Killing vector $\xi^a=(1;0,0,0)^a\rightarrow (1;0,0)^a$ and the azimuthal Killing vector $\psi^a=(0;0,0,1)^a\rightarrow (0;0,1)^a$, which are
\be
\xi_a \frac{dx^a}{d\lambda}=k ; \qquad \psi_a \frac{dx^a}{d\lambda}=\Tilde{k} \quad .
\ee
More explicitly
\be
\frac{dt}{d\lambda}\left(-1+\frac{2m}{r}+\frac{4J^2}{r^4}+\sqrt{\frac{2m}{r}} \frac{dr}{dt}-\frac{2J}{r} \frac{d\phi}{dt}\right)=k
\ee
and
\be
\frac{dt}{d\lambda}\left(-\frac{2J}{r}+r^2\frac{d\phi}{dt}\right)=\Tilde{k} \quad .
\ee
Then if we eliminate $dt/d\lambda$, we find
\be
\Tilde{k}\left(-1+\frac{2m}{r}+\frac{4J^2}{r^4}+\sqrt{\frac{2m}{r}} \frac{dr}{dt}-\frac{2J}{r} \frac{d\phi}{dt}\right)=kr^2\left(\frac{d\phi}{dt}-\frac{2J}{r^3}\right) \quad .
\ee
Now, we can solve this equation for either $dr/dt$ or $d\phi/dt$, then substitute this back into equation \eqref{PGLT_Null_Con} to give a quadratic for either $d\phi/dt$ or $dr/dt$. We can solve for these quadratics explicitly, however these results are rather messy. But, these quadratics become much simpler if instead we give an asymptotic expansion in terms of inverse powers of $r$. These results are still physically relevant since recall that the Lense-Thirring spacetime is a large distance \emph{approximation}. 

Solving for $dr/dt$ we find
\be
\frac{dr}{dt}=-\sqrt{\frac{2m}{r}} P(r) \pm \sqrt{Q(r)}
\ee
where $P(r)$ and $Q(r)$ are given by
\be
P(r)=1-\frac{\Tilde{k}^2}{k^2r^2}+\OO\left(\frac{1}{r^5}\right); \qquad Q(r)=1-\left( 1-\frac{2m}{r}\right)\frac{\Tilde{k}^2}{k^2r^2}+\OO\left(\frac{1}{r^5}\right) \quad .
\ee
Similarly, for $d\phi/dt$ we find
\be
\frac{d\phi}{dt}=\left(\frac{2J}{r^3}-\frac{\Tilde{k}}{kr^2}\right)\Tilde{P}(r) \pm \sqrt{\frac{2m}{r}}\frac{\Tilde{k}}{kr^2}\sqrt{\Tilde{Q}(r)}
\ee
where $\Tilde{P}(r)$ and $\Tilde{Q}(r)$ are given by
\be
\begin{split}
\Tilde{P}(r) & =1-\frac{2\Tilde{k}(Jk+m\Tilde{k})}{k^2r^3}+\OO\left(\frac{1}{r^4}\right); \\
\Tilde{Q}(r) & =1-\frac{\Tilde{k}^2}{k^2r^2}-\frac{2\Tilde{k}(Jk+m\Tilde{k})}{k^2r^3}+\OO\left(\frac{1}{r^5}\right) \quad .
\end{split}
\ee
Overall, these results are tractable. However, the fully explicit results are rather messy, asymptotic expansions does help remedy this. 

\subsection{Non-circular equatorial timelike geodesics}

Overall, our analysis here will be very similar to our analysis of non-circular equatorial null geodesics, but with some deviations. Firstly, we parameterise our curve $x^a(\tau)$ by the proper time parameter $\tau$. The timelike normalisation condition $g_{ab}(dx^a/dt)(dx^b/dt)=-1$ then reads
\be \label{PGLT_Eq_ncirc_TNC}
\left(\frac{dt}{d\tau}\right)^2\left( -1+\left(\frac{dr}{dt}+\sqrt{\frac{2m}{r}}\right)^2+r^2\left(\frac{d\phi}{dt}-\frac{2J}{r^3}\right)^2\right)=-1 \quad .
\ee
We can again construct two conserved quantities from the time translation Killing vector $\xi^a=(1;0,0,0)^a\rightarrow (1;0,0)^a$ and the azimuthal Killing vector $\psi^a=(0;0,0,1)^a\rightarrow (0;0,1)^a$, which are
\be
\xi_a \frac{dx^a}{d\tau}=k ; \qquad \psi_a \frac{dx^a}{d\tau}=\Tilde{k} \quad .
\ee
More explicitly
\be
\frac{dt}{d\tau}\left(-1+\frac{2m}{r}+\frac{4J^2}{r^4}+\sqrt{\frac{2m}{r}}\frac{dr}{dt}-\frac{2J}{r}\frac{d\phi}{dt}\right)=k
\ee
and
\be
\frac{dt}{d\tau}\left(-\frac{2J}{r}+r^2\frac{d\phi}{dt}\right)=\Tilde{k} \quad .
\ee
If we eliminate $dt/d\tau$ in these two equations, we find
\be
\Tilde{k}\left(-1+\frac{2m}{r}+\frac{4J^2}{r^4}+\sqrt{\frac{2m}{r}}\frac{dr}{dt}-\frac{2J}{r}\frac{d\phi}{dt}\right)=kr^2\left(-\frac{2J}{r^3}+r^2\frac{d\phi}{dt}\right) \quad .
\ee
If we eliminate $dt/d\tau$ in the timelike normalisation condition (equation \eqref{PGLT_Eq_ncirc_TNC}), we find
\be \label{PGLT_Eq_ncirc_TNC_2}
\Tilde{k}\left(-1+\left(\frac{dr}{dt}+\sqrt{\frac{2m}{r}}\right)^2+r^2\left(\frac{d\phi}{dt}-\frac{2J}{r^3}\right)^2\right)=-\left(-\frac{2J}{r}+r^2\frac{d\phi}{dt}\right)^2 \quad .
\ee
Now, we can solve this equation for either $dr/dt$ or $d\phi/dt$, then substitute this back into equation \eqref{PGLT_Eq_ncirc_TNC_2} to give a quadratic for either $d\phi/dt$ or $dr/dt$. We can solve for these quadratics explicitly, however, similar to the null geodesics, these results are rather messy. So we instead give these quadratics an asymptotic expansions in terms of inverse powers of $r$. 

Solving for $dr/dt$, we find
\be
\frac{dr}{dt}=-\sqrt{\frac{2m}{r}} P(r) \pm \sqrt{Q(r)}
\ee
where $P(r)$ and $Q(r)$ are given by
\be
P(r)=1-\frac{1}{k^2}+\frac{2m}{k^4r}+\OO\left(\frac{1}{r^2}\right) ; \qquad Q(r)=1-\frac{1}{k^2}+\frac{2m(2-k^2)}{k^4r}+\OO\left(\frac{1}{r^2}\right) \quad .
\ee
Similarly, for $d\phi/dt$ we find
\be
\frac{d\phi}{dt}=\left(\frac{2J}{r^3}-\frac{\Tilde{k}}{kr^2}\right)\Tilde{P}(r) \pm \sqrt{\frac{2m}{r}}\frac{\Tilde{k}}{kr^2}\sqrt{\Tilde{Q}(r)}
\ee
where $\Tilde{P}(r)$ and $\Tilde{Q}(r)$ are given by
\be
\Tilde{P}(r)=1-\frac{2m}{k^2r}+\OO\left(\frac{1}{r^2}\right) ; \qquad \Tilde{Q}(r)=1-\frac{1}{k^2}+\frac{2m(2-k^2)}{k^4r}+\OO\left(\frac{1}{r^2}\right) \quad .
\ee
Similar to the non-circular equatorial null geodesics, while the explicit results are integrable, they are rather messy. However, writing these results in terms of asymptotic expansions does help remedy this.

\section{ISCOs and photon orbits in the equatorial plane}\label{ISCO}

Recall the line element for the Painlev\'{e}-Gullstrand of the Lense-Thirring spacetime is given by
\be
ds^2 = - d t^2 +\left(d r+\sqrt{\frac{2m}{r}} \; dt\right)^2
+ r^2 \left(d\theta^2+\sin^2(\theta)\; \left(d\phi - \frac{2J}{r^3} dt\right) ^2\right) \quad .
\ee
Now consider a tangent vector to a null or timelike geodesic, parameterised by some arbitrary affine parameter $\lambda$, we have
\be
\begin{split}
 g_{ab}\frac{dx^{a}}{d\lambda}\frac{dx^{b}}{d\lambda} = & -\left(\frac{dt}{d\lambda}\right)^{2} + \left(\frac{dr}{d\lambda}+\sqrt{\frac{2m}{r}} \frac{dt}{d\lambda}\right)^{2} \\
 & + r^{2}\left( \left(\frac{d\theta}{d\lambda}\right)^{2}+\sin^{2}(\theta)\left(\frac{d\phi}{d\lambda}-\frac{2J}{r^{3}}\frac{dt}{d\lambda}\right)^{2}\right) \quad .
\end{split}
\ee
Now without loss of generality, we can distinguish between the two physically interesting cases (timelike and null) by defining:
\begin{equation}
    \epsilon = \left\{
    \begin{array}{rl}
    -1 & \qquad\mbox{massive particle, \emph{i.e.} timelike geodesic} \\
     0 & \qquad\mbox{massless particle, \emph{i.e.} null geodesic} .
    \end{array}\right. 
\end{equation}
Such that $g_{ab}(dx^a/d\lambda)(dx^b/d\lambda)=\epsilon$. Now, we consider geodesics on the equatorial plane, that is, we fix $\theta=\pi/2$ such that $d\theta/d\lambda=0$. Physically, these geodesics represent circular orbits confined to the equatorial plane \emph{only}. The timelike/null condition then reads:
\begin{equation}\label{epsilon}
    -\left(\frac{dt}{d\lambda}\right)^{2} + \left(\frac{dr}{d\lambda}+\sqrt{2m\over r}\frac{dt}{d\lambda}\right)^{2} + r^{2}\left(\frac{d\phi}{d\lambda}-\frac{2J}{r^{3}}\frac{dt}{d\lambda}\right)^{2} = \epsilon \quad .
\end{equation}
This spacetime has time translation symmetry and axial symmetry and hence possesses the time translation Killing vector $\xi^a=(1;0,0,0)^a$ and the azimuthal Killing vector $\psi^a=(0;0,0,1)^a$. From these Killing vectors we can construct two conserved quantities, the energy and angular momentum of a test particle (respectively), which are given as
\begin{eqnarray}\label{EL}
    E &=& \left(-1+\frac{2m}{r}+\frac{4J^{2}}{r^{4}}\right) \frac{dt}{d\lambda} + \sqrt{\frac{2m}{r}} \frac{dr}{d\lambda} - \frac{2J}{r} \frac{d\phi}{d\lambda} \ ; \\
    && \nonumber \\
    \label{EG} L &=& r^{2} \frac{d\phi}{d\lambda} - \frac{2J}{r} \frac{dt}{d\lambda} \quad .
\end{eqnarray}
But we can treat equations \eqref{epsilon}, \eqref{EL} and \eqref{EG} as a system of equations in the three unknowns $dt/d\lambda$, $dr/d\lambda$, and $d\phi/d\lambda$. That is, we can analytically solve for $dr/d\lambda$ as a function of $r$, $E$, $L$, and $\epsilon$ \emph{only}. Which yields
\begin{equation}
\frac{dr}{d\lambda} = \pm \sqrt{ 
\left(E+\frac{2JL}{r^3}\right)^2 - \left(1-\frac{2m}{r}\right)\left(\frac{L^2}{r^2}-\epsilon\right)} \quad .
\end{equation}
Similar to what we saw in section \ref{S:Sch_ISCOs}, this equation is what we would expect for a particle with unit mass in usual 1-dimensional, non-relativistic mechanics, under the influence of the following potential
\begin{equation}
V(r)=E^2-\left(\frac{dr}{d\lambda}\right)^2=\left(1-\frac{2m}{r}\right) \left(\frac{L^2}{r^2}-\epsilon\right)+E^2-\left( E+\frac{2JL}{r^3}\right)^2
\end{equation}
We can then use the features of this potential to solve for the radial positions of the innermost stable circular orbit (ISCO) and photon ring of our spacetime. Also notice that in the limit $J\rightarrow 0$, the potential reduces to that of Schwarzschild (equation \eqref{Sch_ISCO_Pot}). Now we consider two cases, the null case where $\epsilon=0$ and the timelike case where $\epsilon=-1$. We start our analysis with the null case. 

\subsection{Null orbits}

For the null case where $\epsilon=0$, our potential reduces to 
\begin{equation}
\begin{split}
V_0(r) & =  \left(1-{2m\over r}\right)  {L^2\over r^2}
 - {4EJL\over r^3} - {4 L^2 J^2\over r^6}\\
& =   {L( L(r-2m)r^3 + 4J(JL+Er^3))\over r^6} \quad .
\end{split}
\end{equation}
Now, the photon ring of our spacetime occurs at the extrema of the potential where $dV_0(r)/dr=0$. That is, the coordinate value of $r$ where we have
\begin{equation}
\frac{dV_0(r)}{dr} = -{2L\over r^7} \left(L r^4 - 3(Lm+2EJ) r^3 -12 J^2 L  \right) = 0
\end{equation}
however, this equation has no analytic solution. But we can obtain an analytic solution if we solve both $V_0(r)=E^2$ and $dV_0(r)/dr=0$ simultaneously. That is, we simultaneously solve the following for $r$
\begin{equation}
L r^4 - 3(Lm+2EJ) r^3 -12 J^2 L = 0 \, ;
\end{equation}
\begin{equation}
E^2 r^6 - L^2 r^4 +2L(Lm+2EJ)r^3+ 4 J^2L^2=0 \quad .
\end{equation}
Eliminating $E$ from these equations yields
\begin{equation}
r^5-6mr^4+9mr^3+72J^2m-36rJ^2=0 \quad .
\end{equation}
(Notice that $L$ has also been eliminated in this process). Now we rearrange:
\begin{equation}
r^3(r-3m)^2 = 36J^2\left(1-\frac{2m}{r}\right) r
\end{equation}
Therefore
\begin{equation}
(r-3m) = \pm {6J\sqrt{1-2m/r}\over r}
\end{equation}
Thence
\begin{equation}
r = 3m \pm {6J\sqrt{1-2m/r}\over r}
\end{equation}
which is still exact. However, we now iterate to find the value of $r$ which corresponds to the photon ring location purely in terms of the parameters which are present in our spacetime. At lowest order:
\begin{equation}
r = 3m + \OO(J)
\end{equation}
Thence, iterating
\begin{equation}
r = 3m \pm {6J\sqrt{1-2/3}\over 3m} + \OO(J^2)
\end{equation}
And finally
\begin{equation}
r = 3m \pm {2J\over \sqrt{3} \; m} + \OO(J^2) \quad .
\end{equation}
Now, in the limit $J\rightarrow 0$, the photon ring location reduces to its known location in Schwarzschild, as we expect. Furthermore, for the Kerr solution, the photon ring occurs at:
\begin{equation}
r_{\text{Kerr}}=2m\left[ 1+\cos\left(\frac{2}{3}\cos^{-1}\left(\pm \frac{J}{m^2}\right)\right)\right] \quad .
\end{equation}
Conducting a Taylor series expansion around $J=0$, yields
\begin{equation}
\begin{split}
r_{\text{Kerr}}(J\rightarrow 0) & =2m\left(\frac{3}{2}\pm \frac{J}{\sqrt{3}\;m^2} + \OO(J^2)\right)\\
& = 3m \pm {2J\over \sqrt{3} \; m} + \OO(J^2) \quad .
\end{split}
\end{equation}
This is exactly the photon ring location in the Lense-Thirring spacetime. This then shows that in the slow-rotation limit, the Kerr solution does indeed reduce to Lense-Thirring as we expect. 

As for the stability of these orbits, we calculate the second derivative of the potential $V_0(r)$, which gives
\begin{equation} \label{ddV0}
\frac{d^2V_0(r)}{dr^2}=- {6L\over r^8} \left( 28 J^2 L +4(mL+2EJ) r^3-Lr^4\right) \quad .
\end{equation}
Notice that here we cannot simply eliminate $L$ as in our previous analysis. We instead solve $dV_0(r)/dr=0$ for $L$, which yields
\begin{equation}
L=- {6 EJ r^3\over 12J^2+3mr^3-r^4} \quad .
\end{equation}
Then substituting this back into equation \eqref{ddV0} we get
\begin{equation}
\frac{d^2V_0(r)}{dr^2} = - {72 E^2 J^2 (r^4+36 J^2)\over r^2 (12J^2+3mr^3-r^4)^2} \quad ,
\end{equation}
which is negative for all values of $r$, therefore all circular equatorial null geodesics are unstable. This means that any photon traveling along a geodesic that passes through the photon ring will continue to spiral towards the event horizon and eventually the singularity at $r=0$.

\subsection{ISCOs}

In the timelike case where $\epsilon=-1$, our potential becomes
\begin{equation}
V_{-1}(r) =  \left(1-{2m\over r}\right) \left(1+{L^2\over r^2}\right) - {4EJL\over r^3} - {4 L^2 J^2\over r^6} \quad .
\end{equation}
The derivative of this is given by
\begin{equation}
\frac{dV_{-1}(r)}{dr} = {2\over r^7} \left(m r^5 - L(L r^4 - 3(Lm+2EJ) r^3 -12 J^2 L)  \right)
\end{equation} 
where, like the null case, has no analytic solution. But if we solve both $V_{-1}(r)=E^2$ and $\frac{dV_{-1}(r)}{dr}=0$ simultaneously, we can begin to form an analytic solution. That is to say, we solve the following equations simultaneously for r:
\begin{equation}
m r^5 - L(L r^4 - 3(Lm+2EJ) r^3 -12 J^2 L) =0 \quad ;
\end{equation}
\begin{equation} \label{E_quad}
 (E^2-1)r^6 +2m r^5 - L^2 r^4 +2L(Lm+2EJ)r^3 + 4 J^2 L^2 =0 \quad .
\end{equation}
Solving for $E$ gives
\begin{equation}
E=- {mr^5 - L^2 r^4+3mL^2 r^3+12J^2L^2\over 6JLr^3} \quad .
\end{equation}
Then if we substitute this back into $\eqref{E_quad}$ we find
\begin{equation} \label{Circ_Con_T}
r^3(3L^2m-L^2r+mr^2)^2  -36J^2L^2(L^2+r^2)(r-2m)=0 \quad .
\end{equation}
This equation states that there exist many circular timelike orbits. However, unlike for null orbits, $L$ is not eliminated, therefore we cannot analytically solve for the ISCO location yet. The second derivative of the potential is given by
\begin{equation}
\frac{d^2V_{-1}(r)}{dr^2} = -{2\over r^8} \left( 2m r^5 -3 L^2 r^4 +12L(mL+2EJ)r^3+84J^2L^2\right) \quad ,
\end{equation}
then substituting our expression for $E$
\begin{equation}
\frac{d^2V_{-1}(r)}{dr^2} = -{2\over r^8} (L^2(r^4+36J^2) -2m r^5) \quad .
\end{equation}
The condition that defines an extremal circular orbit is $\frac{d^2V_{-1}(r)}{dr^2}=0$, that is
\begin{equation}
(L^2(r^4+36J^2) -2m r^5) =0 \quad .
\end{equation}
Using this and our condition for a circular orbit, equation \eqref{Circ_Con_T}, now we can eliminate $L$, which gives
\begin{equation}
m r^6 (r-6m)^2 = 72 J^2 (r^2+mr-10m^2) + 1296J^4(2r-5m)
\end{equation}
Thence
\begin{equation}
(r-6m)^2 = {72 J^2 r^3(r^2+mr-10m^2) + 1296J^4(2r-5m)\over m r^6}
\end{equation}
Thence
\begin{equation}
r=6m \pm {6J\over m r^3} \sqrt{ 2r^3 (r^2+mr-10m^2) + 36J^2(2r-5m)}
\end{equation}
which is still exact. However, we now iterate to find the value of $r$ which corresponds to the ISCO location purely in terms of the parameters which are present in our spacetime. At lowest order:
\begin{equation}
r = 6m +\OO(J)
\end{equation}
Thence, iterating
\begin{equation}
r = 6m +  {4\sqrt{2}\over\sqrt{3}}{J\over m} +  \OO(J^2) \quad .
\end{equation}
Note that when $J\rightarrow 0$, the ISCO simplifies to its known location in Schwar- zschild. Furthermore, for the Kerr solution, the known exact location of the ISCO occurs at:
\begin{equation}
r_{\text{Kerr}} = m \left(3 + Z_2\pm\sqrt{(3-Z_1)((3+Z_1+2Z_2)}\right)
\end{equation}
where
\begin{equation}
Z_1 = 1 + \sqrt[3]{(1-x^2)}\left(\sqrt[3]{(1+x)} +\sqrt[3]{(1-x)}\right); \quad 
Z_2 = \sqrt{3x^2+Z_1^2}
\end{equation}
and
\be
x = J/m^2 \quad .
\ee 
Conducting a Taylor series expansion around $J=0$ we find 
\begin{equation}
r_{\text{Kerr}}(J\rightarrow 0) = 6m +  {4\sqrt{2}\over\sqrt{3}}{J\over m} +  \OO(J^2) \quad .
\end{equation}
this is exactly ISCO location in the Lense-Thirring spacetime. This then shows that in the slow-rotation limit, the Kerr solution does indeed reduce to Lense-Thirring as we expect.

\section{Killing tensor and Carter constant}

Finding a non-trivial Killing tensor in a spacetime is a difficult process as Killing tensors do not arise from the symmetries of the spacetime like Killing vectors do. In fact, there is yet no general algorithm that generates a non-trivial Killing tensor for any given spacetime. However, as shown in \cite{PK:2018, PK:2020}, a non-trivial Killing tensor can be generated if the inverse metric can be written in a very particular form. For the Painlev\'{e}-Gullstard form of the Lense-Thirring spacetime, we find
\begin{equation} \label{PGLT_Killing_Tensor}
K_{ab}=
\begin{bmatrix}
\frac{4J^2\sin^2(\theta)}{r^2} & 0 & 0 & -2Jr\sin^2(\theta)\\
0 & 0 & 0 & 0\\
0 & 0 & r^4 & 0\\
-2Jr\sin^2(\theta) & 0 & 0 & r^4\sin^2(\theta)
\end{bmatrix}
_{ab} \quad .
\end{equation}

\noindent It is then easy to check that $\n_{(c}K_{ab)}=0$, thus proving that $K_{ab}$ as given above is indeed a Killing tensor. Note that in spherical symmetry (i.e. as $J\rightarrow 0$), we have
\begin{equation}
K_{ab}=
\begin{bmatrix}
0 & 0 & 0 & 0\\
0 & 0 & 0 & 0\\
0 & 0 & r^4 & 0\\
0 & 0 & 0 & r^4\sin^2(\theta)
\end{bmatrix}
_{ab} \quad .
\end{equation}
Which is just the non-trivial Killing tensor in Schwarzschild. As shown in appendix \ref{C:Killing}, we can construct conserved quantities from Killing vectors and furthermore from Killing tensors. The conserved quantity that is constructed from the non-trivial Killing tensor of a spacetime is called the Carter constant (we say non-trivial since the metric is always a Killing tensor and admits the conserved quantity $\epsilon$ via $g_{ab}v^av^b=\epsilon$).

The Carter constant of our Lense-Thirring spacetime is given by 
\begin{equation} 
\mathcal{C}=K_{ab}\frac{dx^a}{d\lambda}\frac{dx^b}{d\lambda}=r^4\left[ \left(\frac{d\theta}{d\lambda}\right)^2 + \sin^2(\theta) \left(\frac{d\phi}{d\lambda} - \frac{2J}{r^3}\frac{dt}{d\lambda}\right)^2 \right] \quad .
\end{equation}
We now have 4 conserved quantities defined by the following equations:
\begin{equation}\label{E}
E=\xi_a\dfrac{dx^a}{d\lambda}=\left( 1-\frac{2m}{r}-\frac{4J^2\sin^2(\theta)}{r^4} \right)\frac{dt}{d\lambda}-\sqrt{\frac{2m}{r}}\frac{dr}{d\lambda}+\frac{2J\sin^2(\theta)}{r}\frac{d\phi}{d\lambda} \; ;
\end{equation}
\begin{equation} \label{L}
L=\psi_a\frac{dx^a}{d\lambda}=r^2\sin^2(\theta)\frac{d\phi}{d\lambda}-\frac{2J\sin^2(\theta)}{r}\frac{dt}{d\lambda}  \quad ;
\end{equation}
\begin{equation}\label{Geo_eqn}
\begin{split}
\epsilon=g_{ab}\frac{dx^a}{d\lambda}\frac{dx^b}{d\lambda}= & -\left(\frac{dt}{d\lambda}\right)^2+\left(\frac{dr}{d\lambda}+\sqrt{\frac{2m}{r}}\frac{dt}{d\lambda}\right)^2\\
& + r^2\left[\left(\frac{d\theta}{d\lambda}\right)^2+\sin^2(\theta)\left(\frac{d\phi}{d\lambda}-\frac{2J}{r^3}\frac{dt}{d\lambda}\right)^2\right]   \quad ;
\end{split}
\end{equation}
and of course the definition of the Carter constant above. 

These equations can be greatly simplified:
\begin{equation} \label{L_2}
L=r^2\sin^2(\theta) \left(\frac{d\phi}{d\lambda}-\frac{2J}{r^3}\frac{dt}{d\lambda}\right) \quad ;
\end{equation}
\begin{equation} \label{C_2}
\mathcal{C}=r^4\left(\frac{d\theta}{d\lambda}\right)^2 +{L^2\over \sin^2(\theta)} \quad ;
\end{equation}
\begin{equation}\label{Geo_eqn_2}
\epsilon=  -\left(\frac{dt}{d\lambda}\right)^2+\left(\frac{dr}{d\lambda}+\sqrt{\frac{2m}{r}}\frac{dt}{d\lambda}\right)^2 +{\mathcal{C}\over r^2} \quad ;
\end{equation}
\begin{equation}\label{E_2}
E=\left( 1-\frac{2m}{r}\right)\frac{dt}{d\lambda}-\sqrt{\frac{2m}{r}}\frac{dr}{d\lambda}+\frac{2J}{r^3}L \quad .
\end{equation}
Notice that for any given non-zero values of $\mathcal{C}$ and $L$, equation \eqref{C_2} then gives a range of forbidden azimuthal angles. Since we require that $d\theta/d\lambda$ be real, equation \eqref{C_2} then implies the following condition:
\begin{gather}
\left(r^2\frac{d\theta}{d\lambda}\right)^2=\mathcal{C}-\frac{L^2}{\sin^2(\theta)}>0 \quad ;\\
\Longrightarrow \theta>\sin^{-1}\left(\frac{L}{\sqrt{\mathcal{C}}}\right) \quad .
\end{gather}
Thus giving a range of forbidden azimuthal angles.

\noindent Now, we have 4 equations \eqref{E_2}, \eqref{L_2}, \eqref{Geo_eqn_2} and \eqref{C_2} and 4 unknown functions $dt/d\lambda$, $dr/d\lambda$, $d\theta/d\lambda$ and $d\phi/d\lambda$. So, we can solve for each unknown function analytically, giving:
\begin{equation}
\frac{dt}{d\lambda}=\frac{Er^4+2LJr\pm \sqrt{2mrX(r)}/r^3}{r^3(2m-r)} \quad ;
\label{E:t}
\end{equation}
\begin{equation}
\frac{dr}{d\lambda}=\pm \sqrt{X(r)} \quad ;
\label{E:r}
\end{equation}
\begin{equation}
\frac{d\theta}{d\lambda}=\pm \frac{\sqrt{L^2/\sin^2(\theta)-\mathcal{C}}}{r^2} \quad ;
\label{E:theta}
\end{equation}
and
\begin{equation}
\frac{d\phi}{d\lambda}=
{L\over r^2\sin^2(\theta)} +
\frac{2EJr^4+4LJ^2r\pm 2J\sqrt{2mrX(r)}/r^3}{r^6(2m-r)} \quad ,
\end{equation}
where $X(r)$ is a sextic polynomial given by
\begin{equation} \label{X}
X(r) =(E-2JL/r^3)^2-(1-2m/r)(-\epsilon  +C/r^2) \quad .
\end{equation}
Via equation \eqref{E:r}, we have
\begin{equation} \label{lambda_r}
\lambda(r) = \lambda_0 + \int_{r_0}^r {d\bar r\over\sqrt{X(\bar r)}} \quad .
\end{equation}
We cannot explicitly integrate this equation in closed form, hence this relation cannot be inverted to find $r(\lambda)$. But, we can still attempt to integrate our equations in terms of $r$. From equations \eqref{E:t} and \eqref{E:r}, we find
\begin{equation}
\frac{dt}{dr} = {1\over 1-2m/r} \pm {\sqrt{2m/r} (E-2JL/r^3)\over (1-2m/r)\sqrt{X(r)}} \quad .
\end{equation}
Therefore
\begin{equation}
t(r) = t_0 + \int_{r_0}^r \left( {1\over 1-2m/\bar r} \pm {\sqrt{2m/r} (E-2JL/\bar r^3)\over (1-2m/\bar r)\sqrt{X(\bar r)}} \right) d\bar r \quad .
\end{equation}
Hence
\begin{equation}
t(r) = t_0 + r-r_0 + 2m\ln\left(r-2m\over r_0-2m\right) \pm \int_{r_0}^r \left( {\sqrt{2m/r} (E-2JL/\bar r^3)\over (1-2m/\bar r)\sqrt{X(\bar r)}} \right) d\bar r \quad .
\end{equation}
As before, we cannot explicitly integrate this equation in closed form.

In contrast, for our equation involving $\theta$, we firstly define the critical angle $\theta_*$ given as $\sin^2(\theta_*)=L^2/\mathcal{C}$. Then via equation \eqref{E:theta}, we find
\begin{equation}
\begin{split}
\frac{d\cos(\theta)}{d\lambda} 
& = \mp {\sqrt{\mathcal{C}\sin^2 (\theta)-L^2}\over r^2} \\
& =\mp \sqrt{\mathcal{C}}\; {\sqrt{\sin^2 (\theta)-\sin^2(\theta_*)}\over r^2} \\
& =\mp \sqrt{\mathcal{C}}\; {\sqrt{\cos^2(\theta_*)-\cos^2(\theta)}\over r^2} \quad .
\end{split}
\end{equation}
Rearranging
\begin{equation}
{d\cos(\theta) \over  \sqrt{\cos^2(\theta_*)-\cos^2(\theta)}}  = \mp  \sqrt{\mathcal{C}} {dr\over r^2\sqrt{X(r)}} \quad .
\end{equation}
Then via the chain rule
\begin{equation}
{d\cos^{-1}\left(\frac{\cos(\theta)}{\cos(\theta_*)}\right)} = \pm \sqrt{\mathcal{C}} {dr\over r^2\sqrt{X(r)}}  \quad .
\end{equation}
We now integrate, giving
\begin{equation}
 {\cos^{-1}\left(\frac{\cos(\theta)}{\cos(\theta_*)} \right)} =  {\cos^{-1}\left(\frac{\cos(\theta_0)}{\cos(\theta_*)} \right)}    \pm \sqrt{\mathcal{C}} \int_{r_0}^r {d\bar r\over \bar r^2\sqrt{X(\bar r)}}
\end{equation}
where $\theta_0$ is some arbitrary initial angle. Rearranging for $\cos(\theta)$ gives
\begin{equation}
 {\cos(\theta)} =  \cos(\theta_*) \; \cos\left( {\cos^{-1}\left( \frac{\cos(\theta_0)} {\cos(\theta_*)}\right)}    \pm \sqrt{\mathcal{C}} \int_{r_0}^r {d\bar r\over \bar r^2\sqrt{X(\bar r)}}\right) \quad .
\end{equation}
Hence
\begin{equation}
\begin{split}
\cos(\theta) = &  \cos(\theta_*)\;\left\{  
 \left( \frac{\cos(\theta_0)}{\cos(\theta_*)} \right) \cos \left(\sqrt{\mathcal{C}} \int_{r_0}^r {d\bar r\over \bar r^2\sqrt{X(\bar r)}}\right)\right.\\
 &\left. \pm \sin\left( \cos^{-1}\left( \frac{\cos(\theta_0)}{\cos(\theta_*)}\right)\right)
  \sin\left(\sqrt{\mathcal{C}} \int_{r_0}^r {d\bar r\over \bar r^2\sqrt{X(\bar r)}}\right) \right\} \quad .
\end{split}
\end{equation}
Hence
\begin{equation}
\begin{split}
\cos(\theta) = & \cos(\theta_*)\;\left\{ \left( \frac{\cos(\theta_0)}{\cos(\theta_*)} \right) \cos \left(\sqrt{\mathcal{C}} \int_{r_0}^r {d\bar r\over \bar r^2\sqrt{X(\bar r)}}\right)\right.\\
&\left. \pm \sqrt{1-\left(\cos(\theta_0) \over\cos(\theta_*)\right)^2}
  \sin\left(\sqrt{\mathcal{C}} \int_{r_0}^r {d\bar r\over \bar r^2\sqrt{X(\bar r)}}\right)\right\} \quad .
\end{split}
\end{equation}
Finally
\begin{equation} \label{ctheta}
\begin{split}
\cos(\theta) = & \cos(\theta_0) \;  \cos \left(\sqrt{\mathcal{C}} \int_{r_0}^r {d\bar r\over \bar r^2\sqrt{X(\bar r)}}\right)\\
& \pm  \sqrt{\cos^2(\theta_*)-\cos^2(\theta_0)}\; \sin\left(\sqrt{\mathcal{C}} \int_{r_0}^r {d\bar r\over \bar r^2\sqrt{X(\bar r)}}\right) \quad .
\end{split}
\end{equation}
This shows that $\cos(\theta)$ oscillates in the range $\cos(\theta) \in [-\cos(\theta_*), +\cos(\theta_*)]$. Since $\theta_0$ is some arbitrary initial angle, without any loss of generality we can can choose $\theta_0=\theta_*$. Which then gives
\begin{equation}
 \cos(\theta) = \cos(\theta_*) \;  \cos \left(\sqrt{\mathcal{C}} \int_{r_*}^r {d\bar r\over \bar r^2\sqrt{X(\bar r)}}\right)
\end{equation}
which can be inverted to give $\theta(r)$.

Finally we consider the ordinary differential equation (ODE) for $d\phi/d\lambda$. Which is
\begin{equation}
\frac{d\phi}{d\lambda} = {L\over r^2\sin^2(\theta)} + \frac{2J}{r^3}\frac{dt}{d\lambda} \quad .
\end{equation}
Thence
\begin{equation}
\frac{d\phi}{dr} = \pm{L \over \sqrt{X(r)}\; r^2\sin^2(\theta)} + \frac{2J}{r^3}\frac{dt}{dr} \quad .
\end{equation}
Recall that
\begin{equation}
\frac{dt}{dr} = {1\over 1-2m/r} \pm {\sqrt{2m/r} \; (E-2JL/r^3)\over (1-2m/r)\sqrt{X(r)}} \quad ,
\end{equation}
so we have
\begin{equation}
\frac{d\phi}{dr} = \pm{L \over \sqrt{X(r)}\; r^2\sin^2(\theta)} 
+ \frac{2J}{r^3}\left({1\over 1-2m/r} \pm {\sqrt{2m/r} \;(E-2JL/r^3)\over (1-2m/r)\sqrt{X(r)}}\right) \quad . 
\end{equation}
Integrating
\begin{equation}
\begin{split}
\phi(r)= \phi_0 & \pm \int_{r_0}^r {L \over \sqrt{X(\bar r)}\; \bar r^2\sin^2[\theta(\bar r)]} d\bar r
+  \int_{r_0}^r  \frac{2J}{\bar r^3}{1\over 1-2m/\bar r} d\bar r\\
& \pm  \int_{r_0}^r {2J\sqrt{2m/r} (E-2JL/\bar r^3)\over \bar r^3 (1-2m/\bar r)\sqrt{X(\bar r)}} d\bar r \quad . 
\end{split}
\end{equation}
Now, one of these terms can be explicitly integrated in closed form
\begin{equation}
\int_{r_0}^r  \frac{2J}{\bar r^3}{1\over 1-2m/\bar r} d\bar r = {J\over 2 m^2} \left\{ \ln \left(1-2m/r\over1-2m/r_0\right) + {2m\over r}-{2m\over r_0} \right\} \quad .
\end{equation}
So, in general we have
\begin{equation}
\begin{split}
\phi(r)= \phi_0 & + {J\over 2 m^2} \left\{ \ln \left(1-2m/r\over1-2m/r_0\right) + {2m\over r}-{2m\over r_0} \right\}\\
& \pm L \int_{r_0}^r {1 \over \sqrt{X(\bar r)}\; \bar r^2\sin^2[\theta(\bar r)]} d\bar r
\pm  2J \int_{r_0}^r {\sqrt{2m/r} (E-2JL/\bar r^3)\over \bar r^3 (1-2m/\bar r)\sqrt{X(\bar r)}} d\bar r \quad . 
\end{split}
\end{equation}
\noindent By introducing a fourth constant of the motion we have made our equations of motion integrable. Note that while these equations of motion (generally) cannot be analytically integrated in \emph{closed form}, they can, in principle, be integrated via numerical methods, hence they are integrable in the technical sense. In order to analytically integrate the equations of motion in closed form, we have to impose further conditions.

\section{Carter constant zero}

Our equations greatly simplify if the Carter constant is zero. Firstly, if $\mathcal{C}=0$ then via 
\begin{equation} 
\mathcal{C}=r^4\left(\frac{d\theta}{d\lambda}\right)^2 +{L^2\over \sin^2(\theta)}
\end{equation}
we see that both $L=0$ and $d\theta/d\lambda\equiv 0$, which implies that $\theta(r)=\theta_0$ is constant. Furthermore, our equation for $X(r)$ simplifies drastically, giving
\begin{equation}
X(r) = E^2+\epsilon\left( 1-\frac{2m}{r}\right) \quad .
\end{equation}
Analysing the form of this equation suggests that it would be useful to split our analysis of geodesics with zero Carter constant into the geodesics of photon and the geodesics of massive particles. 

\subsection{Photons with Carter constant zero}

Given that the Carter constant is zero, for photons, we have the following conditions:
\begin{equation}
C=0; \qquad L=0; \qquad  \theta(r)=\theta_0; \qquad
X(r) = E^2.
\end{equation}
Via equation \eqref{lambda_r}, we find
\begin{equation}
\lambda(r) = \lambda_0 + \int_{r_0}^r {d\bar r\over\sqrt{X(\bar r)}}= 
\lambda_0 + \int_{r_0}^r {d\bar r\over E}= \lambda_0 + {r-r_0\over E}
\end{equation}
and furthermore
\begin{equation}
t(r) = t_0 + r-r_0 + 2m\ln\left(r-2m\over r_0-2m\right) \pm \int_{r_0}^r {\sqrt{2m/\bar r}\over 1-2m/\bar r} d\bar r \quad .
\end{equation}
We can integrate this explicitly to find
\begin{equation}
\begin{split}
t(r) = t_0 & + r-r_0 + 2m\ln\left(r-2m\over r_0-2m\right)\\
& \pm \left\{2(\sqrt{2mr}-\sqrt{2mr_0}) 
+ 2m \ln\left( {1-\sqrt{2m/r}\over1+\sqrt{2m/r}} \;\;{1+\sqrt{2m/r_0}\over1-\sqrt{2m/r_0}} \right) \right\} \quad .
\end{split}
\end{equation}
Finally we have
\begin{equation}
\begin{split}
\phi(r)= \phi_0 & +{J\over 2 m^2} \left\{ \ln \left(1-2m/r\over1-2m/r_0\right) + {\frac{2m}{r}}-{\frac{2m}{r_0}} \right\} \\
& \pm  2J \int_{r_0}^r {\sqrt{2m/\bar r} \over \bar r^3 (1-2m/\bar r)} d\bar r \quad . 
\end{split}
\end{equation}
Which we can integrate explicitly to find
\begin{equation}
\begin{split}
\phi(r)= \phi_0 & +{J\over 2 m^2} \left\{ \ln \left(1-2m/r\over1-2m/r_0\right) + {\frac{2m}{r}}-{\frac{2m}{r_0}} \right\}\\
& \pm  2J \left\{ {1\over2m^2} \left[\sqrt{\frac{2m}{r}}-\sqrt{\frac{2m}{r_0}}\right] +{1\over6m^2} \left[\left(\frac{2m}{r}\right)^{3/2}-\left(\frac{2m}{r_0}\right)^{3/2}\right] \right.\\
& \hspace{1cm} \left. + {1\over4m^2} \ln\left( {1-\sqrt{2m/r}\over1+\sqrt{2m/r}} \;\;{1+\sqrt{2m/r_0}\over1-\sqrt{2m/r_0}} \right)  \right\}  \quad .
\end{split}
\end{equation}
These equations are rather complex but fully explicit. We also note that in the limit where $r\rightarrow r_0$ our equations of motion have sensible limits.

\subsection{Massive particles with Carter constant zero}

The form of $X(r)$ suggests that we should split our analysis into two parts, firstly where $E=1$ since in this case $X(r)$ simplifies significantly to $2m/r$ and secondly where $E\neq 1$. Physically, the geodesics with $E=1$ represent particles that begin their trajectory with zero initial velocity. We can gain additional insight if we appeal to Newtonian gravity. In Newtonian gravity, particles with $E=0$ will follow elliptic orbits, particles with $E<0$ will follow parabolic orbits and particles with $E>0$ will follow hyperbolic orbits. 

For $E=1$, we have the following conditions:
\be
C=0; \quad L=0; \quad \theta(r)=\theta_0; \quad X(r)=\frac{2m}{r} \quad .
\ee
So we have
\be \label{lamda(r)_E_eq_1}
\begin{split}
\lambda(r) & = \lambda_0+\int_{r_0}^r \frac{d\Bar{r}}{\sqrt{X(\Bar{r})}} \\
& =\lambda_0+\int_{r_0}^r \frac{d\Bar{r}}{\sqrt{2m/\Bar{r}}} \\
& =\lambda_0+\frac{1}{3}\sqrt{\frac{2}{m}}\left(r^{3/2}-r_0^{3/2}\right) \quad .
\end{split}
\ee
Recall that in general we have
\be
t(r) = t_0+r-r_0+2m\ln\left(\frac{r-2m}{r_0-2m}\right)\pm \int_{r_0}^r \frac{\sqrt{2m/\Bar{r}}(E-2JL/\Bar{r}^3)}{(1-2m/\Bar{r})\sqrt{X(\Bar{r})}}\, d\Bar{r}
\ee
So given the conditions stated above where $E=1$ and $L=0$, we have
\be
\begin{split} 
t(r) & = t_0+r-r_0+2m\ln\left(\frac{r-2m}{r_0-2m}\right)\pm \int_{r_0}^r \frac{\sqrt{2m/\Bar{r}}}{(1-2m/\Bar{r})\sqrt{2m/\Bar{r}}}\, d\Bar{r} \\
& = t_0+r-r_0+2m\ln\left(\frac{r-2m}{r_0-2m}\right)\pm \int_{r_0}^r \frac{d\Bar{r}}{(1-2m/\Bar{r})} \\
& = t_0+r-r_0+2m\ln\left(\frac{r-2m}{r_0-2m}\right)\pm\left(r-r_0+2m\ln\left(\frac{r-2m}{r_0-2m}\right)\right) \quad .
\end{split}
\ee
However, notice that when we take the minus sign we get $t(r)=t_0$, which is unphysical. Hence, here we are forced to take the plus sign, yielding
\be \label{t(r)_E_eq_1}
t(r)=t_0+2\left(r-r_0+2m\ln\left(\frac{r-2m}{r_0-2m}\right)\right) \quad .
\ee
Now solving for $\phi(r)$ we have
\be
\begin{split}
\phi(r) = \phi_0 & +\frac{J}{2m^2}\left(\ln\left(\frac{1-2m/r}{1-2m/r_0}\right)+\frac{2m}{r}-\frac{2m}{r_0}\right) \\
& \pm 2J\int_{r_0}^r \frac{\sqrt{2m/\Bar{r}}(E-2JL/\Bar{r}^3)}{\Bar{r}^3(1-2m/\Bar{r})\sqrt{X(\Bar{r})}}\, d\Bar{r}
\end{split}
\ee
\clearpage
thence
\be
\begin{split}
\phi(r) = \phi_0 & +\frac{J}{2m^2}\left(\ln\left(\frac{1-2m/r}{1-2m/r_0}\right)+\frac{2m}{r}-\frac{2m}{r_0}\right) \\
& \pm 2J\int_{r_0}^r \frac{d\Bar{r}}{\Bar{r}^3(1-2m/\Bar{r})} 
\end{split}
\ee
so we have
\be \label{E_eq_1_phi}
\begin{split}
\phi(r) =\phi_0 & +\frac{J}{2m^2}\left(\ln\left(\frac{1-2m/r}{1-2m/r_0}\right)+\frac{2m}{r}-\frac{2m}{r_0}\right) \\
& \pm \frac{J}{2m^2}\left(\ln\left(\frac{1-2m/r}{1-2m/r_0}\right)+\frac{2m}{r}-\frac{2m}{r_0}\right)
\end{split}
\ee
The $\pm$ sign here does makes physical sense. The $\pm$ indicates if the particle is moving with or against the rotation of the central mass, plus if it moves with and minus if it moves against. If the particle moves against the rotation, according to \eqref{E_eq_1_phi}, we find that the $\phi$ coordinate is constant along the geodesic.

Now, For $E\neq 1$, we have the following conditions:
\be
C=0; \quad L=0; \quad \theta(r)=\theta_0; \quad X(r)=E^2-1+\frac{2m}{r} \quad .
\ee
So we have
\be
\begin{split}
\lambda(r) & = \lambda_0+\int_{r_0}^r \frac{d\Bar{r}}{\sqrt{X(\Bar{r})}} \\
& =\lambda_0+\int_{r_0}^r \frac{d\Bar{r}}{\sqrt{E^2-1+2m/\Bar{r}}} \\
& = \lambda_0 + \frac{r\sqrt{E^2-1+2m/r}-r_0\sqrt{E^2-1+2m/r_0}}{E^2-1} \\
& \hspace{1cm} -\frac{m}{(E^2-1)^{3/2}}\ln\left(\frac{r(E^2-1+m/r+\sqrt{E^2-1+2m/r})}{r_0(E^2-1+m/r_0+\sqrt{E^2-1+2m/r_0})}\right)
\end{split}
\ee
If we conduct a Taylor series expansion around $E=1$, we find
\be
\lambda(r) = \frac{1}{3}\sqrt{\frac{2}{m}}\left(r^{3/2}-r_0^{3/2}\right) + \OO(E-1) \quad .
\ee
So in the limit where $E\rightarrow 1$, our expression for $\lambda(r)$ reduces to \eqref{lamda(r)_E_eq_1}, as we expect.

For our expression for $t(r)$ is now given by
\be
\begin{split}
t(r) &  = t_0+r-r_0+2m\ln\left(\frac{r-2m}{r_0-2m}\right) + \int_{r_0}^r \frac{E\sqrt{2m/\Bar{r}}}{(1-2m/\Bar{r})\sqrt{E^2-1+2m/\Bar{r}}}\, d\Bar{r} \\
& = t_0+r-r_0+2m\ln\left(\frac{r-2m}{r_0-2m}\right)\\
& \hspace{0.9cm} + \frac{2E\sqrt{2m}}{E^2-1}\left( r\sqrt{E^2-1+\frac{2m}{r}}-r_0\sqrt{E^2-1+\frac{2m}{r_0}}\right) \\
& \hspace{0.9cm} + 2m\ln\left(\frac{2E\sqrt{m}-\sqrt{2r(E^2-1+2m/r)}}{2E\sqrt{m}+\sqrt{2r(E^2-1+2m/r)}}\right) \\
& \hspace{0.9cm} + 2m\ln\left(\frac{2E\sqrt{m}+\sqrt{2r_0(E^2-1+2m/r_0)}}{2E\sqrt{m}-\sqrt{2r_0(E^2-1+2m/r_0)}}\right) \quad .
\end{split}
\ee
If we conduct a Taylor series expansion around $E=1$, we find
\be
t(r) = t_0+2\left(r-r_0+2m\ln\left(\frac{r-2m}{r_0-2m}\right)\right)+\OO(E-1) \quad .
\ee
So in the limit where $E\rightarrow 1$, our expression for $t(r)$ reduces to \eqref{t(r)_E_eq_1}, as we expect.

Now lastly, for our expression for $\phi(r)$ is now given by
\be
\begin{split}
\phi(r) & = \phi_0+\frac{J}{2m^2}\left(\ln\left(\frac{1-2m/r}{1-2m/r_0}\right)+\frac{2m}{r}-\frac{2m}{r_0}\right) \\
& \hspace{0.97cm} \pm 2J\int_{r_0}^r \frac{E\sqrt{2m/\Bar{r}}}{\Bar{r}^3(1-2m/\Bar{r})\sqrt{E^2-1+2m/\Bar{r}}}\, d\Bar{r} \quad .
\end{split}
\ee
Explicitly, we have
\be
\begin{split} 
\phi(r) & = \phi_0+\frac{J}{2m^2}\left(\ln\left(\frac{1-2m/r}{1-2m/r_0}\right)+\frac{2m}{r}-\frac{2m}{r_0}\right) \\
& \hspace{0.97cm} \pm \frac{J}{2m^2}\left[\frac{1}{E}\ln\left(\frac{2E\sqrt{m}-\sqrt{2r(E^2-1+2m/r)}}{2E\sqrt{m}+\sqrt{2r(E^2-1+2m/r)}}\right)\right. \\
& \hspace{2.6cm} + \frac{1}{E}\ln\left(\frac{2E\sqrt{m}+\sqrt{2r_0(E^2-1+2m/r_0)}}{2E\sqrt{m}-\sqrt{2r_0(E^2-1+2m/r_0)}}\right) \\
& \hspace{2.6cm} + \frac{E^2-3}{2}\ln\left(\frac{2\sqrt{m}-\sqrt{2r(E^2-1+2m/r)}}{2\sqrt{m}+\sqrt{2r(E^2-1+2m/r)}}\right) \\
& \hspace{2.6cm} + \frac{E^2-3}{2}\ln\left(\frac{2\sqrt{m}+\sqrt{2r_0(E^2-1+2m/r_0)}}{2\sqrt{m}-\sqrt{2r_0(E^2-1+2m/r_0)}}\right) \\
& \left. \hspace{2.6cm} + \sqrt{2m}\left(\sqrt{\frac{E^1-1+2m/r}{r}}-\sqrt{\frac{E^2-1+2m/r_0}{r_0}}\right)\right] \quad .
\end{split}
\ee
If we conduct a Taylor series expansion around $E=1$, we find
\be
\begin{split}
\phi(r) = \phi_0 & +\frac{J}{2m^2}\left(\ln\left(\frac{1-2m/r}{1-2m/r_0}\right)+\frac{2m}{r}-\frac{2m}{r_0}\right) \\
& \pm \frac{J}{2m^2}\left(\ln\left(\frac{1-2m/r}{1-2m/r_0}\right)+\frac{2m}{r}-\frac{2m}{r_0}\right)+\OO(E-1) \quad .
\end{split}
\ee
So in the limit where $E\rightarrow 1$, our expression for $\phi(r)$ reduces to \eqref{E_eq_1_phi}, as we expect.

Overall, while these equations of motion are rather lengthy, they are fully explicit and yield the required limits when we let $E\rightarrow 1$.

\section{General circular orbits}

Now that we have an expression for the Carter constant, we now have enough constants of the motion to integrate more general geodesics such as the geodesics that correspond to general circular orbits in our spacetime. That is, circular orbits that are no longer constrained to the equatorial plane. Since we are dealing with circular orbits, we fix our $r$ coordinate to some fixed value $r=r_0$, which implies that $dr/d\lambda=0$. This condition simplifies our 4 constants of the motion:

\begin{equation} \label{L_C}
L=r_0^2\sin^2(\theta) \left(\frac{d\phi}{d\lambda}-\frac{2J}{r_0^3}\frac{dt}{d\lambda}\right) \quad ;
\end{equation}
\begin{equation} 
\mathcal{C}=r_0^4\left(\frac{d\theta}{d\lambda}\right)^2 +{L^2\over \sin^2(\theta)} \quad ;
\end{equation}
\begin{equation}\label{Geo_eqn_C}
\epsilon=  -\left(1-{2m\over r_0}\right) \left(\frac{dt}{d\lambda}\right)^2
+{\mathcal{C}\over r_0^2} \quad ;
\end{equation}
\begin{equation}\label{E_C}
E=\left( 1-\frac{2m}{r_0}\right)\frac{dt}{d\lambda}
+\frac{2J}{r_0^3}L \quad .
\end{equation}
Solving for $dt/d\lambda$ shows that it is constant
\begin{equation} \label{Circ_dt_dl}
{dt\over d\lambda} = {E-2JL/r_0^3\over1-2m/r_0} \quad .
\end{equation}
Hence, we find that we also have the constraint
\begin{equation}
\epsilon= -{(E-2JL/r_0^3)^2\over1-2m/r_0}+{\C\over r_0^2}
\end{equation}
which implies
\begin{equation}
\left(\epsilon-{\C\over r_0^2}\right) \left(1-\frac{2m}{r_0}\right)  =  -{\left( E-\frac{2JL}{r_0^3}\right)^2} \quad .
\end{equation}
However, via equation \eqref{X} this constraint implies that $X(r_0)=0$.\\

\noindent The two remaining equations of motion are 
\begin{equation}\label{circ_dphi_dl}
\frac{d\phi}{d\lambda}= \frac{2J}{r_0^3}\; {E-2JL/r_0^3\over1-2m/r_0} + { L\over r_0^2 \sin^2(\theta)}
\end{equation}
and
\begin{equation} \label{circ_dtheta_dl}
\frac{d\theta}{d\lambda} =  {1\over r_0^2} \sqrt{\C-{L^2\over\sin^2(\theta)}} \quad .
\end{equation}
Slightly modifying our previous derivation of equation \eqref{ctheta}, we find the solution to equation \eqref{circ_dtheta_dl} to be
\begin{equation}
\begin{split}
{\cos(\theta(\lambda))} = &   
 \cos(\theta_0) \;  \cos \left(\sqrt{\C} \; {\lambda-\lambda_0\over r_0^2} \right) \\
& \pm  \sqrt{\cos^2(\theta_*)-\cos^2(\theta_0)}\; 
  \sin\left(\sqrt{\C}\; {\lambda-\lambda_0\over r_0^2} \right) .
\end{split}
\end{equation}
which is equivalent to
\begin{equation}
 {\cos(\theta)} =  \cos(\theta_*) \; \cos\left( {\cos^{-1}\left( \cos(\theta_0) \over \cos(\theta_*)\right)}    \pm \sqrt{\C} \;{\lambda-\lambda_0\over r_0^2}\right) \quad .
\end{equation}
Now to solve equation \eqref{circ_dphi_dl}, qualitatively, we assert that
\begin{equation}
\sin(\theta) = A \sin( B+ E\lambda)
\end{equation}
where $A$ and $B$ are arbitrary constants. So we have
\begin{equation}
\int {d\lambda \over \sin^2\theta}  = \int {d\lambda \over (A \sin( B+ E\lambda))^2} 
= - {\cot(B+E\lambda)\over A^2 E} \quad .
\end{equation}
Thence
\begin{equation}
\begin{split}
\phi = \phi_0 & +  \frac{2J}{r_0^3}\; {E-2JL/r_0^3\over1-2m/r_0} (\lambda-\lambda_0)- {L\over \cos(\theta_*^2) \sqrt{\C} }  \cot\left( {\sin^{-1}\left( \sin(\theta_0) \over \cos(\theta_*)\right)}   \right)\\
& +{L\over \cos(\theta_*^2) \sqrt{\C} }  \cot\left( {\sin^{-1}\left( \sin(\theta_0) \over \cos(\theta_*)\right)}    \pm \sqrt{C} \;{\lambda-\lambda_0\over r_0^2}\right) \quad .
\end{split}
\end{equation}
Some of these equations of motion are rather complicated, but they simplify significantly in special cases. We now analyse one such case.

\subsection{Circular orbits with $L=0$}

We now consider the special case where $L=0$. In doing so, our equations of motion simplify further to
\begin{equation}
\left(\frac{d\phi}{d\lambda}\right) = \frac{2J}{r_0^3}\; {E\over1-2m/r_0}
\end{equation}
and
\begin{equation} 
\left(\frac{d\theta}{d\lambda}\right) =  {\sqrt{C}\over r_0^2} \quad .
\end{equation}
So we have
\begin{equation}
\phi=\phi_0 + \frac{2J}{r_0^3}\; {E\over1-2m/r_0}(\lambda-\lambda_0);
\qquad
\theta=\theta_0+  {\sqrt{C}\over r_0^2} (\lambda-\lambda_0).
\end{equation}
The periodicities of the above equations are related since
\begin{equation}
\epsilon= -{E^2\over1-2m/r_0}+{C\over r_0^2}
\end{equation}
and from equation \eqref{Circ_dt_dl}, we now also have
\begin{equation}
{dt\over d\lambda} = {E\over1-2m/r_0}
\end{equation}
therefore we find
\begin{equation}
\phi=\phi_0 + \frac{2J}{r_0^3}\; (t-t_0);
\qquad
\theta=\theta_0+  {\sqrt{C}(1-2m/r_0)\over E r_0^2}\; (t-t_0).
\end{equation}

Overall, we have seen that once we are given the Carter constant of a spacetime, the geodesic equations become integrable. Hence, we can then solve for a myriad of different types of physically interesting geodesics.

\chapter{Conclusions}\label{C:con}

In the first few chapters we set out the mathematical framework to describe Einstein's theory of general relativity, this framework was that of differential geometry. We started by describing the structure of spacetime as a 4-dimensional manifold, this structure allowed us to define vectors, dual vectors and tensors in our spacetime. We then developed the notion of a derivative operator in this manifold, we used the non-commutativity of the derivative operator to define curvature on our manifold which is represented by the Riemann tensor $R^a{}_{bcd}$. We used this notion of curvature to show that the paths that particles travel along, geodesics, are curved when the Riemann tensor is non-zero (i.e. in a curved spacetime). We used this idea of geodesics being curved due to the curvature of spacetime as a description of gravity. Instead of describing gravity via a force between massive objects, we say that the presence of mass-energy in our spacetime causes our spacetime to be curved which then causes geodesics to be curved. The field equation that relates the mass-energy in a spacetime to the curvature is Einstein's equation, equation \eqref{Ein_Eq}. 

The first solution of the vacuum Einstein equation, equation \eqref{Vac_Ein_Eq}, that we studied was the Schwarzschild solution. This spacetime physically represents a massive, non-rotating, central mass in a vacuum. We saw that this solution can be used to model non-rotating stars, planets and black holes. However, the metric components as written in Hilbert's coordinates were singular at two distinct values of $r$: $r=0$ and $r=2m$. But by examination of a curvature invariant such as the Kretschmann scalar ($R_{abcd}R^{abcd}$) showed that the singularity at $r=0$ is a true physical singularly while the singularity at $r=2m$ is a merely a coordinate singularly. By performing a coordinate transformation, we were able to rid ourselves of this coordinate singularity. Furthermore, the resulting metric had two useful qualities: the 3-metric was diagonal and the metric was unit-lapse. This coordinate system we had now written the Schwarzschild solution in is the Painlev\'{e}-Gullstrand coordinate system. 

We then turned our attention to another solution of the vacuum Einstein equation, the Kerr solution. The Kerr solution is a highly non-trivial generalisation to the Schwarzschild solution where we allow our central mass to rotate. This makes the physics involved significantly more difficult, in particular calculating the geodesics of the Kerr solution is considerably harder than for the Schwarzschild solution. However, we saw that if we can recast our metric in unit-lapse form, calculation of some geodesics (notably the rain geodesics) become much simpler. There already existed two well known unit-lapse forms of the Kerr metric: the Doran metric, equation \eqref{Kerr_Doran}, and the Nat\'{a}rio metric, equation \eqref{Kerr_Natario}. By analysing the conditions for a metric to be in unit-lapse form, we were able to generate two new explicit unit-lapse forms of the Kerr metric: the Boyer-Lindquist-rain metric, equation \eqref{Kerr_BL-rain}, and the Eddington-Finkelstein-rain metric, equation \eqref{Kerr_BL-rain}. We were also able to show that there exists a general infinite class of unit-lapse forms of the Kerr metric.  We then wished to simplify the Kerr metric further by diagonalising the 3-metric of a unit-lapse form of the Kerr metric to then put the Kerr metric into Painlev\'{e}-Gullstrand form. We attempted to do this by assuming our metric was already in unit-lapse form, then we made a general coordinate transformation that retained both unit-lapse and axisymmetry and then demanded the off diagonal components of the 3-metric to vanish. However, we found a consistency condition that could not be satisfied, equation \eqref{PG_Kerr_Const_Con_2}. Hence, there does not exist a Painlev\'{e}-Gullstrand form of the Kerr metric. 
\enlargethispage{20pt}

Lastly, we analysed the Lense-Thirring spacetime. The Lense-Thirring spacetime is an \emph{approximate} solution to the vacuum Einstein equation, this means that the spacetime is not Ricci flat, but the components of the Ricci tensor go to 0 as $r\rightarrow \infty$. The Lense-Thirring spacetime approximates Kerr for slow rotations ($J\rightarrow 0$) and at large distances ($r\rightarrow \infty$). This spacetime has useful advantages over the Kerr spacetime. Firstly, the Lense-Thirring metric is much simpler than the Kerr metric, hence further analysis of the spacetime is much simpler in the Lense-Thirring spacetime. Secondly, the Lense-Thirring metric can be put into Painlev\'{e}-Gullstrand form, equation \eqref{PGLT_Met}. It is useful to note that while the Lense-Thirring spacetime is an \emph{approximate} solution to the vacuum Einstein equation, there exists no Birkhoff theorem for axisymmetric solutions in $3+1$ dimensions. Hence, the Kerr solution does not \emph{perfectly} describe any physical rotating star or planet in the universe, due to the non-trivial mass multipole moments of these objects. The Kerr solution will only approximately model these objects at large distances (i.e. large $r$), which is where the Lense-Thirring spacetime approximates Kerr. We then used this new form of the Lense-Thirring spacetime to generate a rather simple tetrad, which allowed us the express curvature tensors in a simple form. This new form of the metric allowed the geodesics of the spacetime to be calculated in a simple manner. We calculated a large array of geodesics, the now simple rain geodesics, on-axis geodesics, general equatorial geodesics, ISCOs and photon orbits. We also found a non-trivial Killing tensor for this spacetime, equation \eqref{PGLT_Killing_Tensor}, which allowed us to generate the Carter constant, the fourth constant of the motion. Now that we had 4 constants of the motion, this made the geodesic equations integrable. Hence, we were able to give more explicit results for general geodesics of the spacetime. Looking at more specific examples, such as geodesics with Carter constant zero and general circular orbits, allowed us to give fully explicit geodesics.\\

\noindent Overall, the work given herein has many interesting mathematical and physical applications. Firstly, unit-lapse spacetimes are quite common and occur rather naturally for many specific analogue spacetimes \cite{unexpected, visser:1997, visser:1998, volovik:1999, stone:2001, visser:2001, fischer:2002, novello:2002, probing, Visser:vortex, LRR, Liberati:2005, Weinfurtner:2005, visser:2010, Visser:2013, Liberati:2018, Schuster:2018}. In the context of an analogue spacetime, the unit-lapse condition physically corresponds to constant signal propagation speed (for example, this holds to a high degree of approximation for sound waves in water). Analogue spacetimes are physically interesting as they give good physical intuition into more complex spacetimes, for example they can be used to help model infall and accretion. In an astrophysical context, unit-lapse forms of stationary spacetimes are rather useful since they allow for very simple and immediate calculation of a large class of timelike geodesics, the rain geodesics. Physically these geodesics represent zero angular momentum observers (ZAMOs), with zero initial velocity that are dropped from spatial infinity and are a rather tractable probe of the physics occurring in the spacetime. Mathematically, improved coordinate systems of the Kerr spacetime are rather important since they give a better understanding of the rather complicated and challenging Kerr spacetime, see the discussion in reference \cite{kerr-intro}. These improved coordinate systems, for example, can be applied to the attempts at finding a ``Gordon form" of the Kerr spacetime \cite{Gordon-form} and can also be applied to attempts at upgrading the ``Newman-Janis trick" from an ansatz to a full algorithm \cite{ansatz}. Also, these new forms of the Kerr metric allows for a greater observational ability to differentiate exact Kerr black holes from ``black hole mimickers" \cite{small-dark-heavy, BH-in-GR, phenomenology, viability, geodesicaly-complete, pandora, causal, LISA, Simpson:2020, Simpson:2019, Simpson:2018, Boonserm:2018, Simpson:2019-core,  
Berry:2020-core-1, Berry:2020-core-2, Bardeen:1968, Hayward:2005, Frolov:2014, Frolov:2017}. 

The Painlev\'{e}-Gullstrand form of the Lense-Thirring spacetime metric is a very tractable model for the vacuum regions outside rotating stars and planets. Furthermore, if we make the replacement $m\rightarrow m(r)$, that is we have
\be 
ds^2=-dt^2+\left(dr+\sqrt{\frac{2m(r)}{r}}\, dt\right)^2+r^2\left( d\theta^2+\sin^2(\theta)\left( d\phi-\frac{2J}{r^3}\, dt\right)^2\right)
\ee 
the Lense-Thirring metric can be generalised to model spherically symmetric dark matter halos and can hence be used to approximately model the gravitational fields of spiral galaxies. Given that the Lense-Thirring spacetime can be applied to many astrophysically interesting situations, and also given that the Painlev\'{e}-Gullstrand form of the Lense-Thirring spacetime is rather easy to work with (compared to the Kerr solution), for some astrophysically interesting situations, it may prove rather useful to use the Painlev\'{e}-Gullstrand form of the Lense-Thirring spacetime as opposed to using the full Kerr solution.




\appendix
\chapter{Killing vectors and tensors}\label{C:Killing}

Killing vectors and Killing tensors are incredibly useful mathematical objects that are present in all spacetimes. Knowledge of these objects is paramount in order to conduct further analysis of these spacetimes, a good example of this is calculating the ISCOs of various spacetimes. We start by defining some maps on manifolds, then relate these to the symmetries present in the spacetime and then construct the Killing vector from this symmetry map and discuss its properties.\\

\noindent Let $\M$ and $\mathcal{N}$ be manifolds. A map $\phi:\M \rightarrow \mathcal{N}$ is a diffeomorphism iff it is a $C^{\infty}$ map, 1-to-1, onto and has a $C^{\infty}$ inverse map $\phi^{-1}$. Consider a one-parameter group of diffeomorphisms, that is, the map $\phi_t: \R \times \M \rightarrow \M$. For a fixed point $p\in \M$, $\phi_t(p): \R \rightarrow \M$ is a curve, which we call an orbit of $\phi_t$ which passes though $p$ at $t=0$. We then define $v|_p$ to be the tangent vector to this curve at $t=0$. Hence, we see that every diffeomorphism in our spacetime generates a vector field $v$.\\

\noindent A diffeomorphism is an isometry if $\phi g_{ab}=g_{ab}$, i.e. if the metric is invariant under the action of the diffeomorphism $\phi$, then $\phi$ is an isometry. Naturally, we can view isometries as symmetry maps on our spacetime. For example, in flat spacetime we have a set of ten symmetry maps. We have four from the set of translations (one for each coordinate: $t$, $x$, $y$, $z$ if we are using Cartesian coordinates), three for rotations and three for boosts (or Lorentz transformations), which defines the Poincar\'{e} group of symmetry transformations in flat spacetime. In general relativity, we deal with spacetimes which may have symmetries, but usually not all symmetries in the Poincar\'{e} group. Above, we saw that every diffeomorphism in our spacetime generates a vector field, hence every isometry in our spacetime generates a vector field, which we call a Killing vector field, $\xi^a$. From the condition $\phi g_{ab}=g_{ab}$, the necessary and sufficient condition for $\xi^a$ to be a Killing vector is if it satisfies Killing's equation
\be 
\n_a\xi_b+\n_b\xi_a=0
\ee 
or equivalently
\be \label{Kill_Eqn}
\n_{(a}\xi_{b)}=0 \quad .
\ee
An extremely useful property of Killing vectors is that they generate conserved quantities.\\

\noindent\textbf{Proposition A.1} \\

\noindent Let $\xi^a$ be a Killing vector and let $\gamma$ be a geodesic with tangent vector $T^a$. Then $\xi_aT^a$ is constant along $\gamma$.\\

\noindent \emph{Proof}\\
We have
\be
T^b\n_b(\xi_aT^a)=T^bT^a\n_b\xi_a+\xi_aT^b\n_bT^a=0 \quad .
\ee
The first term is zero due to Killing's equation, equation \eqref{Kill_Eqn}, while the second term is zero due to the geodesic equation, equation \eqref{Geo_Eqn_Kill}.\\
$\square$\\

\noindent We can think of Killing vectors generating conserved quantities as Noether's theorem applied to general relativity. Some examples of conserved quantities are: energy, which arises from time-translation symmetry and generates the following Killing vector $\xi^a=(-1;0,0,0)$ and angular momentum which arises from spherical or axial symmetry and generates the Killing vector $\psi^a=(0;0,0,1)$.\\

\noindent We can generalise the notion of Killing vectors to Killing tensors. A Killing tensor is a symmetric tensor of type $(0, l)$ which satisfies the following condition
\be \label{Killing_Tensor_Def}
\n_{(b}K_{a_1...a_l)}=0 \quad .
\ee
Similar to Killing vectors, Killing tensors also give rise to conserved quantities. \\

\noindent\textbf{Proposition A.2} \\

\noindent Let $K_{a_1...a_l}$ be a Killing tensor and let $\gamma$ be a geodesic with tangent vector $T^a$. Then $K_{a_1...a_l}T^a...T^{a_l}$ is constant along $\gamma$.\\

\noindent \emph{Proof}\\
We have
\be
\begin{split}
T^b\n_b(K_{a_1...a_l}T^{a_1}...T^{a_l})= & T^{a_1}...T^{a_l}T^b\n_bK_{a_1...a_l}+K_{a_1...a_l}T^{a_2}...T^{a_l}T^b\n_bT^{a_1} \\
& +\sum_{i=2}^{l-1}K_{a_1...a_l}T^{a_1}...T^{a_{i-1}}T^{a_{i+1}}...T^{a_l}T^b\n_bT^{a_i} \\
& +K_{a_1...a_l}T^{a_1}...T^{a_{l-1}}T^b\n_bT^{a_l} 
\end{split}
\ee
The first term is zero via equation \eqref{Killing_Tensor_Def}, while the remaining terms are zero due to the geodesic equation, equation \eqref{Geo_Eqn_Kill}.\\
$\square$\\

\noindent However, while Killing tensors are similar to Killing vectors in this respect, Killing tensors do not naturally arise from isometries in our spacetime. However, the existence of a Killing tensor in a spacetime gives rise to other conserved quantities in the spacetime which is extremely useful. For example this allows us to explicitly integrate the geodesic equations in rather complex spacetimes.\\

\noindent Since Killing tensors do not naturally arise from isometries in a spacetime, there exists no algorithm for generating non-trivial Killing tensors for any general spacetime. However, if the inverse metric of a spacetime can be put into a particular form, then a non-trivial Killing tensor can be generated. As shown in \cite{PK:2018, PK:2020}, if the inverse metric of an axisymmetric spacetime can be written in the following form
\be \label{PK_Inverse_Metric}
g^{ab}=\frac{1}{A_1(r)+B_1(\theta)}
\begin{bmatrix}
A_2(r) & 0 & 0 & 0 \\
0 & B_2(\theta) & 0 & 0 \\
0 & 0 & A_3(r)+B_3(\theta) & A_4(r)+B_4(\theta) \\
0 & 0 & A_4(r)+B_4(\theta) & A_5(r)+B_5(\theta)
\end{bmatrix}^{ab}
\ee
note that here we are now using the following coordinate system $(r, \theta, \phi, t)$. Then, the corresponding contravariant non-trivial Killing tensor is then given by
\be \label{PK_Killing_tensor}
K^{ab}=\frac{1}{A_1(r)+B_1(\theta)}
\begin{bmatrix}
A_2(r)B_1(\theta) & 0 & 0 & 0 \\
0 & -B_2(\theta)A_1(r) & 0 & 0 \\
0 & 0 & K^{33} & K^{34} \\
0 & 0 & K^{34} & K^{44}
\end{bmatrix}^{ab}
\ee
where 
\be
K^{33}=B_1(\theta)A_3(r)-A_1(r)B_3(\theta)
\ee
\be
K^{34}=B_1(\theta)A_4(r)-A_1(r)B_4(\theta)
\ee
\be
K^{44}=B_1(\theta)A_5(r)-A_1(r)B_5(\theta) \quad .
\ee
Hence, to find a non-trivial Killing tensor for general spacetime, we first look if the inverse metric can be recast in the form given by equation \eqref{PK_Inverse_Metric}. If so, then we generate the non-trivial Killing tensor given by \eqref{PK_Killing_tensor}. Once given the contravariant Killing tensor, we can lower the indices via the metric and then recast the Killing tensor in any coordinate system via equation \eqref{ten_trans}. This is the approach we used when generating the non-trivial Killing tensor of the Painlev\'{e}-Gullstrand form of the Lense-Thirring spacetime.

\chapter{Publications}

\begin{enumerate}[label={[\arabic*]}]
  \item Joshua Baines, Thomas Berry, Alex Simpson and Matt Visser,\\
  ``Painlev\'{e}-Gullstrand form of the Lense-Thirring spacetime",\\
  $\left.\right.$[\href{https://arxiv.org/abs/2006.14258v2}{arXiv:2006.14258v2 [gr-qc]}]
  \item Joshua Baines, Thomas Berry, Alex Simpson and Matt Visser,\\
  ``Unit-lapse versions of the Kerr spacetime",\\
  Class. Quant. Grav. \textbf{38} (2021) 5, 055001 \href{https://doi.org/10.1088/1361-6382/abd071}{doi:10.1088/1361-6382/abd071} [\href{https://arxiv.org/abs/2008.03817}{arXiv:2008.03817 [gr-qc]}]
  \item Joshua Baines, Thomas Berry, Alex Simpson and Matt Visser,\\
  ``Darboux diagonalization of the spatial 3-metric in Kerr spacetime",\\
  Gen.Rel.Grav. \textbf{53} (2021) 1, 3 \href{https://doi.org/10.1007/s10714-020-02765-0}{doi:10.1007/s10714-020-02765-0}\\ $\left.\right.$[\href{https://arxiv.org/abs/2009.01397}{arXiv:2009.01397 [gr-qc]}]
\end{enumerate}


\end{document}